\documentclass[twocolumn]{aastex63}

\usepackage[T1]{fontenc}
\usepackage[utf8]{inputenc}
\usepackage{multirow}
\usepackage{xspace}
\usepackage{grffile}
\usepackage{xstring}

\graphicspath{{./}{figures/}}

\newcommand{\kms}{km\,s$^{-1}$\xspace}

\newcommand{\pcm}[1]{cm$^{-#1}$\xspace}

\newcommand{\mum}{$\mu$m\xspace}
\newcommand{\Msun}{M$_\odot$\xspace}
\newcommand{\Msunyr}{M$_\odot$\,yr$^{-1}$\xspace}

\newcommand{\xclass}{\textsc{xclass}\xspace}
\renewcommand{\arcsec}{\ensuremath{^{\prime\prime}}\xspace}

\newcommand{\co}{CO(3--2)\xspace}
\newcommand{\hcn}{HCN(4--3)\xspace}
\newcommand{\hco}{HCO$^+$(4--3)\xspace}
\newcommand{\cs}{CS(7--6)\xspace}

\newcommand{\Leroy}[1]{\IfEqCase{#1}{
        {p}{\citepalias{2018ApJ...869..126L}\xspace}
        {t}{\citetalias{2018ApJ...869..126L}\xspace}
    }[\PackageError{leroy}{Undefined option to Leroy: #1}{}]
}

\received{-}
\revised{-}
\accepted{10 June 2020}
\submitjournal{ApJ}

\shorttitle{Molecular ISM in the SSCs of NGC~253}
\shortauthors{Krieger et al.}

\begin{document}

\title{The Molecular ISM in the Super Star Clusters of the Starburst NGC~253}

\correspondingauthor{Nico Krieger}
\email{krieger@mpia.de}

\author[0000-0003-1104-2014]{Nico Krieger}
    \affiliation{Max-Planck-Institut f\"ur Astronomie, K\"onigstuhl 17, 69120 Heidelberg, Germany}
\author{Alberto D. Bolatto}
    \affiliation{Department of Astronomy, University of Maryland, College Park, MD 20742, USA}
\author{Adam K. Leroy}
    \affiliation{Department of Astronomy, The Ohio State University, 4055 McPherson Laboratory, 140 West 18th Ave, Columbus, OH 43210, USA}
\author{Rebecca C. Levy}
    \affiliation{Department of Astronomy, University of Maryland, College Park, MD 20742, USA}
\author{Elisabeth A.C. Mills}
    \affiliation{Physics Department, Brandeis University, 415 South Street, Waltham, MA 02453}
\author{David S. Meier}
    \affiliation{New Mexico Institute of Mining and Technology, 801 Leroy Place, Socorro, NM 87801, USA}
    \affiliation{National Radio Astronomy Observatory, P.O. Box O, 1003 Lopezville Road, Socorro, NM 87801, USA}
\author{J\"urgen Ott}
    \affiliation{National Radio Astronomy Observatory, P.O. Box O, 1003 Lopezville Road, Socorro, NM 87801, USA}
\author{Sylvain Veilleux}
    \affiliation{Department of Astronomy, University of Maryland, College Park, MD 20742, USA}
\author{Fabian Walter}
    \affiliation{Max-Planck-Institut f\"ur Astronomie, K\"onigstuhl 17, 69120 Heidelberg, Germany}
    \affiliation{National Radio Astronomy Observatory, P.O. Box O, 1003 Lopezville Road, Socorro, NM 87801, USA}
\author{Axel Wei\ss}
    \affiliation{Max-Planck-Institut f\"ur Radioastronomie, Auf dem H\"ugel 69, 53121 Bonn, Germany}

%% Note that the \and command from previous versions of AASTeX is now
%% depreciated in this version as it is no longer necessary. AASTeX 
%% automatically takes care of all commas and "and"s between authors names.

%% AASTeX 6.2 has the new \collaboration and \nocollaboration commands to
%% provide the collaboration status of a group of authors. These commands 
%% can be used either before or after the list of corresponding authors. The
%% argument for \collaboration is the collaboration identifier. Authors are
%% encouraged to surround collaboration identifiers with ()s. The 
%% \nocollaboration command takes no argument and exists to indicate that
%% the nearby authors are not part of surrounding collaborations.

\begin{abstract}
We present submillimeter spectra of the (proto-)super star cluster (SSC) candidates in the starbursting center of the nearby galaxy NGC~253 identified by \citet{2018ApJ...869..126L}. The 2.5\,pc resolution of our ALMA cycle 3 observations approach the size of the SSCs and allows the study of physical and chemical properties of the molecular gas in these sources. 
In the 14 SSC sources and in the frequency ranges $342.0-345.8$\,GHz and $353.9-357.7$\,GHz we detect 55 lines belonging to 19 different chemical species. The SSCs differ significantly in chemical complexity, with the richest clusters showing 19 species and the least complex showing 4 species. We detect HCN isotopologues and isomers (H$^{13}$CN, HC$^{15}$N, H$^{15}$NC), abundant HC$_3$N, SO and S$^{18}$O, SO$_2$, and H$_2$CS. The gas ratios CO/HCN, CO/HCO$^+$ are low, $\sim 1-10$, implying high dense gas fractions in the SSCs.
Line ratio analyses suggests chemistry consistent with photon-dominated regions and mechanical heating. 
None of the SSCs near the galaxy center show line ratios that imply an X-ray dominated region, suggesting that heating by any (still unknown) AGN does not play a major role.
The gas temperatures are high in most sources, with an average rotational temperature of $\sim 130$\,K in SO$_2$. 
The widespread existence of vibrationally excited HCN and HC$_3$N transitions implies strong IR radiation fields, potentially trapped by a greenhouse effect due to high continuum opacities.
\end{abstract}


%%%%%%%%%%%%%%%%%%%%%%%%%%%%%%%%%%%%%%%%%%%%%%%%%%%%%%%%%%%%%%%%%%%%%%%%%%%%%%%%%%%%%%%%%%%%%%%%%%%%

\section{Introduction} \label{section: introduction}

\defcitealias{2018ApJ...869..126L}{L18}

% SSCs
Super star clusters (SSCs) are massive ($\mathrm{M}_* > 10^5$\,\Msun), compact ($\mathrm{R} \sim 1$\,pc) clusters of stars. They are frequently found in starbursts in the centers of galaxies or galaxy mergers, such as M~82, NCG~253 or the Antennae galaxies \citep[e.g.][]{1992AJ....103..691H,2003dhst.symp..153W,2005ApJ...621..278M,2010ARA&A..48..431P,2018ApJ...869..126L}. The stellar properties of SSCs are similar to Galactic globular clusters and thus SSCs might represent a younger generation of the same sort of stellar systems \citep[e.g.][]{2001ApJ...554L..29G,2010ARA&A..48..431P}. The extreme conditions under which SSCs form are rare in the present-day universe but are thought to be common around the peak of the cosmic star formation rate history, the era when most of today's globular clusters formed. Hence, observations of forming SSCs might offer a glimpse into the physics of a mode of star formation common in the early universe.

% NGC~253
\object{NGC253} is one of the nearest starburst systems, at a distance of 3.5\,Mpc \citep{Rekola:2005ha}. It is considered one of the prototypical starburst galaxies, with a star formation rate (SFR) of $\sim 2$\,\Msunyr in its center \citep{Ott:2005il,Leroy:2015ds,2015MNRAS.450L..80B}. NGC~253 has a prominent bar that feeds gas to the nuclear starburst \citep{2000PASJ...52..785S,2004ApJ...611..835P}. The gas flows lead to intense star formation, which creates feedback driving outflows that have been detected across the spectrum from X-ray to radio wavelengths in ionized, neutral and molecular gas \citep{Turner:1985iy,2000ApJS..129..493H,Strickland:2000wd,Strickland:2002kp,Sharp:2010jl,Sturm:2011jb, Westmoquette:2011bp,2013Natur.499..450B,2017ApJ...835..265W,2019ApJ...881...43K}. 

% SSCs in NGC~253
It has been known for a while now that NGC~253 hosts an SSC. \citet{Watson:1996dn} and \citet{Kornei:2009ee} detected a young, deeply embedded SSC in HST imaging of the nuclear region. Hints of further SSCs were discovered in radio observations \citep{1997ApJ...488..621U} but do not show obvious counterparts in optical or near-IR imaging \citep{2017ApJ...835..265W}. The massive and dense molecular clouds in NGC~253 seen in the millimeter and sub-millimeter \citep[e.g.][]{Sakamoto:2011et,Leroy:2015ds,2015ApJ...801...63M} provide an ideal environment for SSC formation. \citet{2017ApJ...849...81A} showed that massive star formation (SF) is indeed present in small ($<10$\,pc) gas clumps identified from ALMA observations. Utilizing even higher resolution ALMA observations, \citet[][hereafter \citetalias{2018ApJ...869..126L}]{2018ApJ...869..126L} characterized 14 proto-SSCs still deeply embedded in their natal gas and dust clouds. At least some of these SSCs are very young \citep[$<1$\,Myr;][]{2020MNRAS.491.4573R}, with many showing roughly equal, but still uncertain, masses of young stars and gas \Leroy{p}.

% observations
Due to its proximity, NGC~253 is an ideal target for high-resolution studies of the physics and chemistry of the star-forming gas in starbursts. Several studies of the chemical environment in NGC~253 have been carried out but none of them had the resolution to approach the scale of stellar clusters \citep[e.g.][]{2006ApJS..164..450M,2015ApJ...801...63M,2015A&A...579A.101A,2019ApJ...871..170M}. In this article, we utilize high-resolution (2.5\,pc, 0.15\arcsec) ALMA observations in band~7 ($\sim 350$\,GHz) to study the physical and chemical environment of individual embedded (proto-)SSC candidates. 

In deep imaging, we detect up to 19 molecular species with up to 55 spectral lines over 7.6\,GHz bandwidth ($342.0-345.8$\,GHz, $353.9-357.7$\,GHz) in each SSC.
Among the detected species are commonly used dense gas tracers, potential PDR tracers, optically thin isotopologues and vibrationally excited species. 
These spectral lines and the ratios among them are sensitive to the physical and chemical ISM properties. They depend on heating and cooling of the gas, the energy source, ionization or radiation properties such as the IR field through non-thermal excitation or radiative pumping. Together, the inferred properties determine the state of the natal gas clouds of the SSCs and the early feedback exerted on them.

% structure
This article is structured as follows: Section~\ref{section: data} describes the observations, data reduction and line identification. The procedure of fitting the spectra is laid out in Section~\ref{section: spectral line fitting}, which also shows the fitted spectra. The results are interpreted and discussed in Section~\ref{section: discussion}. We conclude with a summary (Section~\ref{section: summary}). Appendices list the details of spectral fitting, present the fitted spectra of all sources and list obtained quantities.

%%%%%%%%%%%%%%%%%%%%%%%%%%%%%%%%%%%%%%%%%%%%%%%%%%%%%%%%%%%%%%%%%%%%%%%%%%%%%%%%%%%%%%%%%%%%%%%%%%%%

\section{Data reduction}\label{section: data}

\subsection{Observations, calibration and Imaging}

Data reduction and imaging of our ALMA cycle~4 observations are described in detail in \citet{2019ApJ...881...43K}. Though that study presented only the narrow frequency range containing the \co line, we applied the same data reduction described there to the whole observed lower and upper sidebands (LSB, USB). The sidebands are tuned to $342.0-345.8$\,GHz (LSB) and $353.9-357.7$\,GHz (USB), yielding a combined 7.6\,GHz total bandwidth. 

The pipeline-calibrated and continuum-subtracted visibilities are imaged in 2.5\,MHz channels ($\sim 2.5$\,\kms at $\sim 350$\,GHz) using Briggs weighting (robust parameter of 0.5). The synthesized beam varies only slightly between LSB and USB ($\sim 2$\% linear deviation, $\sim 4$\% beam area deviation), so both bands are restored with the same beam of $0.17\arcsec \times 0.13\arcsec$. This offers the benefit of identical resolution in both bands. The per-channel noise is 0.37\,K.
% Synthesized beam sizes: LSB: $0.172294\arcsec \times 0.132516\arcsec$, USB: $0.1684115\arcsec \times 0.129755\arcsec$

Line crowding left us with limited bandwidth to fit the continuum, and as a result the initial continuum subtraction in the $u,v$ plane left a low level of residual continuum in the USB for some sources (SSCs~2, 5, 8, 10, 13, 14). To account for this, we also carry out am image-plane continuum subtraction for the affected spectra. As the line-free ranges are small in some sources in the USB, the errors in the continuum fits can be substantial, up to $\sim 0.2$~K. This can be relevant for our faintest detected lines, which have brightness $\lesssim 2$\,K. In this case, the uncertainty due to the continuum is of the same order as the uncertainty as the flux calibration. For brighter lines and the LSB, the uncertainty in the continuum subtraction is negligible.

\subsection{Spectra}\label{section: spectra}

We aim to constrain the physical and chemical properties of the SSCs identified by \Leroy{t}. They found the sizes of the SSCs to be of order the beam size (deconvolved sizes of $\sim 1.5-4$\,pc). Therefore, we extract single pixel ($0.1\arcsec \times 0.1\arcsec$) spectra at the recorded center positions.

For easier handling of spectra, we shift the spectral axis from the observed frequency to the rest frame frequency. The \cs line provides a good estimate for the systemic velocity of the SSCs. It is clearly detected in all sources and does not show complex line profiles. In some sources, \cs has multiple velocity components but one component clearly dominates. We use this dominant component as the reference. We estimate the velocity from a Gaussian fit and apply the Doppler correction to shift to the rest frequency. Our estimates of the source velocities are consistent within $<5$\,\kms with \Leroy{t}, who used an independently imaged version of the same observations. 

In SSC\,3, two peaks of close to equal peak intensity do not allow for a robust determination of the source velocity. There we obtain a combined estimate from the H$^{15}$NC, SO and H$^{13}$CN lines. Including these lines neither improves, nor worsens the velocity estimate for the other sources.

Note that the spectral fits described in section~\ref{section: spectral line fitting} still have the velocity centroid as a free parameter. This allows us to detect shifts of lines relative to each other.

Figure~\ref{figure: fitted bands} shows an overview of the spectra. Zoom-ins are given in Figure~\ref{figure: sample spectrum} and Figure Set 2.

\subsection{Line Identification}\label{section: line ID}
We identified the spectral lines in our spectra with the help of Splatalogue\footnote{\url{https://www.cv.nrao.edu/php/splat}} \citep{2007AAS...21113211R} using data from the Cologne Database for Molecular Spectroscopy \citep[CDMS\footnote{\url{https://cdms.astro.uni-koeln.de/}},][]{2005JMoSt.742..215M} and
Jet Propulsion Laboratory \citep[JPL,][]{1998JQSRT..60..883P} catalogs. Spectra centered on the bright continuum sources were examined independently by four investigators and guided by lower resolution observations in the literature. The lines consistently identified by all investigators form the final line list (Table~\ref{table: intensities}) that we use in this article.
We mark vibrationally excited spectral lines by an asterisk or give the exact vibrational state.

Note that our line identification is conservative. Some spectra show faint features not included as identified lines (cf. Figure~\ref{figure: sample spectrum}). We do not analyze these further in this paper.

\begin{figure*}
    \centering
    \includegraphics[height=0.9\textheight]{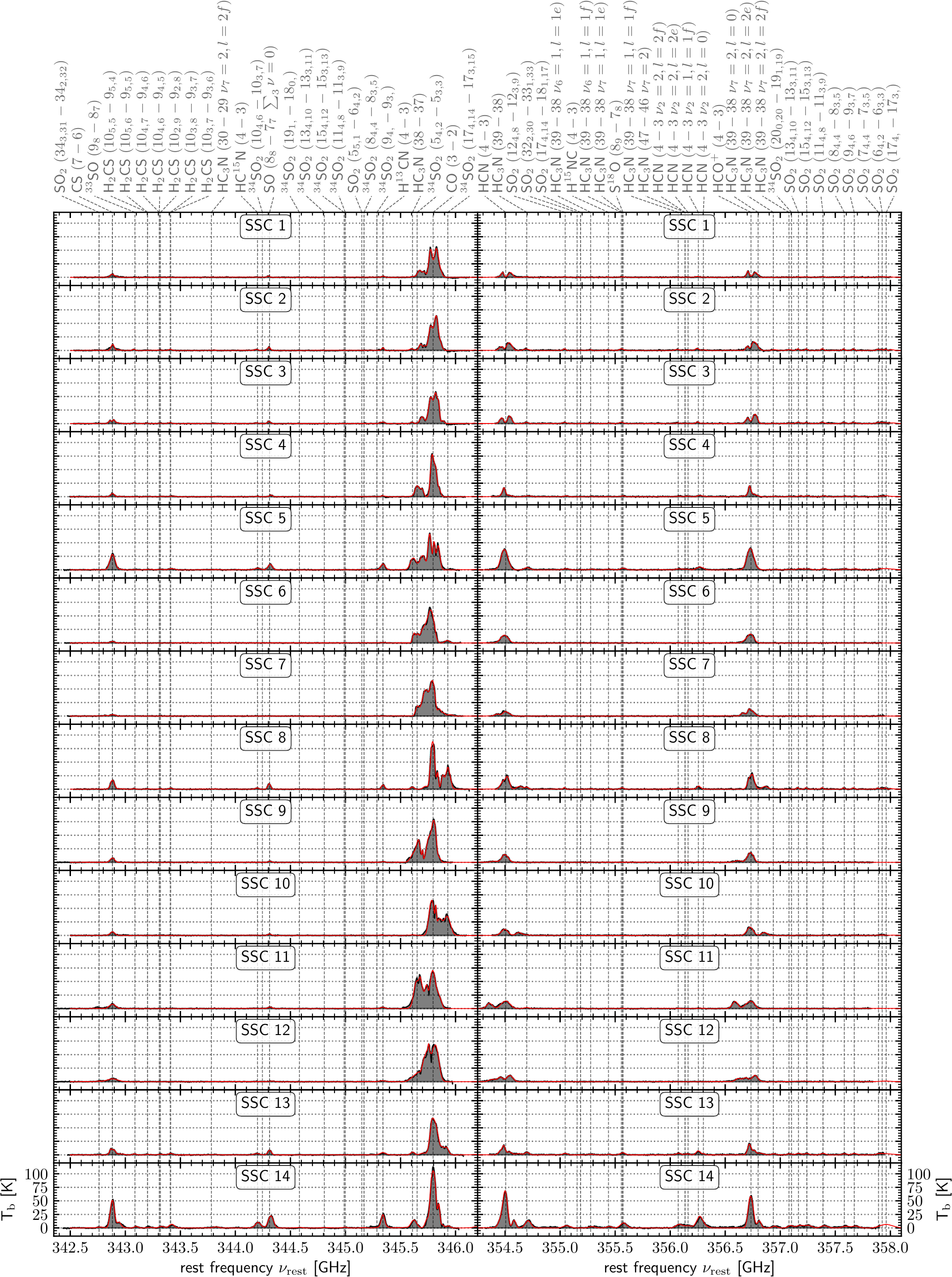}
    \caption{Overview and relative comparison of the spectra (black) and \xclass fits (red) of the 14 SSCs. Note that the frequency axis is presented as rest frequency to allow for easier interpretation of the spectra (cf. Section~\ref{section: spectra})}. The detected species are labeled at the top. A zoom-in of this figure is given in Figure Set 2 and Figure~\ref{figure: sample spectrum} for SSC~14.
    \label{figure: fitted bands}
\end{figure*}

\vspace{\baselineskip}
\newpage
% figure set
\figsetstart
\figsetnum{2}
\figsettitle{Spectra of the LSB (\emph{top}) and USB (\emph{bottom}) for the 14 SSCs. These plots show a zoom into Figure~\ref{figure: fitted bands}. The observed spectrum (black) sits on top of a grey band indicating the $16^\mathrm{th}$ to $84^\mathrm{th}$ percentiles of the noise added for error estimation in the fit (cf. Section~\ref{section: error estimation}). The red lines represent the median fit obtained by \xclass using the fitting procedure described in Section~\ref{section: spectral fitting procedure}. The Monte Carlo-estimated errors are shown as a red band ($16^\mathrm{th}$ to $84^\mathrm{th}$ percentiles) that is hardly visible due to its small size relative to the strong spectral lines.}
    
\figsetgrpstart
\figsetgrpnum{2.1}
\figsetgrptitle{SSC 1}
\figsetplot{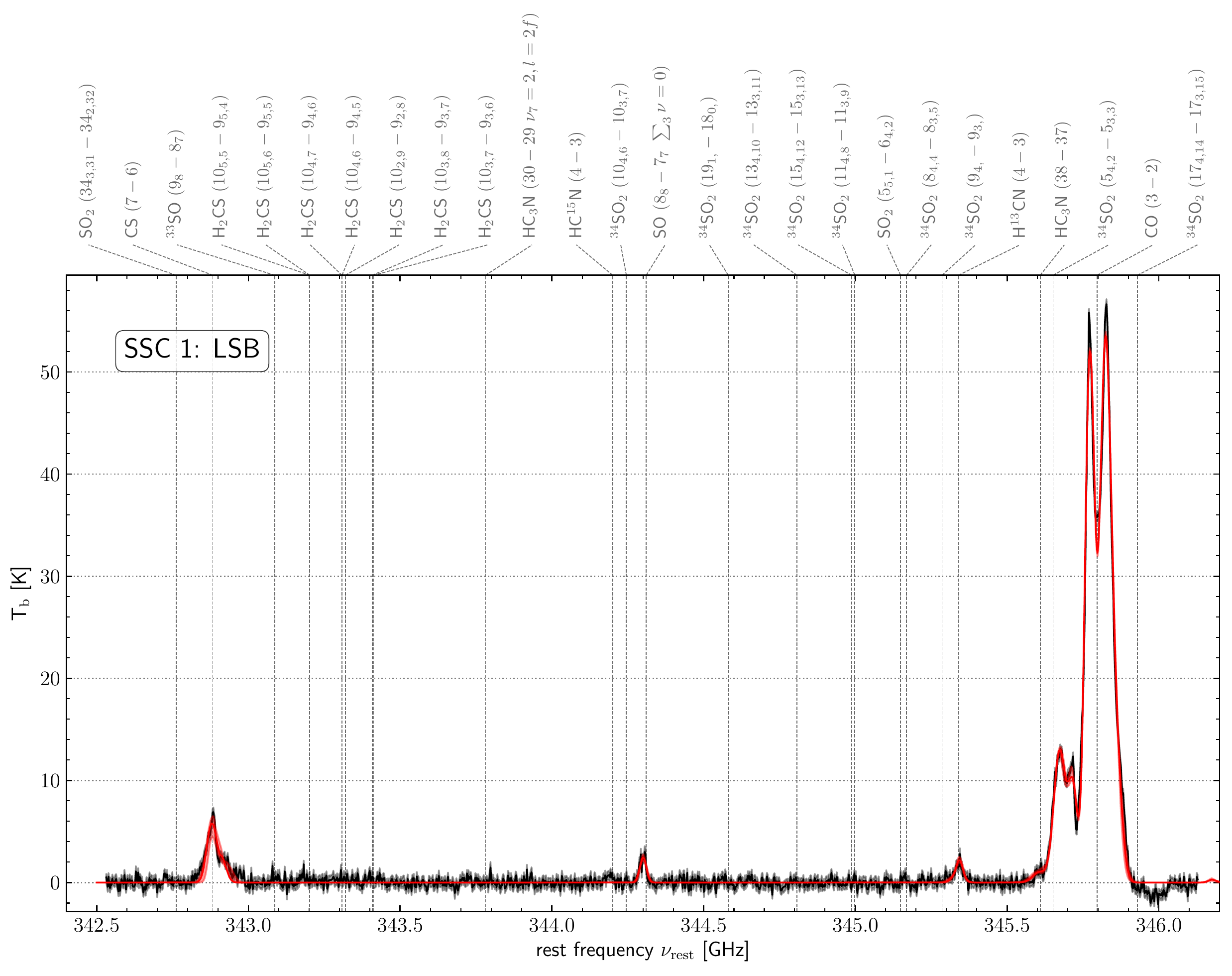}
\figsetplot{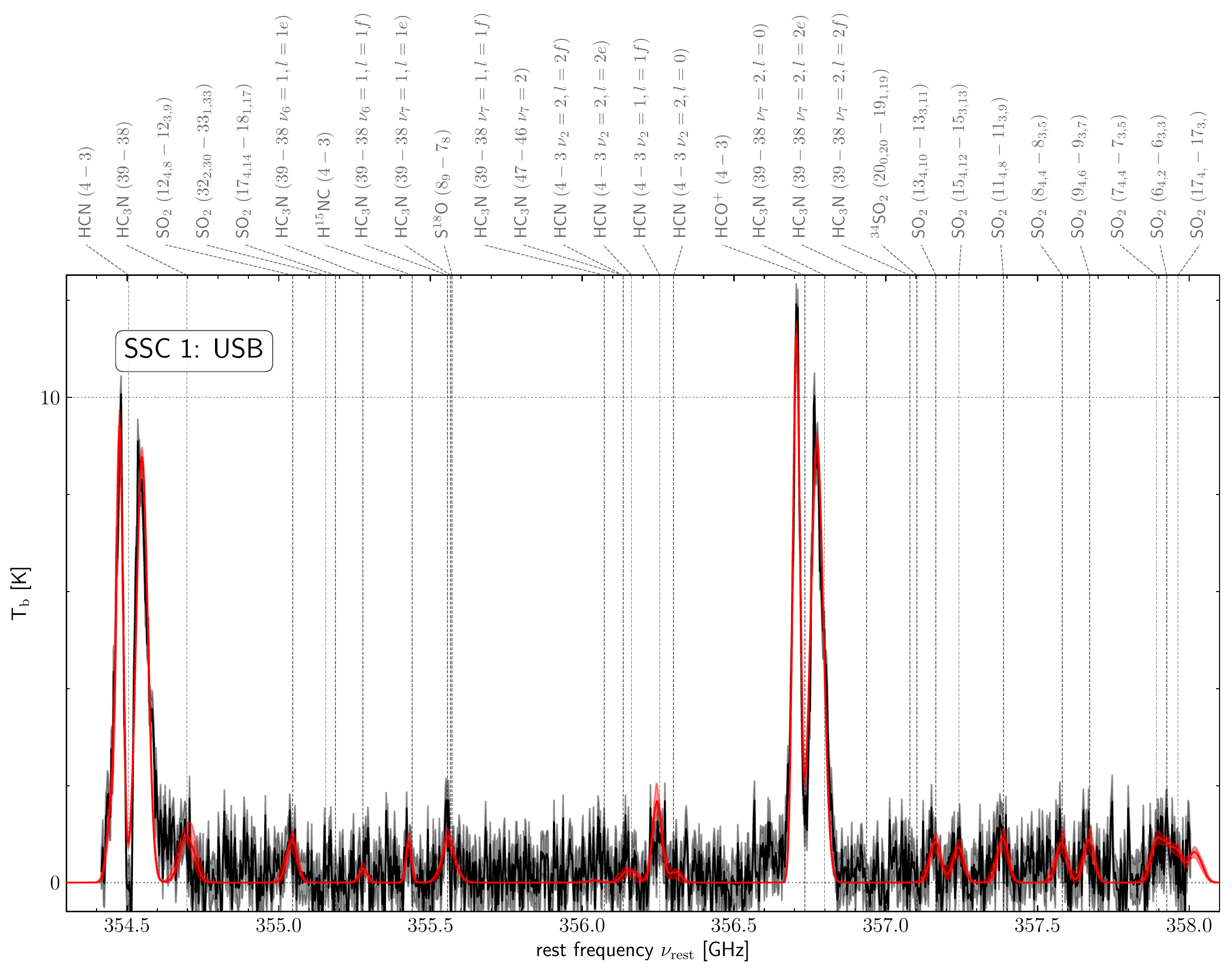}
\figsetgrpnote{Spectra of the LSB (\emph{top}) and USB (\emph{bottom}) in SSC~1. The observed spectrum (black) sits on top of a grey band indicating the $16^\mathrm{th}$ to $84^\mathrm{th}$ percentiles of the noise added for error estimation in the fit (cf. Section~\ref{section: error estimation})). The red lines represent the median fit obtained by \xclass using the fitting procedure described in Section~\ref{section: spectral fitting procedure}. The Monte Carlo-estimated errors are shown as a red band ($16^\mathrm{th}$ to $84^\mathrm{th}$ percentiles) that is hardly visible due to its small size relative to the strong spectral lines.}
\figsetgrpend
    
\figsetgrpstart
\figsetgrpnum{2.2}
\figsetgrptitle{SSC 2}
\figsetplot{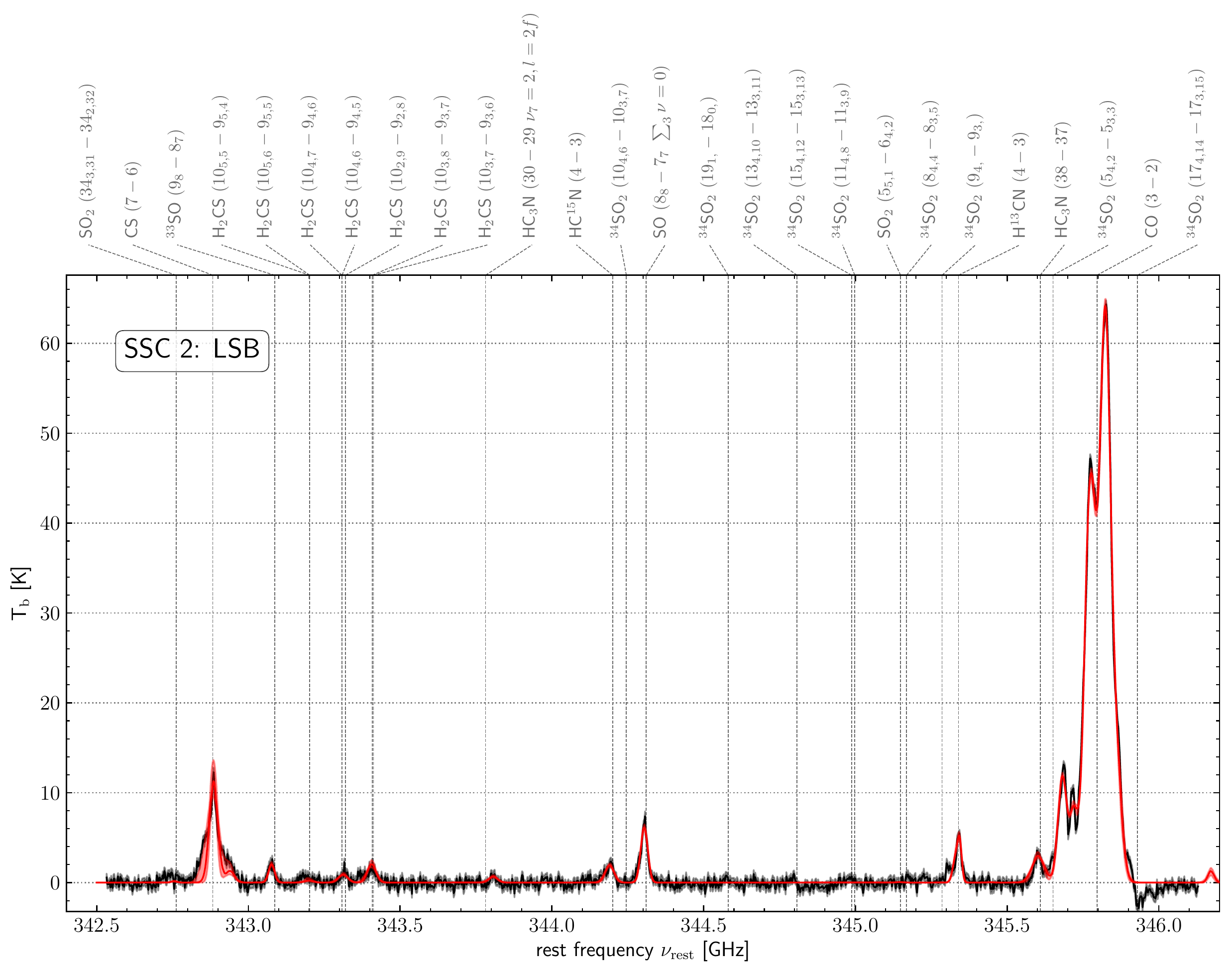}
\figsetplot{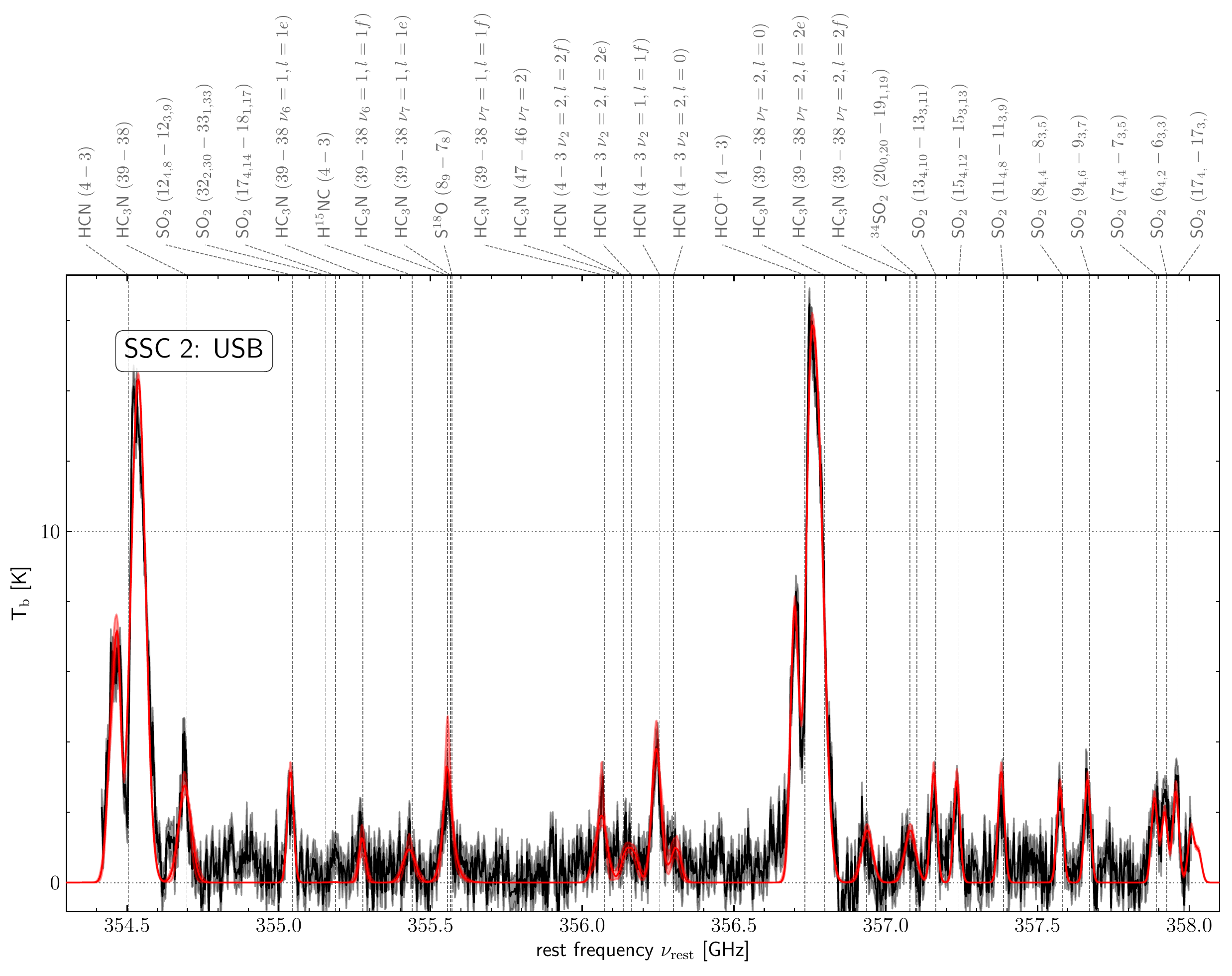}
\figsetgrpnote{Spectra of the LSB (\emph{top}) and USB (\emph{bottom}) in SSC~2. The observed spectrum (black) sits on top of a grey band indicating the $16^\mathrm{th}$ to $84^\mathrm{th}$ percentiles of the noise added for error estimation in the fit (cf. Section~\ref{section: error estimation})). The red lines represent the median fit obtained by \xclass using the fitting procedure described in Section~\ref{section: spectral fitting procedure}. The Monte Carlo-estimated errors are shown as a red band ($16^\mathrm{th}$ to $84^\mathrm{th}$ percentiles) that is hardly visible due to its small size relative to the strong spectral lines.}
\figsetgrpend

\figsetgrpstart
\figsetgrpnum{2.3}
\figsetgrptitle{SSC 3}
\figsetplot{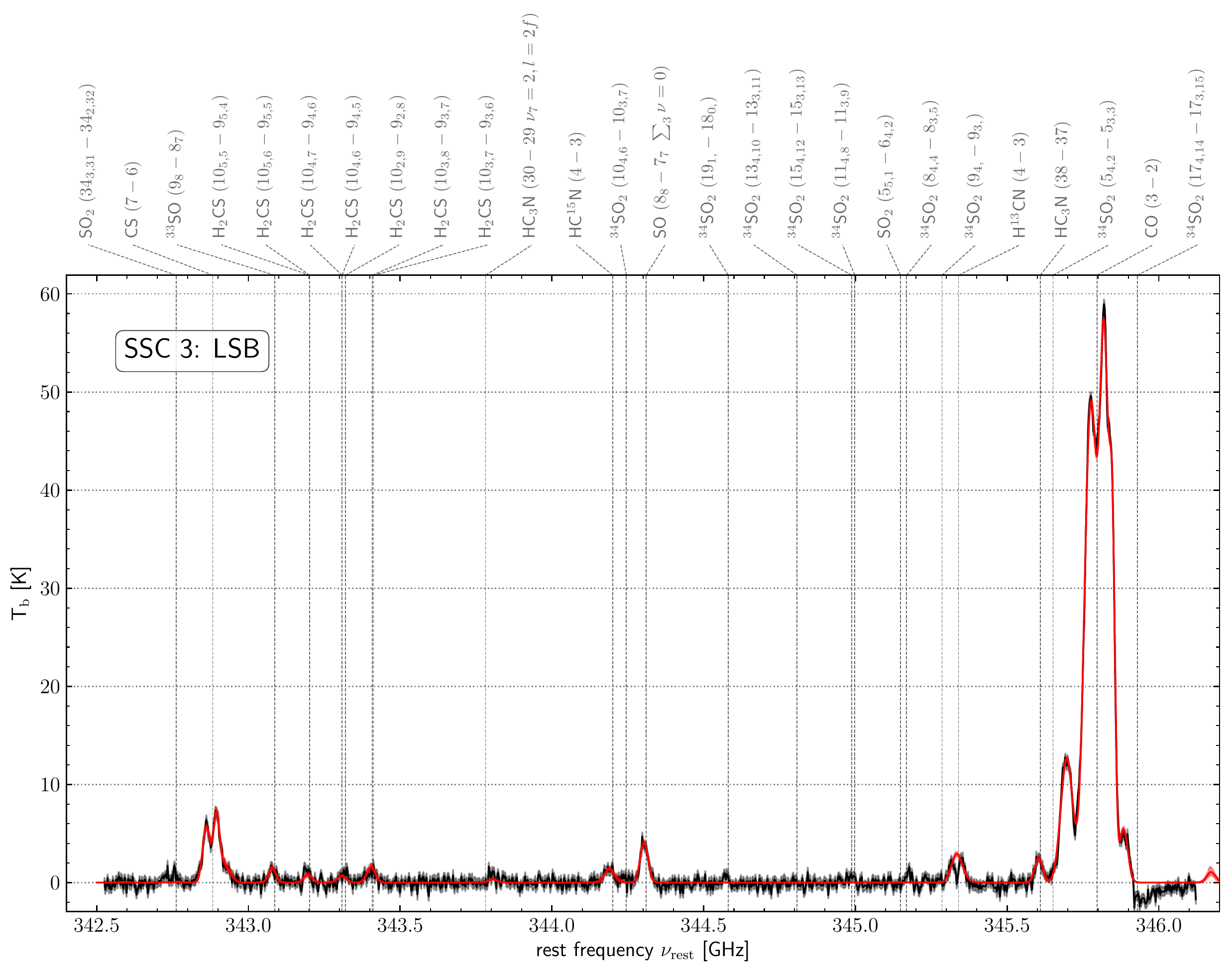}
\figsetplot{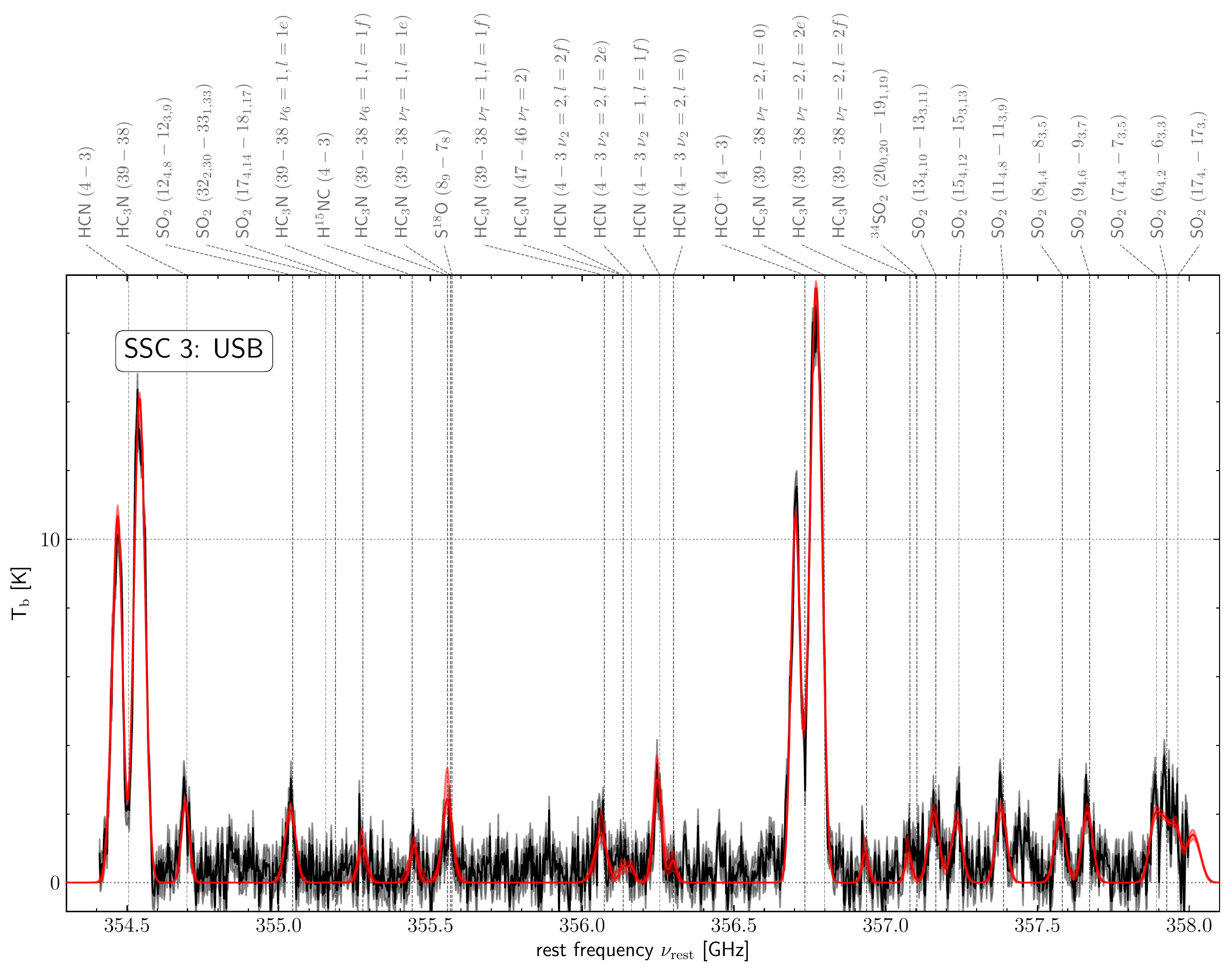}
\figsetgrpnote{Spectra of the LSB (\emph{top}) and USB (\emph{bottom}) in SSC~3. The observed spectrum (black) sits on top of a grey band indicating the $16^\mathrm{th}$ to $84^\mathrm{th}$ percentiles of the noise added for error estimation in the fit (cf. Section~\ref{section: error estimation})). The red lines represent the median fit obtained by \xclass using the fitting procedure described in Section~\ref{section: spectral fitting procedure}. The Monte Carlo-estimated errors are shown as a red band ($16^\mathrm{th}$ to $84^\mathrm{th}$ percentiles) that is hardly visible due to its small size relative to the strong spectral lines.}
\figsetgrpend

\figsetgrpstart
\figsetgrpnum{2.4}
\figsetgrptitle{SSC 4}
\figsetplot{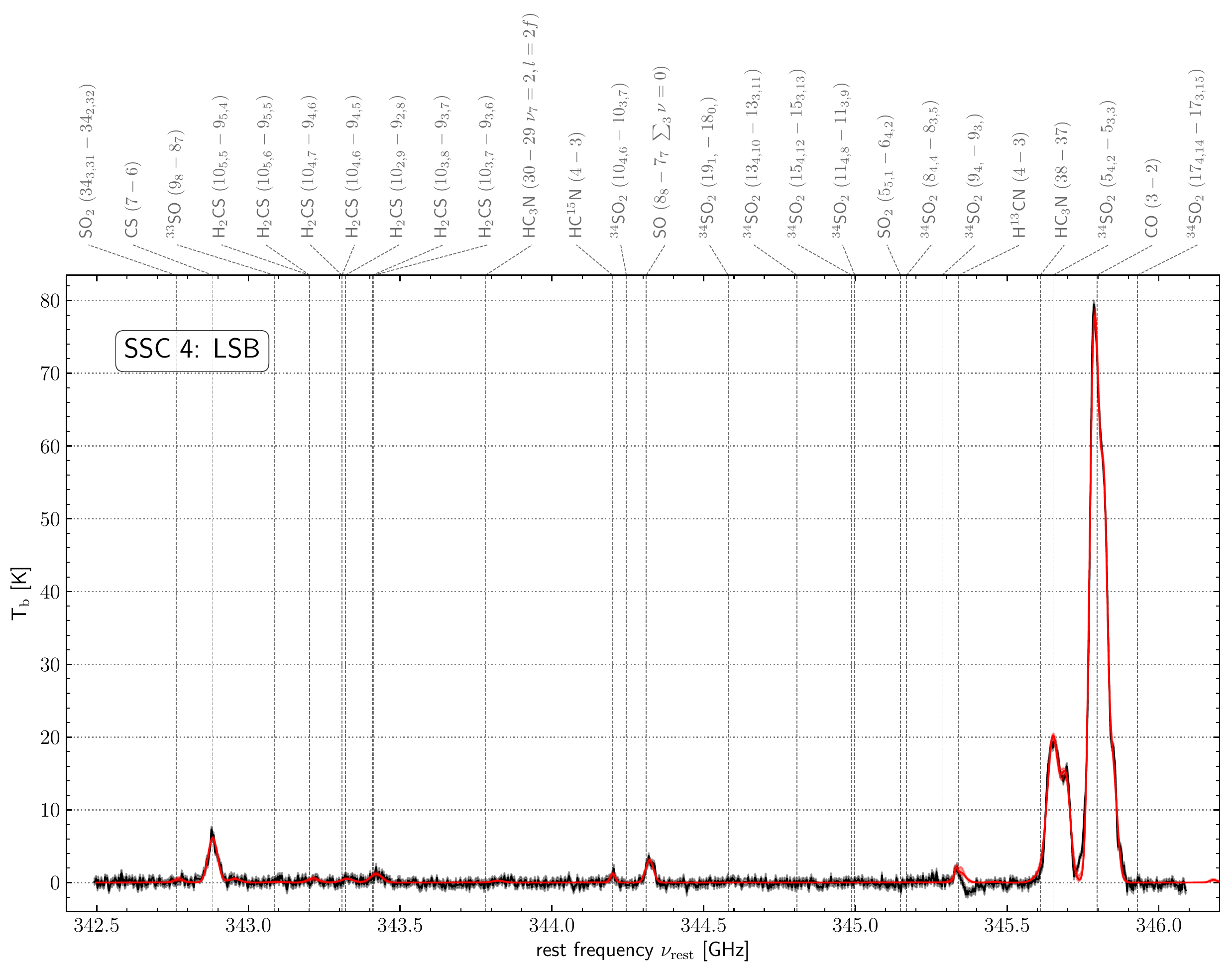}
\figsetplot{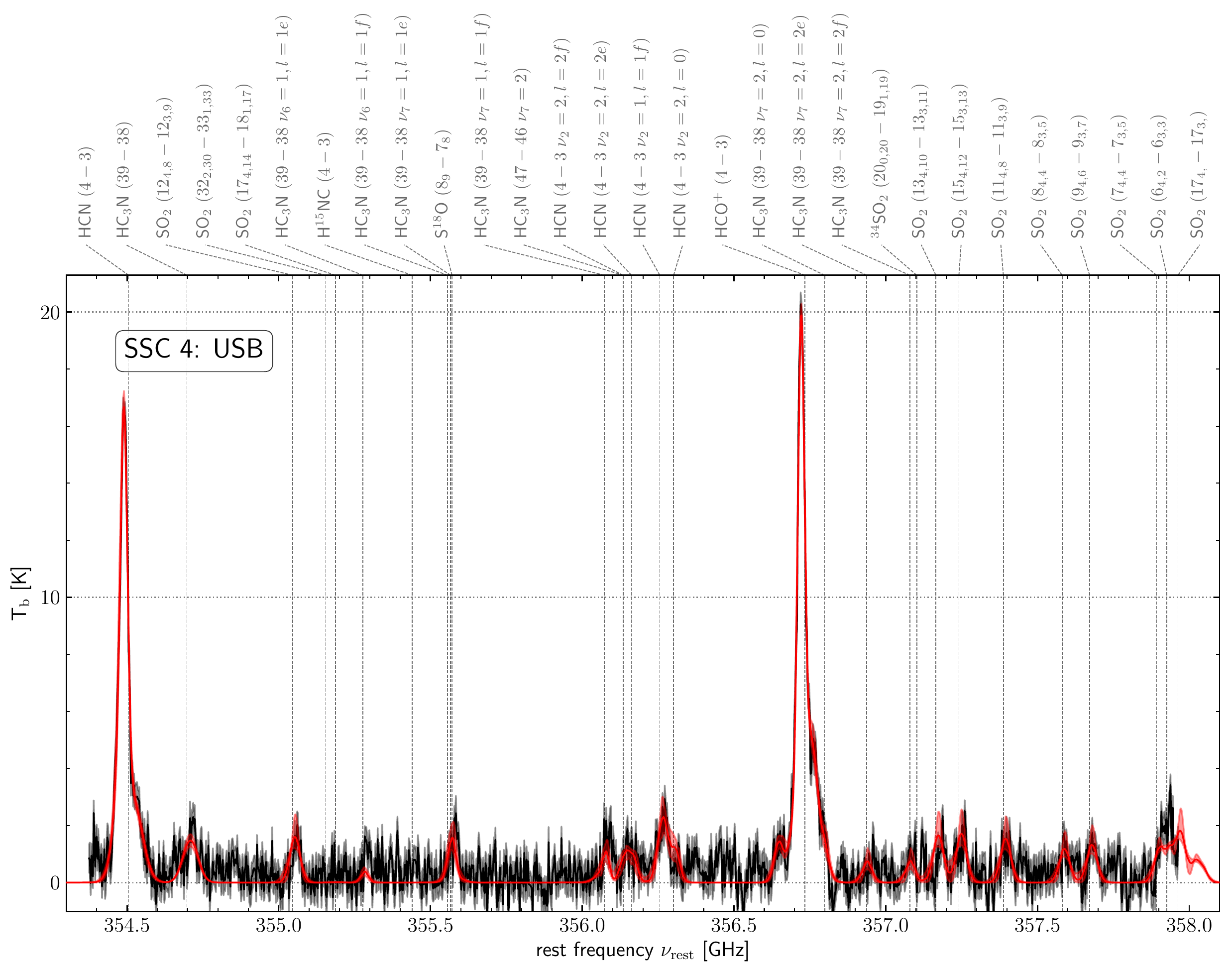}
\figsetgrpnote{Spectra of the LSB (\emph{top}) and USB (\emph{bottom}) in SSC~4. The observed spectrum (black) sits on top of a grey band indicating the $16^\mathrm{th}$ to $84^\mathrm{th}$ percentiles of the noise added for error estimation in the fit (cf. Section~\ref{section: error estimation})). The red lines represent the median fit obtained by \xclass using the fitting procedure described in Section~\ref{section: spectral fitting procedure}. The Monte Carlo-estimated errors are shown as a red band ($16^\mathrm{th}$ to $84^\mathrm{th}$ percentiles) that is hardly visible due to its small size relative to the strong spectral lines.}
\figsetgrpend

\figsetgrpstart
\figsetgrpnum{2.5}
\figsetgrptitle{SSC 5}
\figsetplot{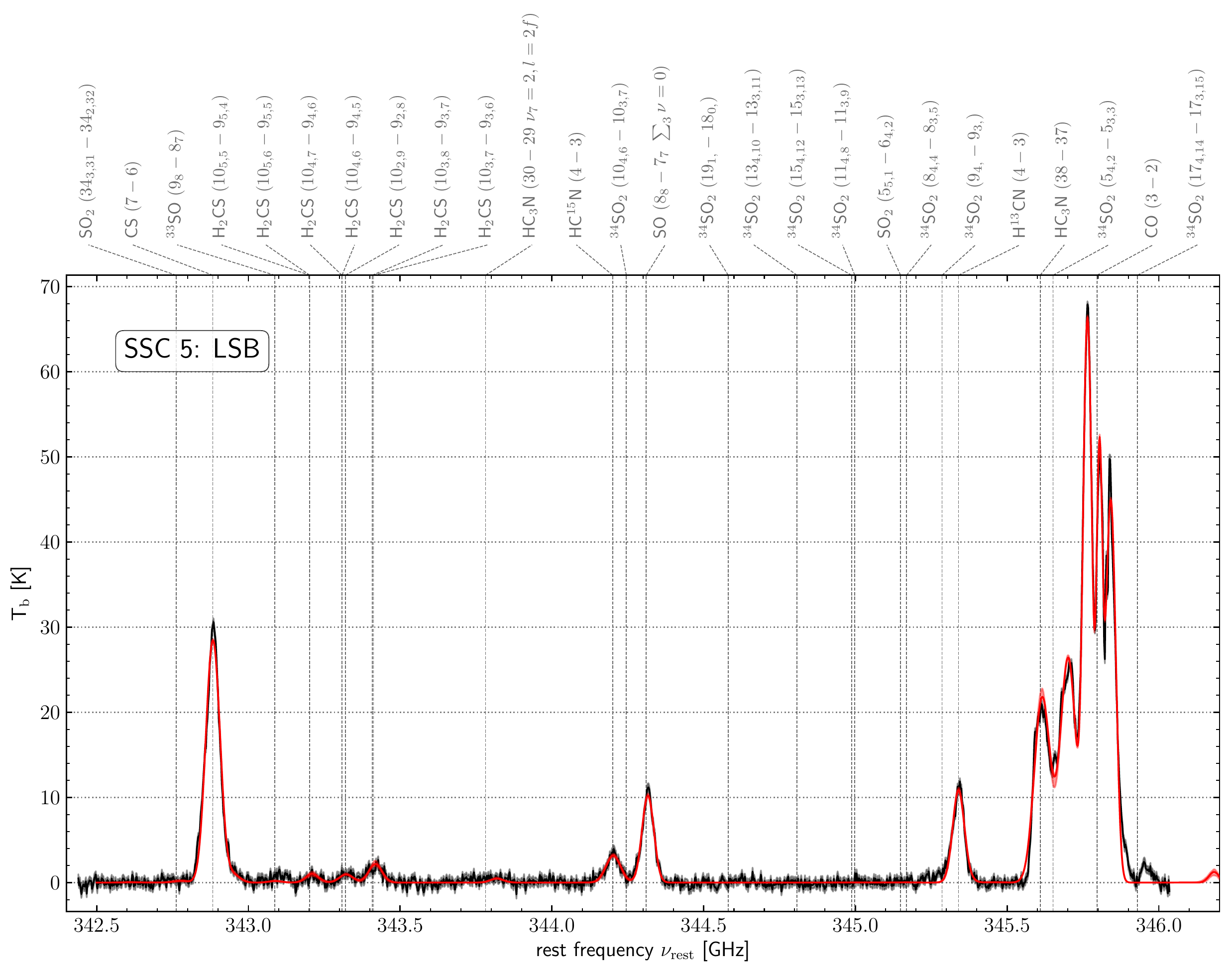}
\figsetplot{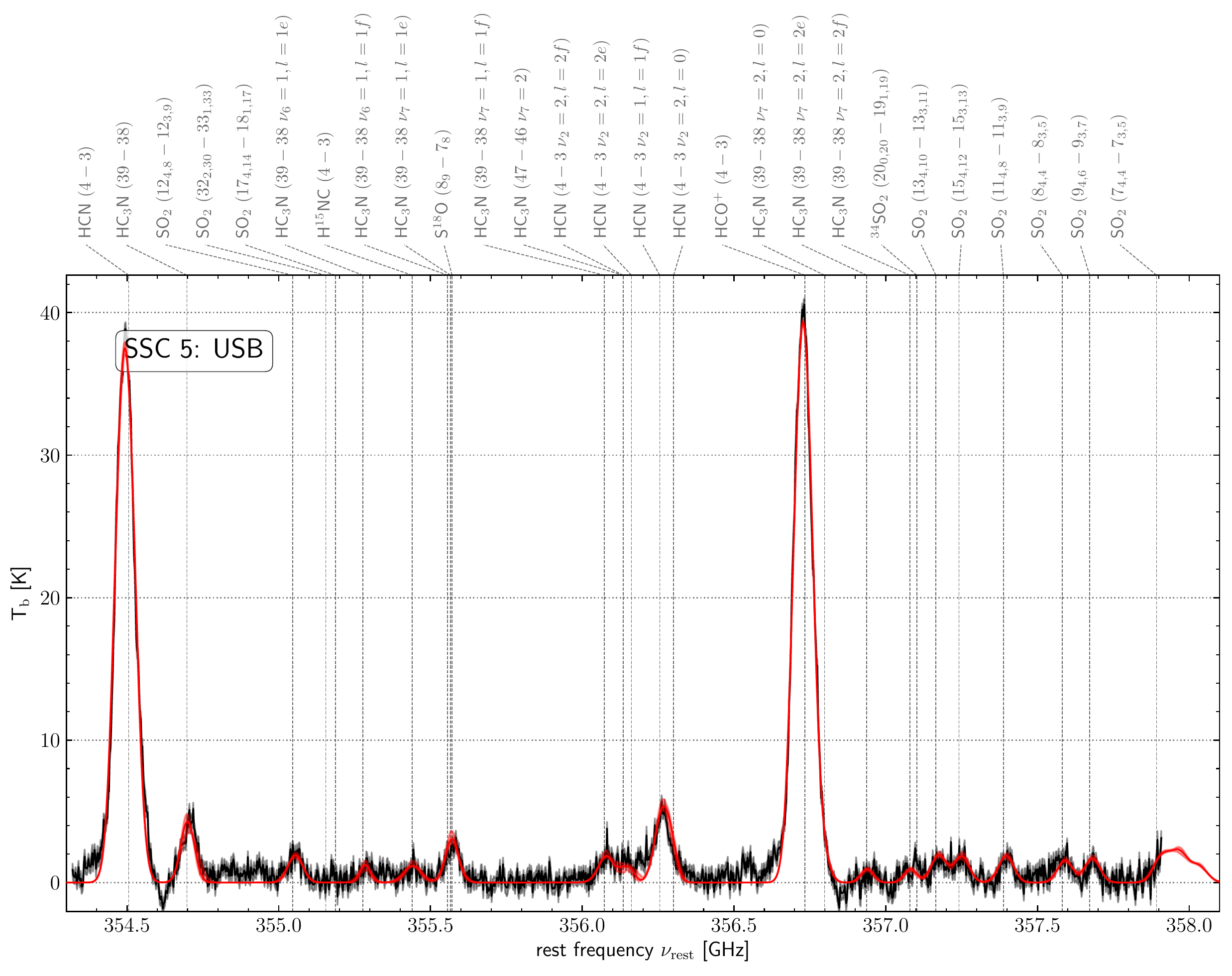}
\figsetgrpnote{Spectra of the LSB (\emph{top}) and USB (\emph{bottom}) in SSC~5. The observed spectrum (black) sits on top of a grey band indicating the $16^\mathrm{th}$ to $84^\mathrm{th}$ percentiles of the noise added for error estimation in the fit (cf. Section~\ref{section: error estimation})). The red lines represent the median fit obtained by \xclass using the fitting procedure described in Section~\ref{section: spectral fitting procedure}. The Monte Carlo-estimated errors are shown as a red band ($16^\mathrm{th}$ to $84^\mathrm{th}$ percentiles) that is hardly visible due to its small size relative to the strong spectral lines.}
\figsetgrpend

\figsetgrpstart
\figsetgrpnum{2.6}
\figsetgrptitle{SSC 6}
\figsetplot{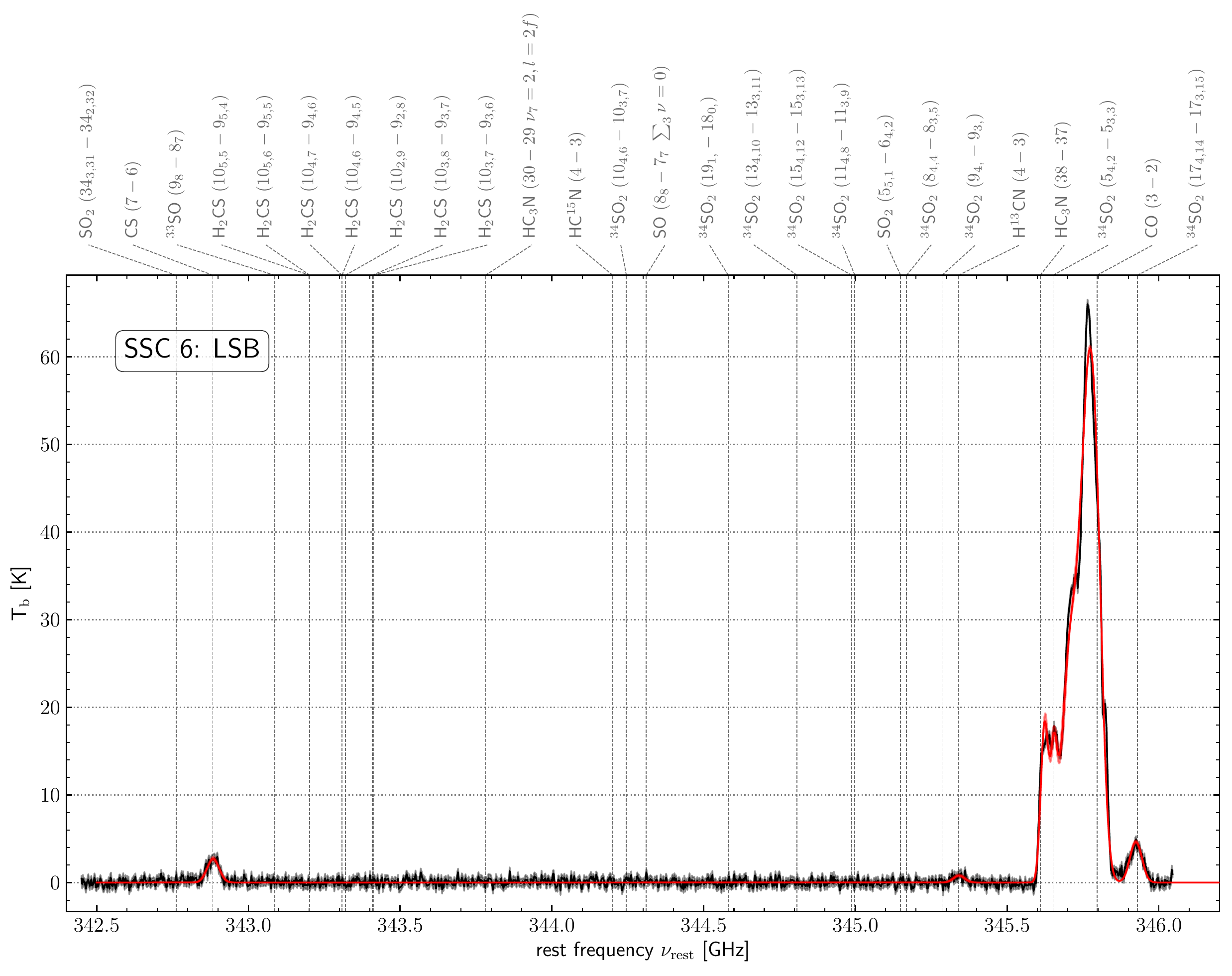}
\figsetplot{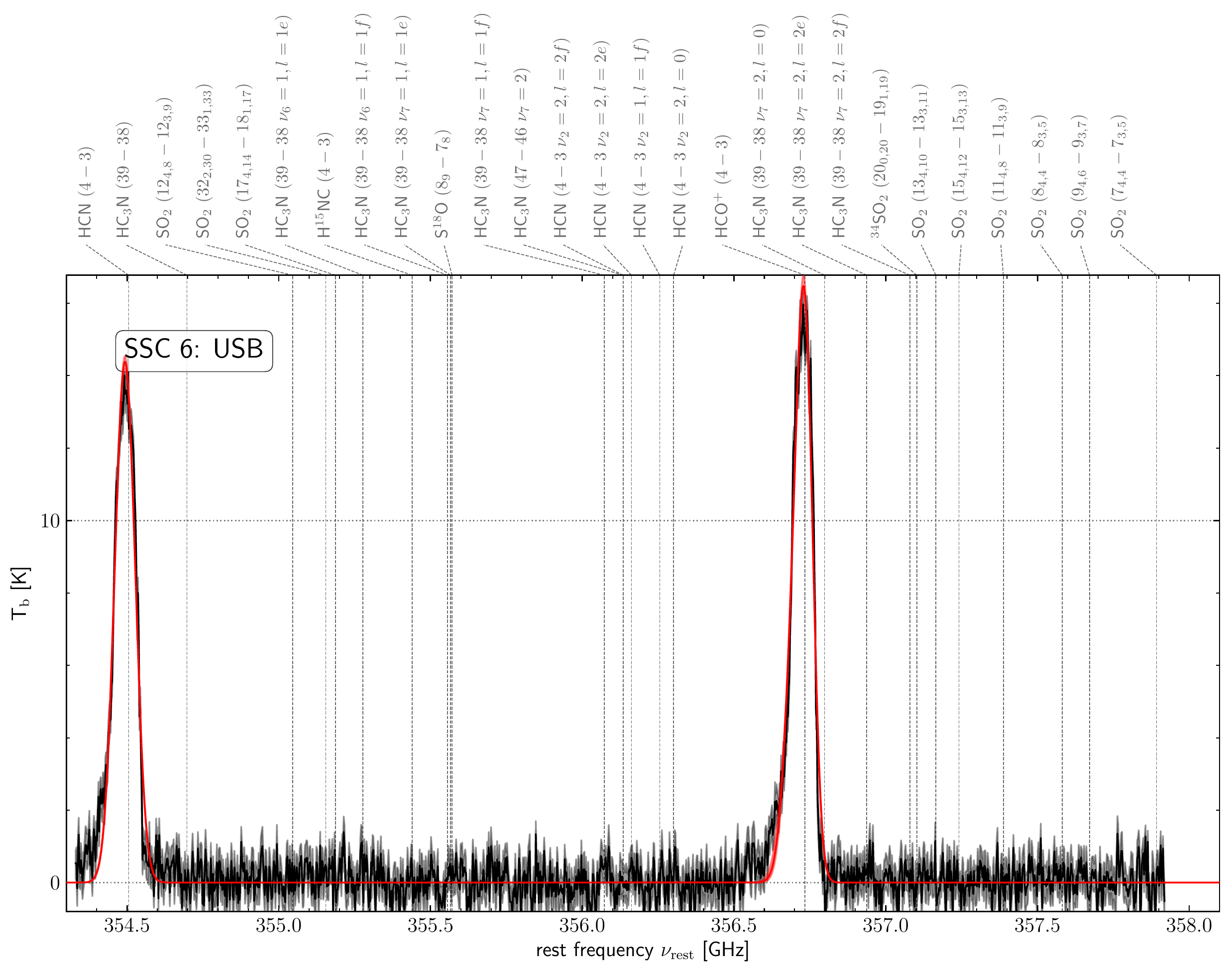}
\figsetgrpnote{Spectra of the LSB (\emph{top}) and USB (\emph{bottom}) in SSC~6. The observed spectrum (black) sits on top of a grey band indicating the $16^\mathrm{th}$ to $84^\mathrm{th}$ percentiles of the noise added for error estimation in the fit (cf. Section~\ref{section: error estimation})). The red lines represent the median fit obtained by \xclass using the fitting procedure described in Section~\ref{section: spectral fitting procedure}. The Monte Carlo-estimated errors are shown as a red band ($16^\mathrm{th}$ to $84^\mathrm{th}$ percentiles) that is hardly visible due to its small size relative to the strong spectral lines.}
\figsetgrpend

\figsetgrpstart
\figsetgrpnum{2.7}
\figsetgrptitle{SSC 7}
\figsetplot{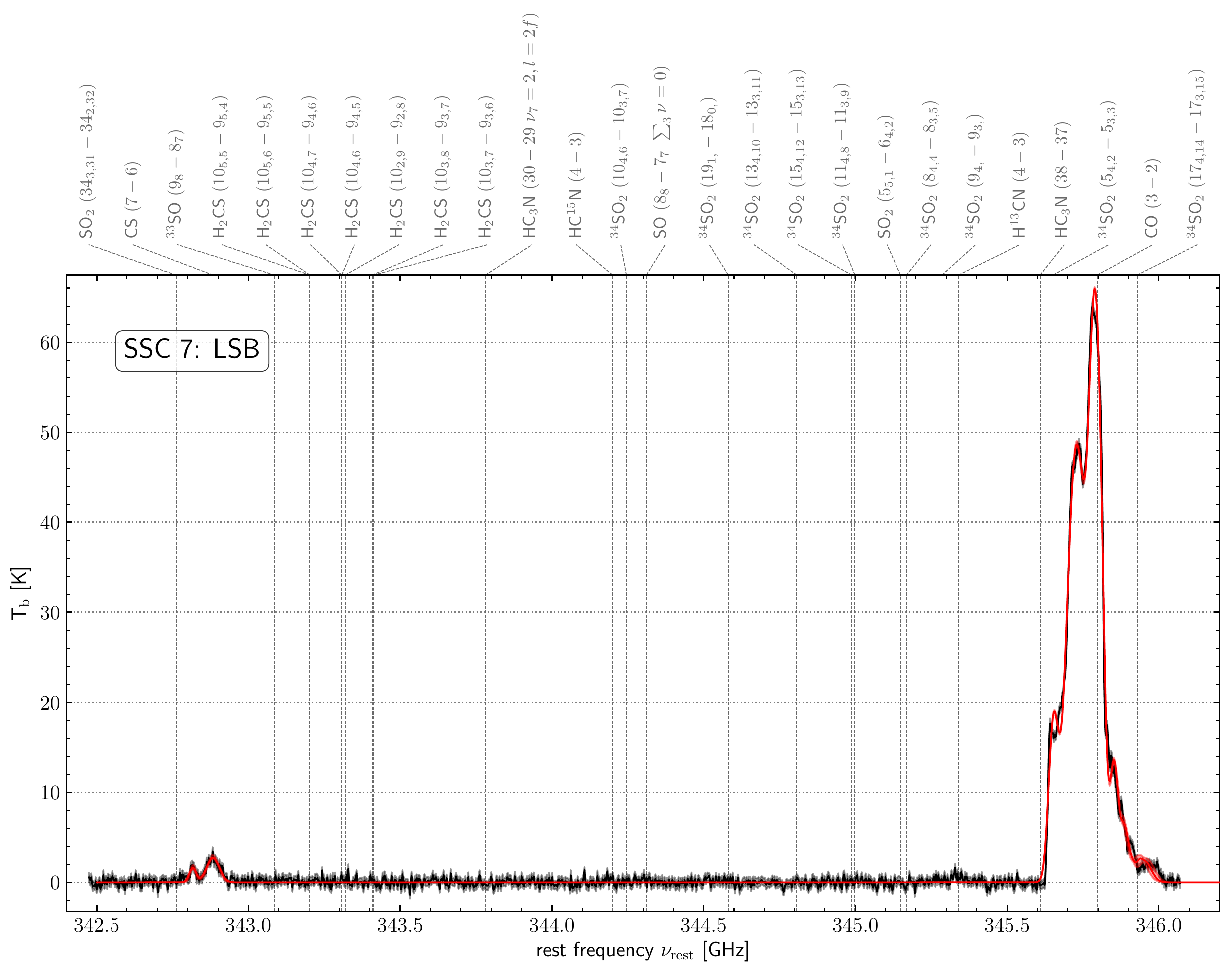}
\figsetplot{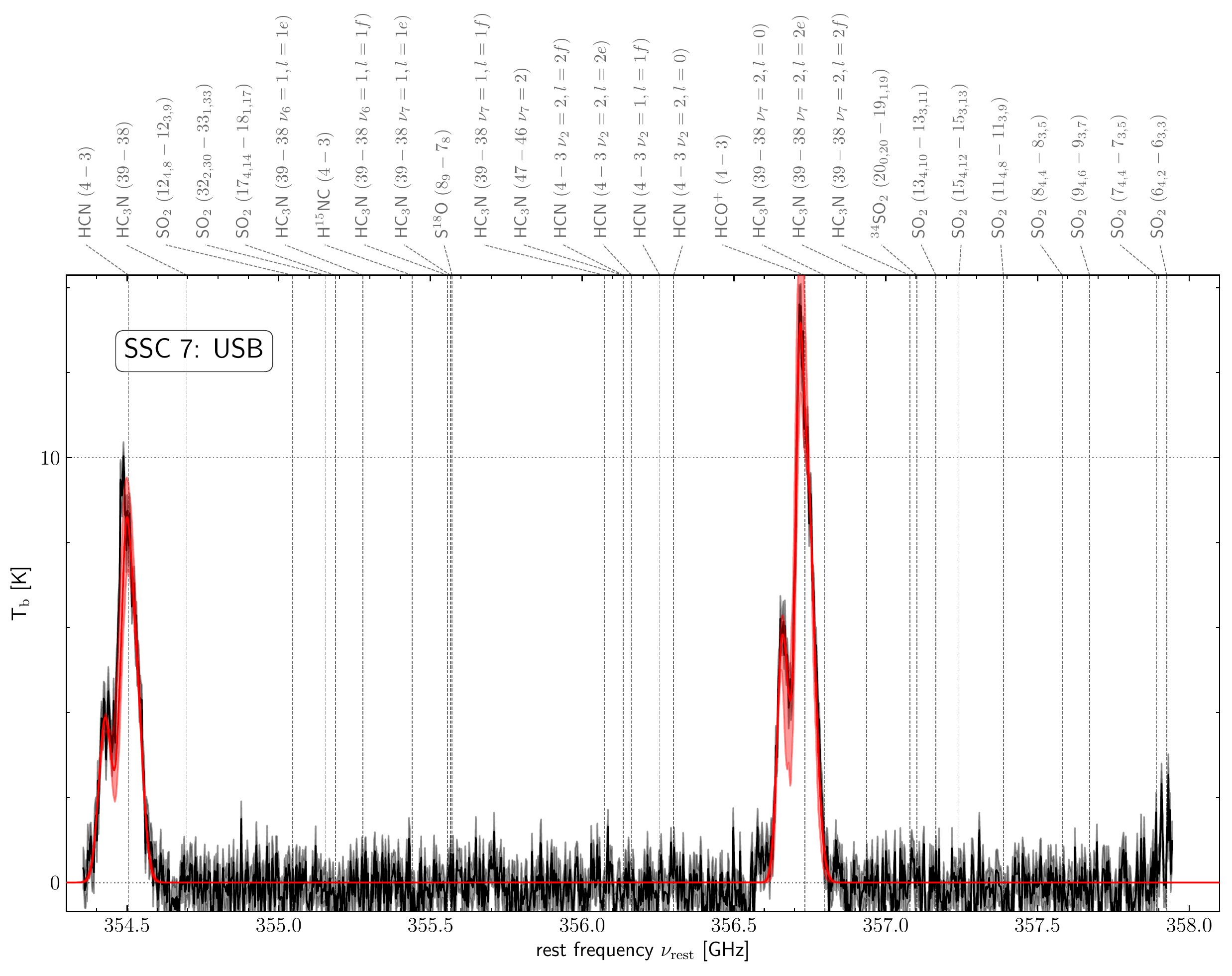}
\figsetgrpnote{Spectra of the LSB (\emph{top}) and USB (\emph{bottom}) in SSC~7. The observed spectrum (black) sits on top of a grey band indicating the $16^\mathrm{th}$ to $84^\mathrm{th}$ percentiles of the noise added for error estimation in the fit (cf. Section~\ref{section: error estimation})). The red lines represent the median fit obtained by \xclass using the fitting procedure described in Section~\ref{section: spectral fitting procedure}. The Monte Carlo-estimated errors are shown as a red band ($16^\mathrm{th}$ to $84^\mathrm{th}$ percentiles) that is hardly visible due to its small size relative to the strong spectral lines.}
\figsetgrpend

\figsetgrpstart
\figsetgrpnum{2.8}
\figsetgrptitle{SSC 8}
\figsetplot{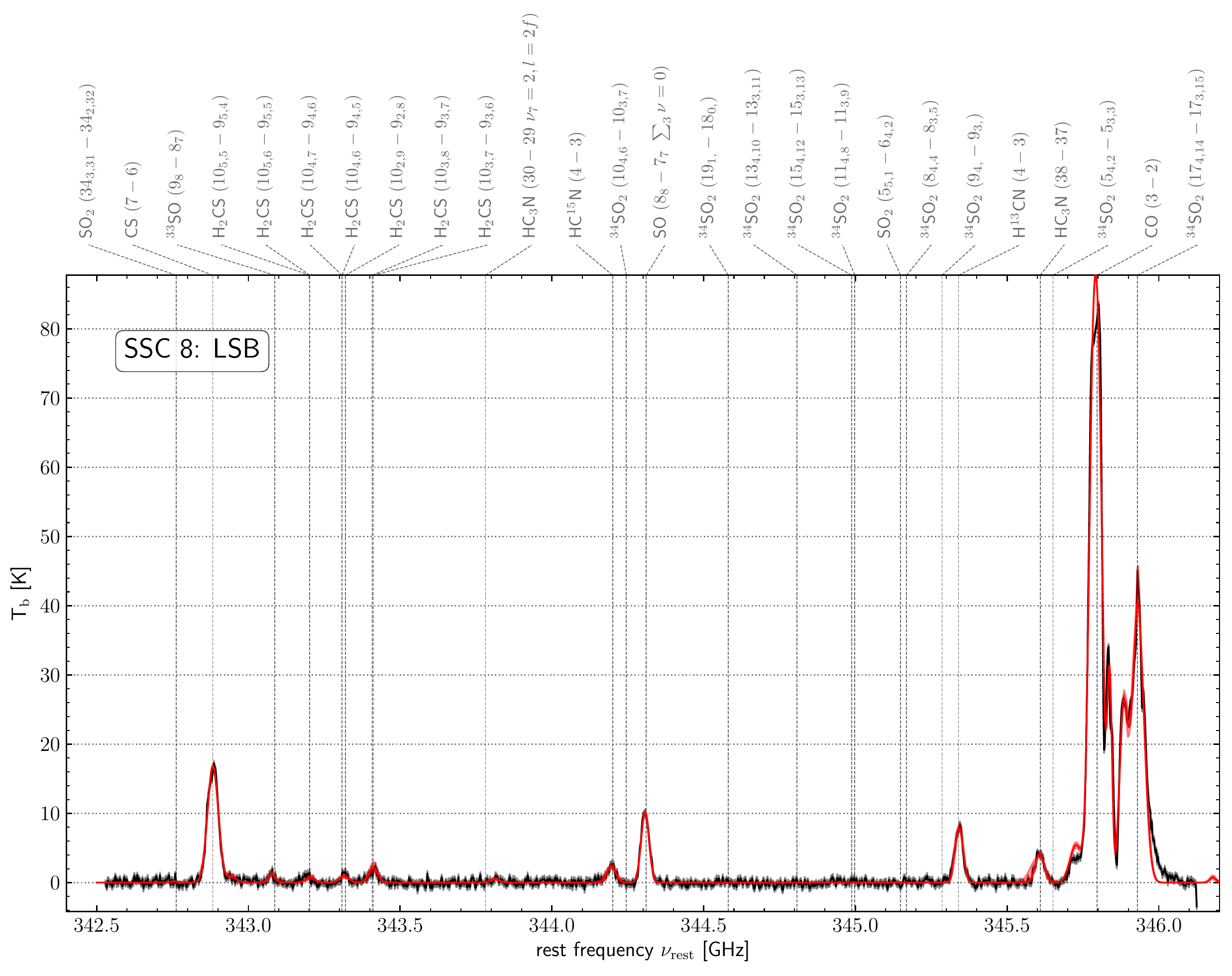}
\figsetplot{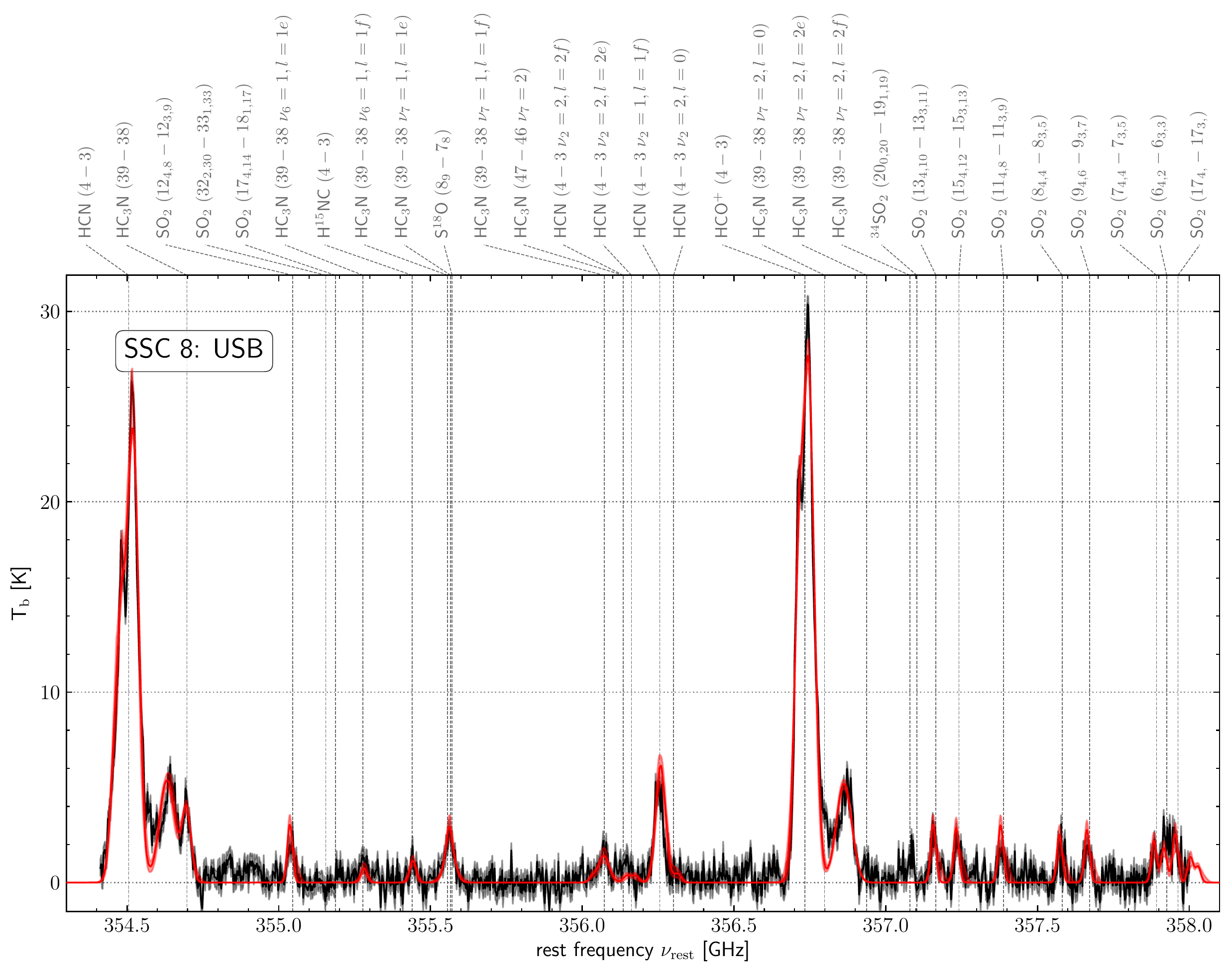}
\figsetgrpnote{Spectra of the LSB (\emph{top}) and USB (\emph{bottom}) in SSC~8. The observed spectrum (black) sits on top of a grey band indicating the $16^\mathrm{th}$ to $84^\mathrm{th}$ percentiles of the noise added for error estimation in the fit (cf. Section~\ref{section: error estimation})). The red lines represent the median fit obtained by \xclass using the fitting procedure described in Section~\ref{section: spectral fitting procedure}. The Monte Carlo-estimated errors are shown as a red band ($16^\mathrm{th}$ to $84^\mathrm{th}$ percentiles) that is hardly visible due to its small size relative to the strong spectral lines.}
\figsetgrpend

\figsetgrpstart
\figsetgrpnum{2.9}
\figsetgrptitle{SSC 9}
\figsetplot{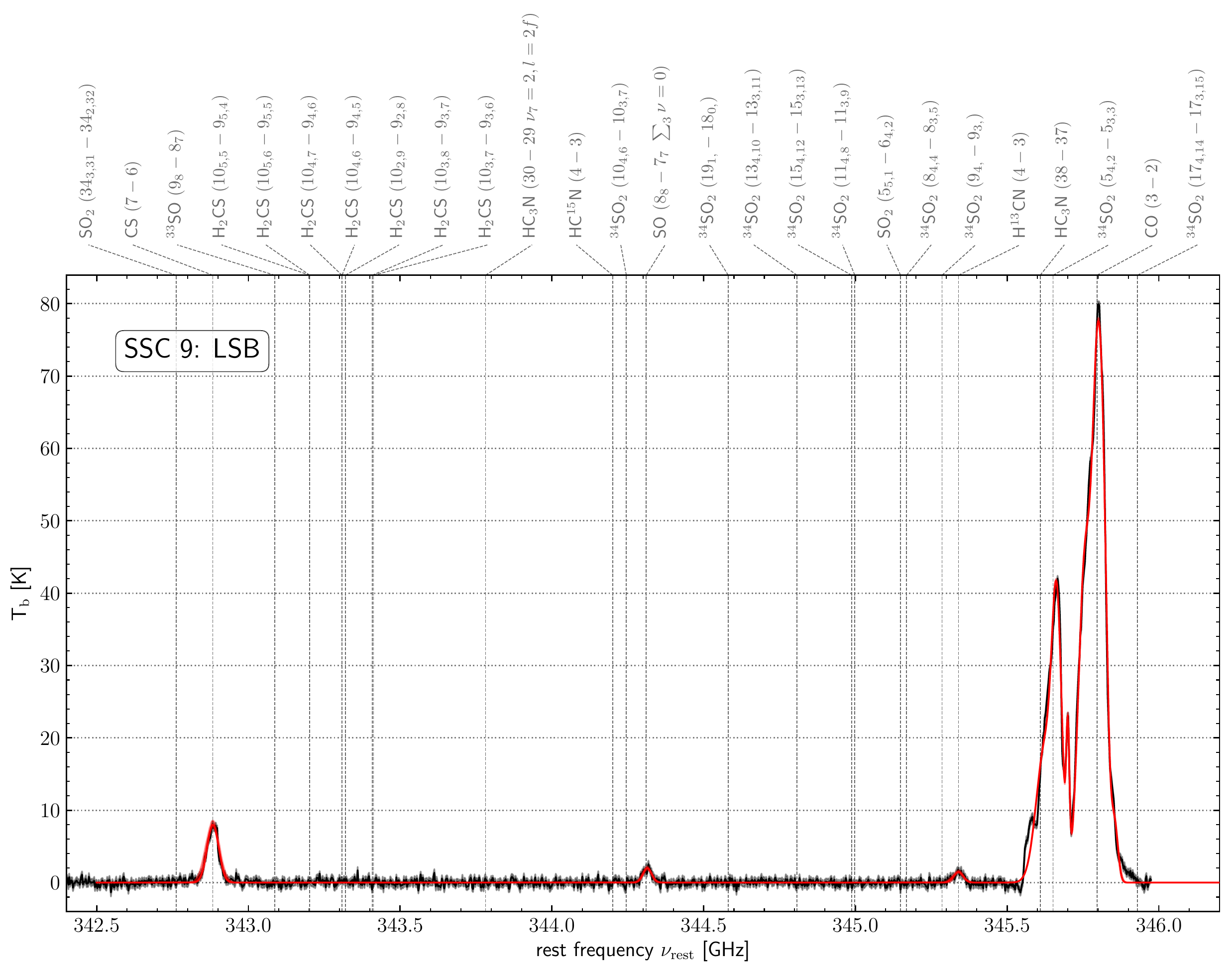}
\figsetplot{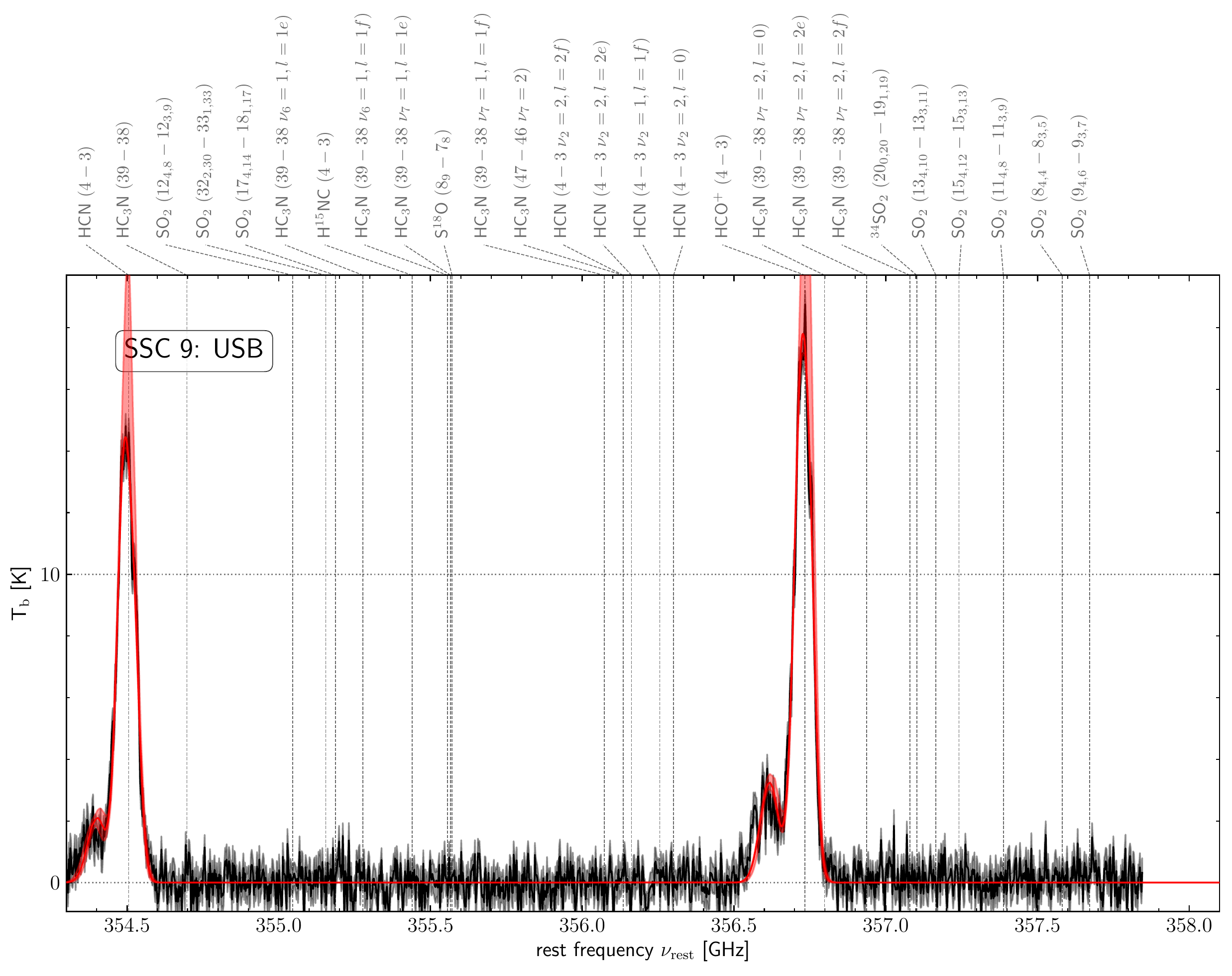}
\figsetgrpnote{Spectra of the LSB (\emph{top}) and USB (\emph{bottom}) in SSC~9. The observed spectrum (black) sits on top of a grey band indicating the $16^\mathrm{th}$ to $84^\mathrm{th}$ percentiles of the noise added for error estimation in the fit (cf. Section~\ref{section: error estimation})). The red lines represent the median fit obtained by \xclass using the fitting procedure described in Section~\ref{section: spectral fitting procedure}. The Monte Carlo-estimated errors are shown as a red band ($16^\mathrm{th}$ to $84^\mathrm{th}$ percentiles) that is hardly visible due to its small size relative to the strong spectral lines.}
\figsetgrpend

\figsetgrpstart
\figsetgrpnum{2.10}
\figsetgrptitle{SSC 10}
\figsetplot{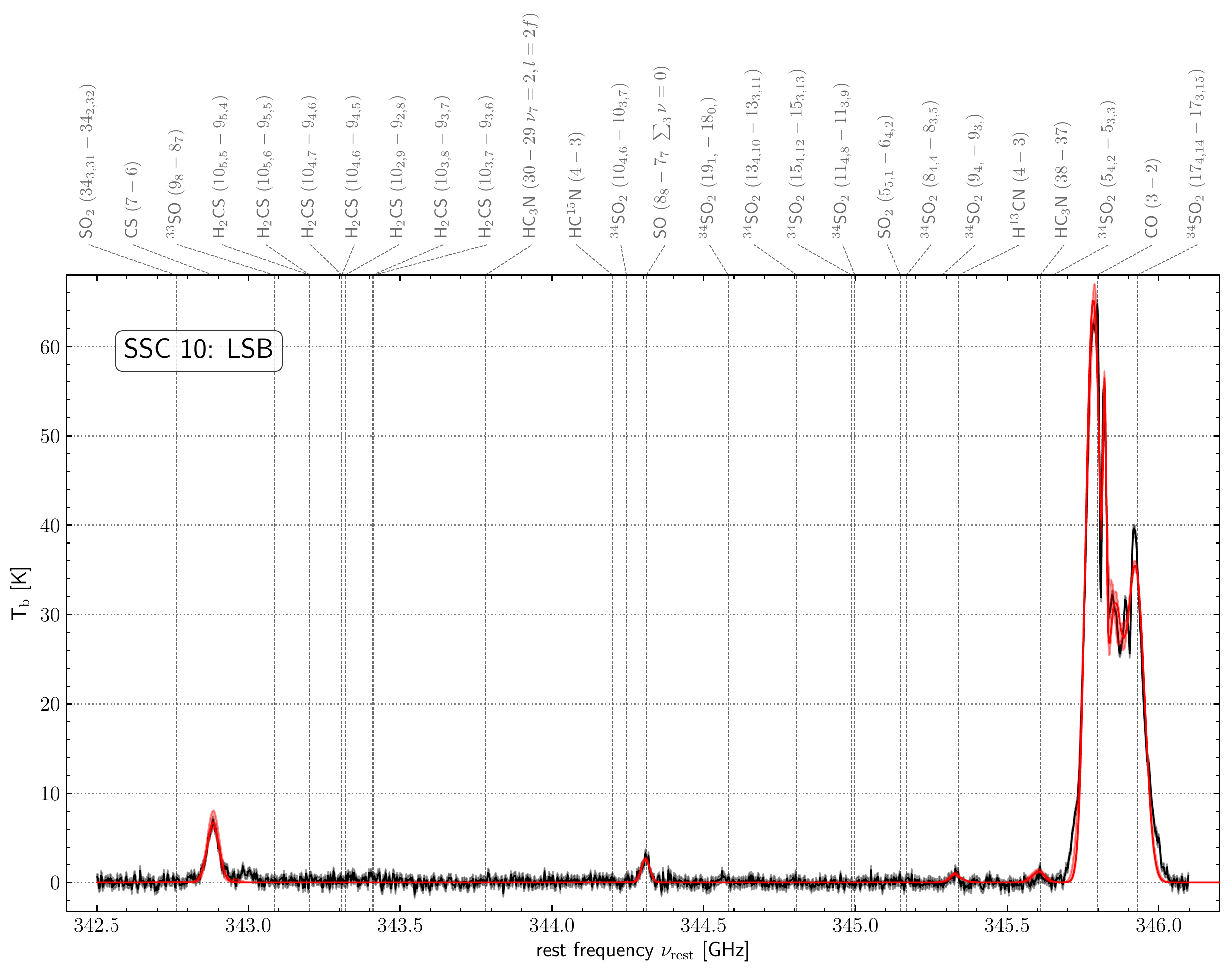}
\figsetplot{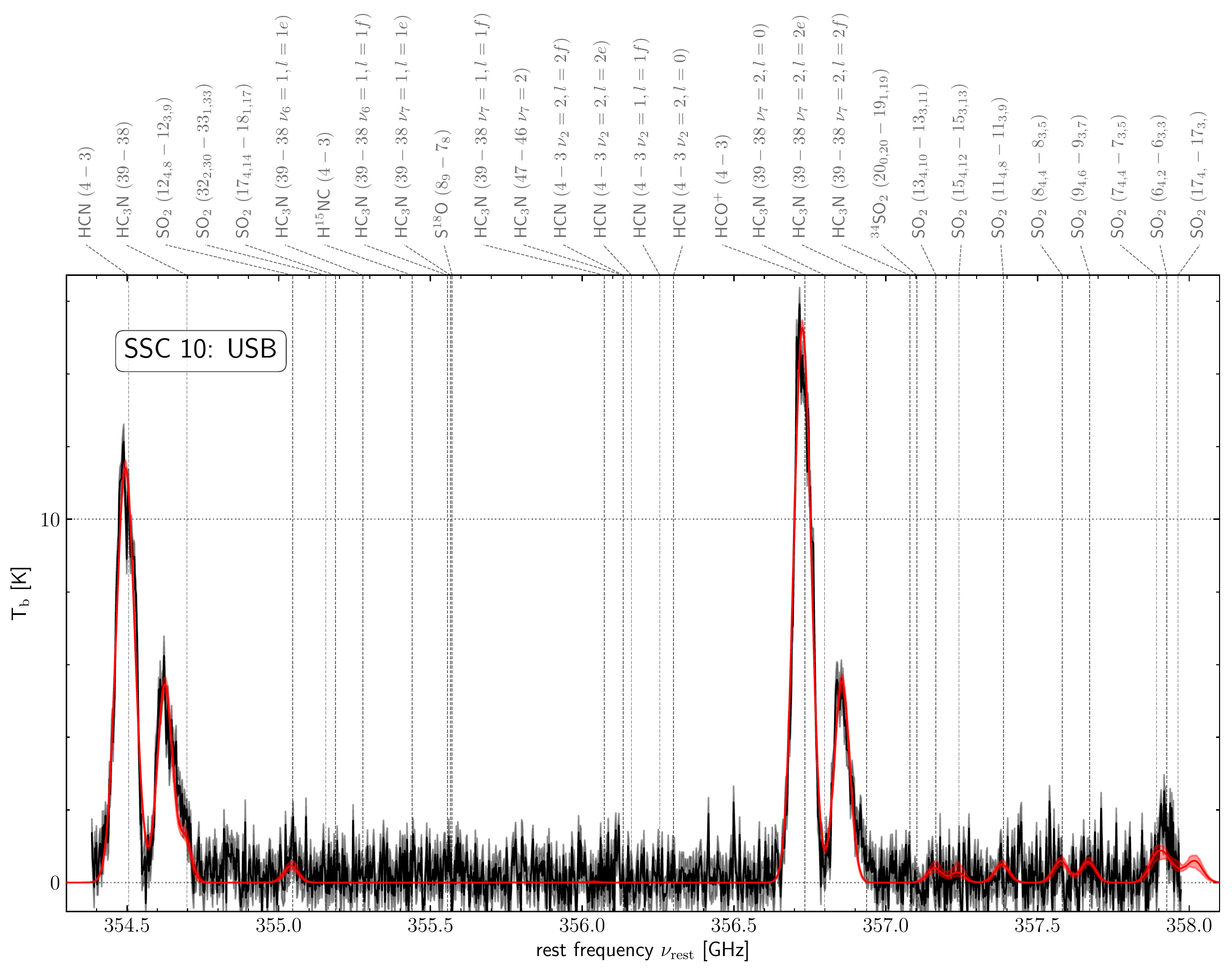}
\figsetgrpnote{Spectra of the LSB (\emph{top}) and USB (\emph{bottom}) in SSC~10. The observed spectrum (black) sits on top of a grey band indicating the $16^\mathrm{th}$ to $84^\mathrm{th}$ percentiles of the noise added for error estimation in the fit (cf. Section~\ref{section: error estimation})). The red lines represent the median fit obtained by \xclass using the fitting procedure described in Section~\ref{section: spectral fitting procedure}. The Monte Carlo-estimated errors are shown as a red band ($16^\mathrm{th}$ to $84^\mathrm{th}$ percentiles) that is hardly visible due to its small size relative to the strong spectral lines.}
\figsetgrpend

\figsetgrpstart
\figsetgrpnum{2.11}
\figsetgrptitle{SSC 11}
\figsetplot{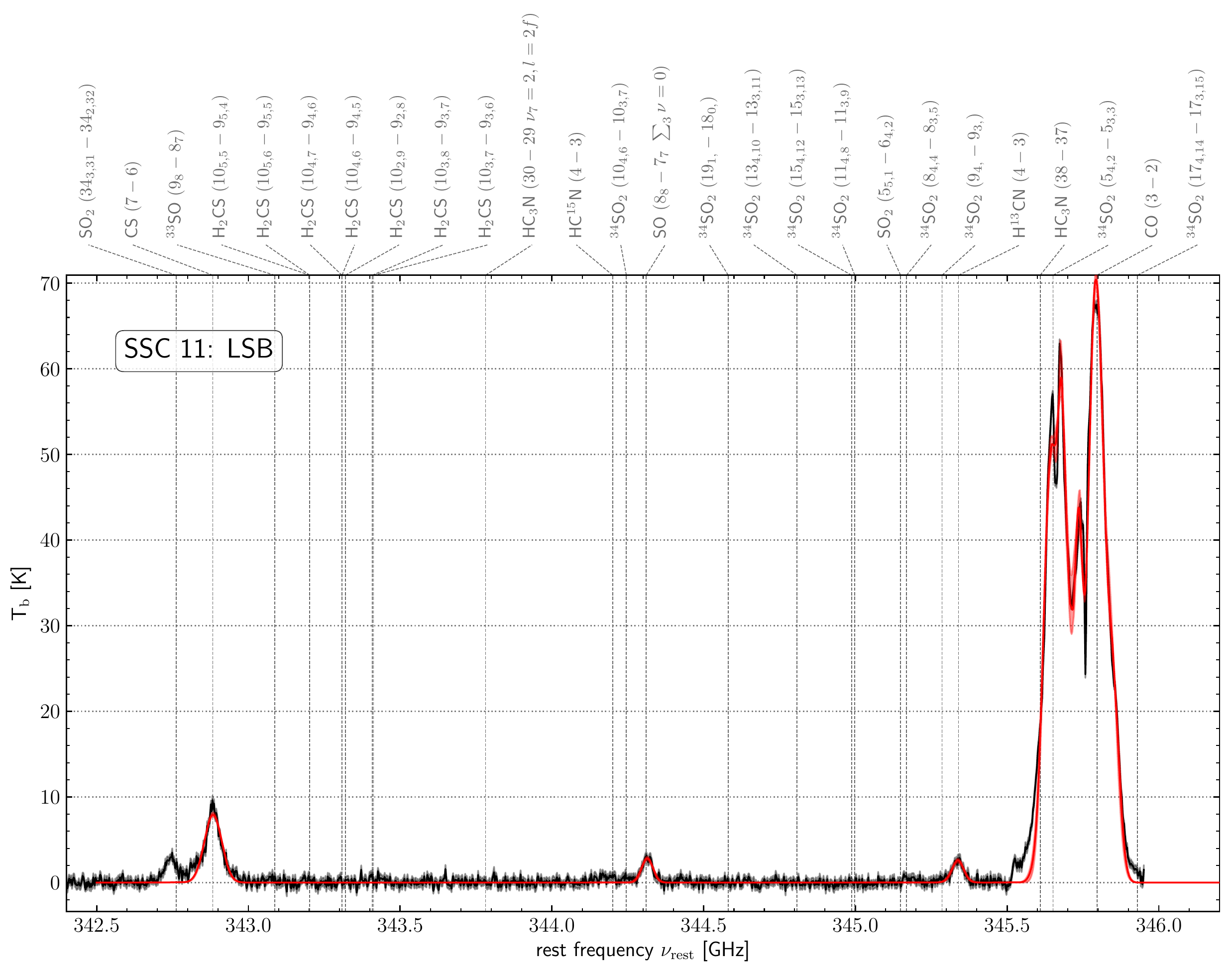}
\figsetplot{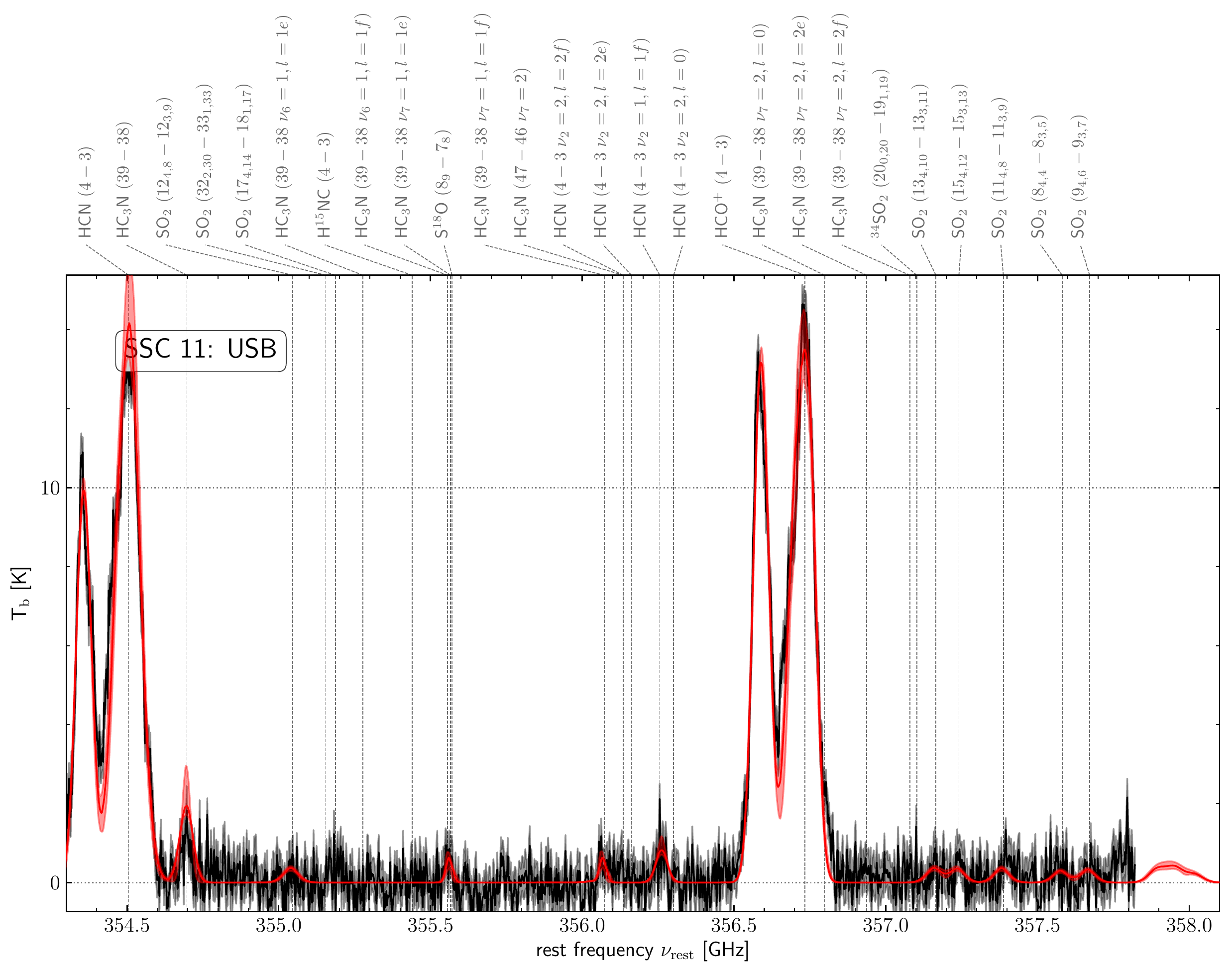}
\figsetgrpnote{Spectra of the LSB (\emph{top}) and USB (\emph{bottom}) in SSC~11. The observed spectrum (black) sits on top of a grey band indicating the $16^\mathrm{th}$ to $84^\mathrm{th}$ percentiles of the noise added for error estimation in the fit (cf. Section~\ref{section: error estimation})). The red lines represent the median fit obtained by \xclass using the fitting procedure described in Section~\ref{section: spectral fitting procedure}. The Monte Carlo-estimated errors are shown as a red band ($16^\mathrm{th}$ to $84^\mathrm{th}$ percentiles) that is hardly visible due to its small size relative to the strong spectral lines.}
\figsetgrpend

\figsetgrpstart
\figsetgrpnum{2.12}
\figsetgrptitle{SSC 12}
\figsetplot{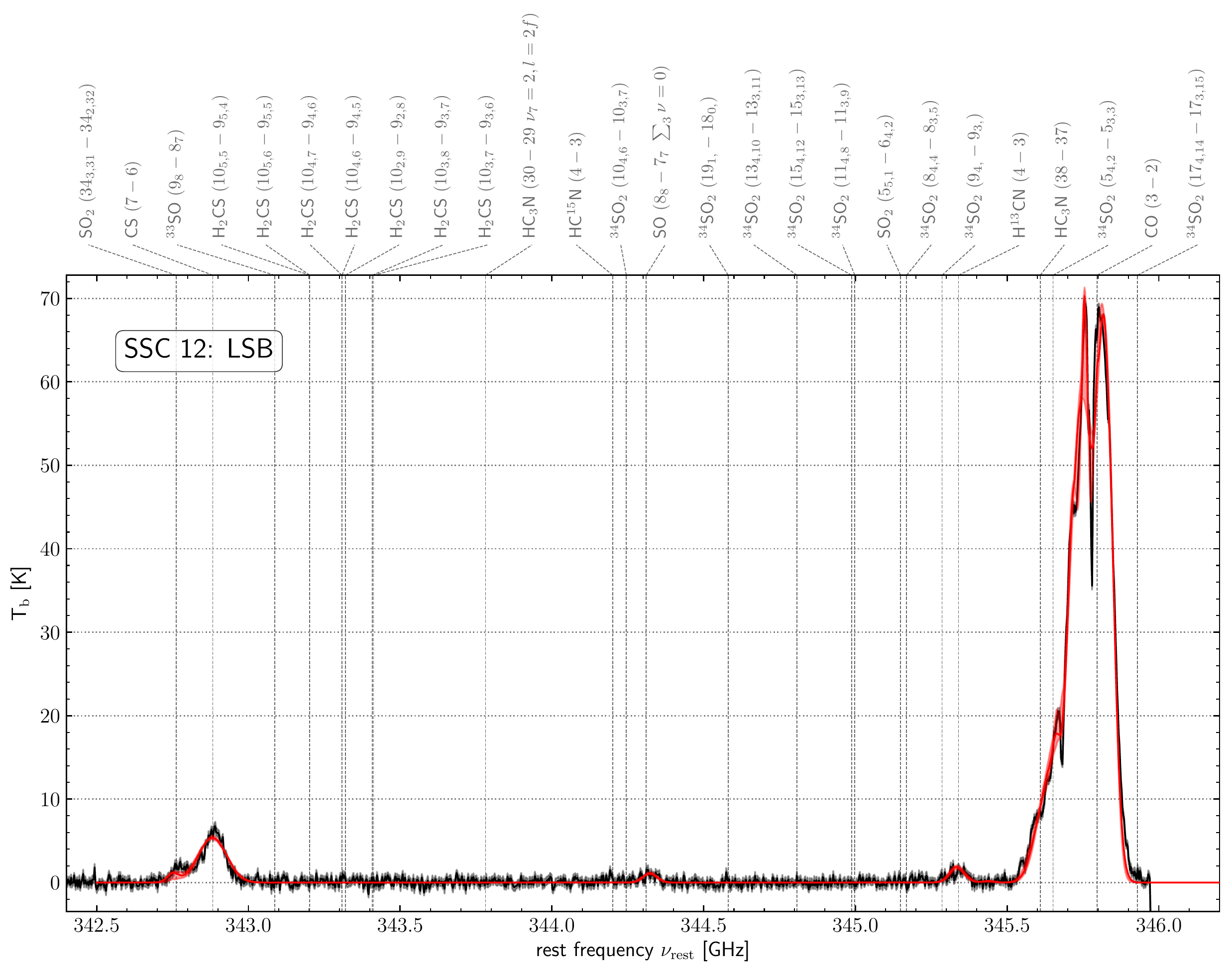}
\figsetplot{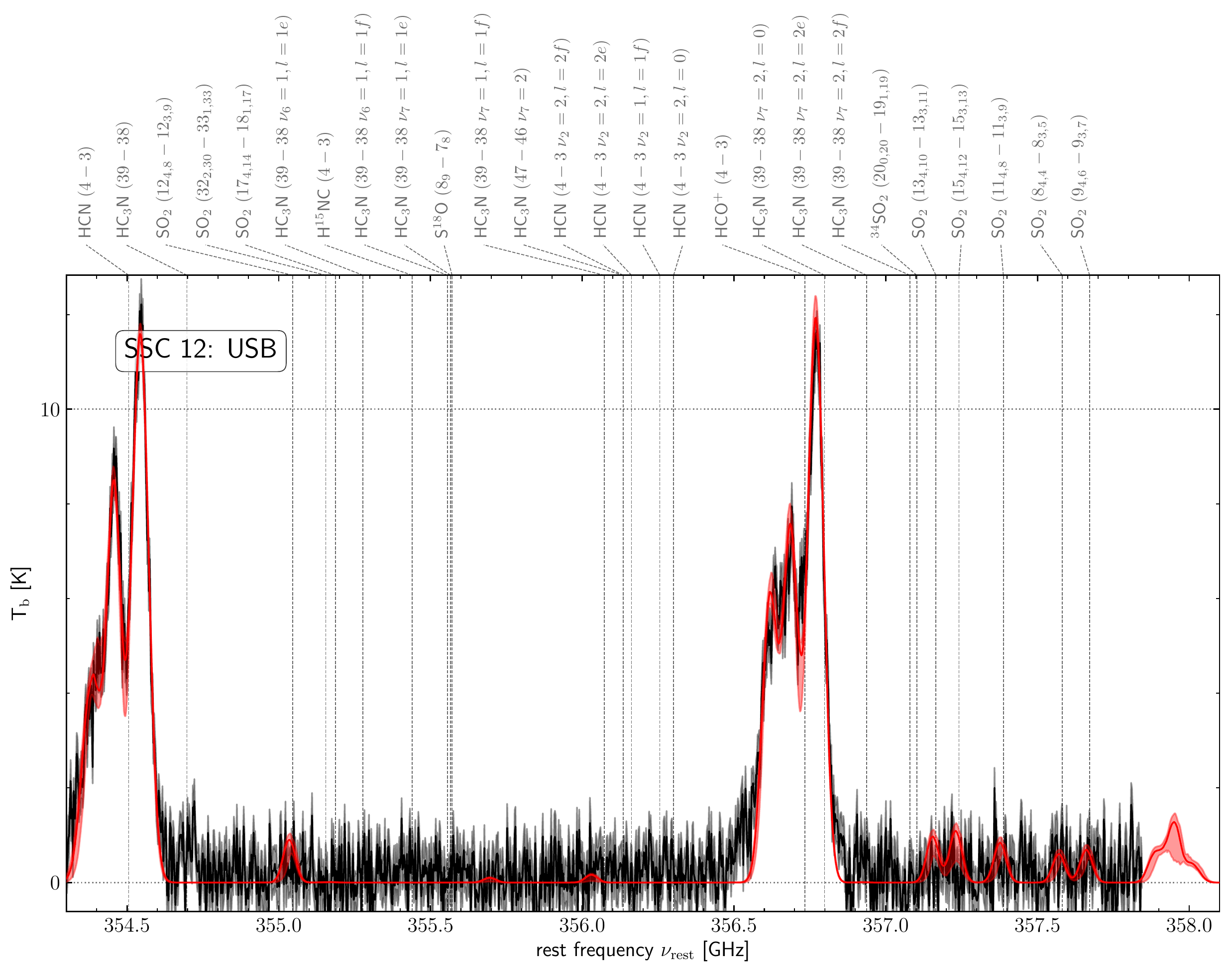}
\figsetgrpnote{Spectra of the LSB (\emph{top}) and USB (\emph{bottom}) in SSC~12. The observed spectrum (black) sits on top of a grey band indicating the $16^\mathrm{th}$ to $84^\mathrm{th}$ percentiles of the noise added for error estimation in the fit (cf. Section~\ref{section: error estimation})). The red lines represent the median fit obtained by \xclass using the fitting procedure described in Section~\ref{section: spectral fitting procedure}. The Monte Carlo-estimated errors are shown as a red band ($16^\mathrm{th}$ to $84^\mathrm{th}$ percentiles) that is hardly visible due to its small size relative to the strong spectral lines.}
\figsetgrpend

\figsetgrpstart
\figsetgrpnum{2.13}
\figsetgrptitle{SSC 13}
\figsetplot{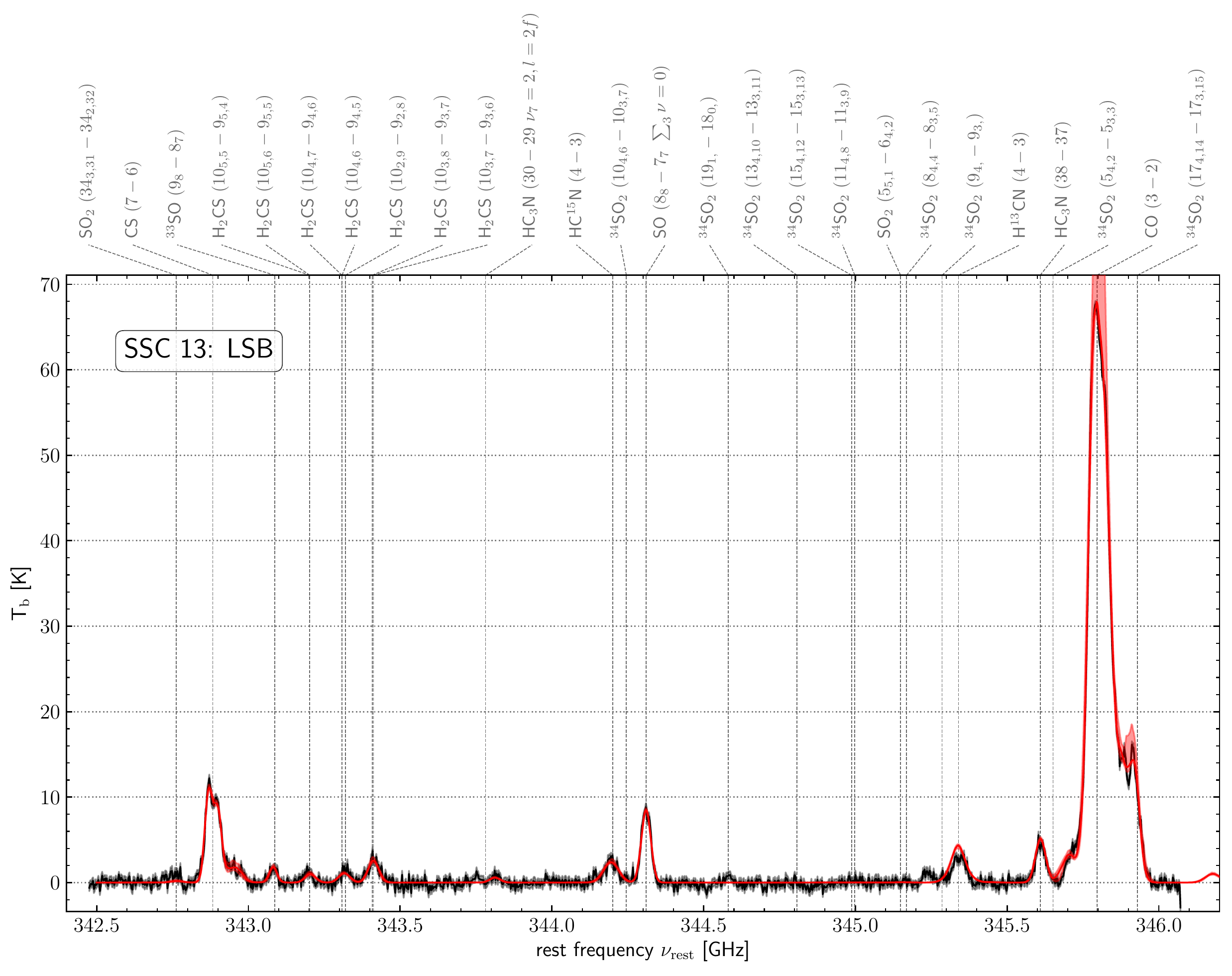}
\figsetplot{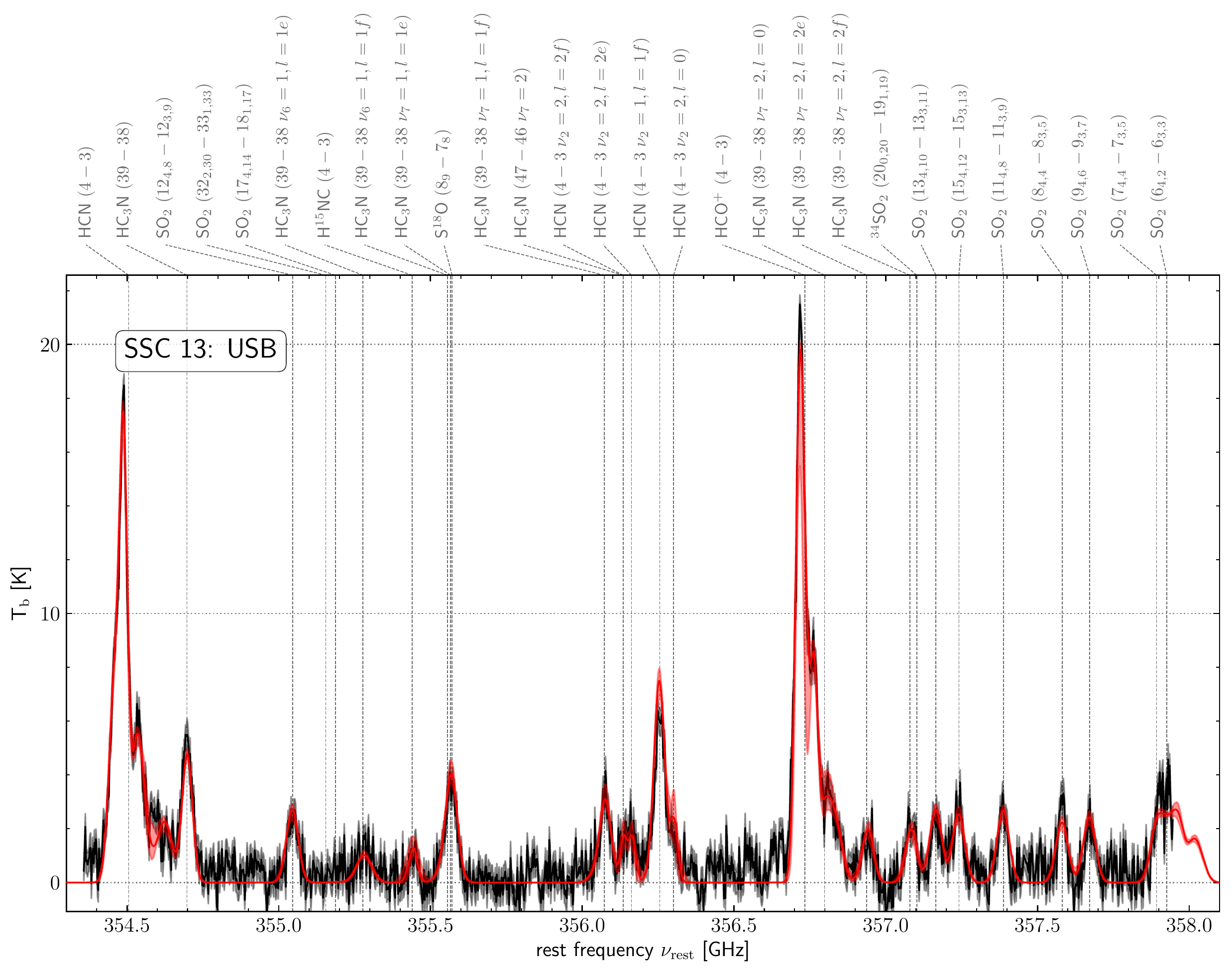}
\figsetgrpnote{Spectra of the LSB (\emph{top}) and USB (\emph{bottom}) in SSC~13. The observed spectrum (black) sits on top of a grey band indicating the $16^\mathrm{th}$ to $84^\mathrm{th}$ percentiles of the noise added for error estimation in the fit (cf. Section~\ref{section: error estimation})). The red lines represent the median fit obtained by \xclass using the fitting procedure described in Section~\ref{section: spectral fitting procedure}. The Monte Carlo-estimated errors are shown as a red band ($16^\mathrm{th}$ to $84^\mathrm{th}$ percentiles) that is hardly visible due to its small size relative to the strong spectral lines.}
\figsetgrpend

\figsetgrpstart
\figsetgrpnum{2.14}
\figsetgrptitle{SSC 14}
\figsetplot{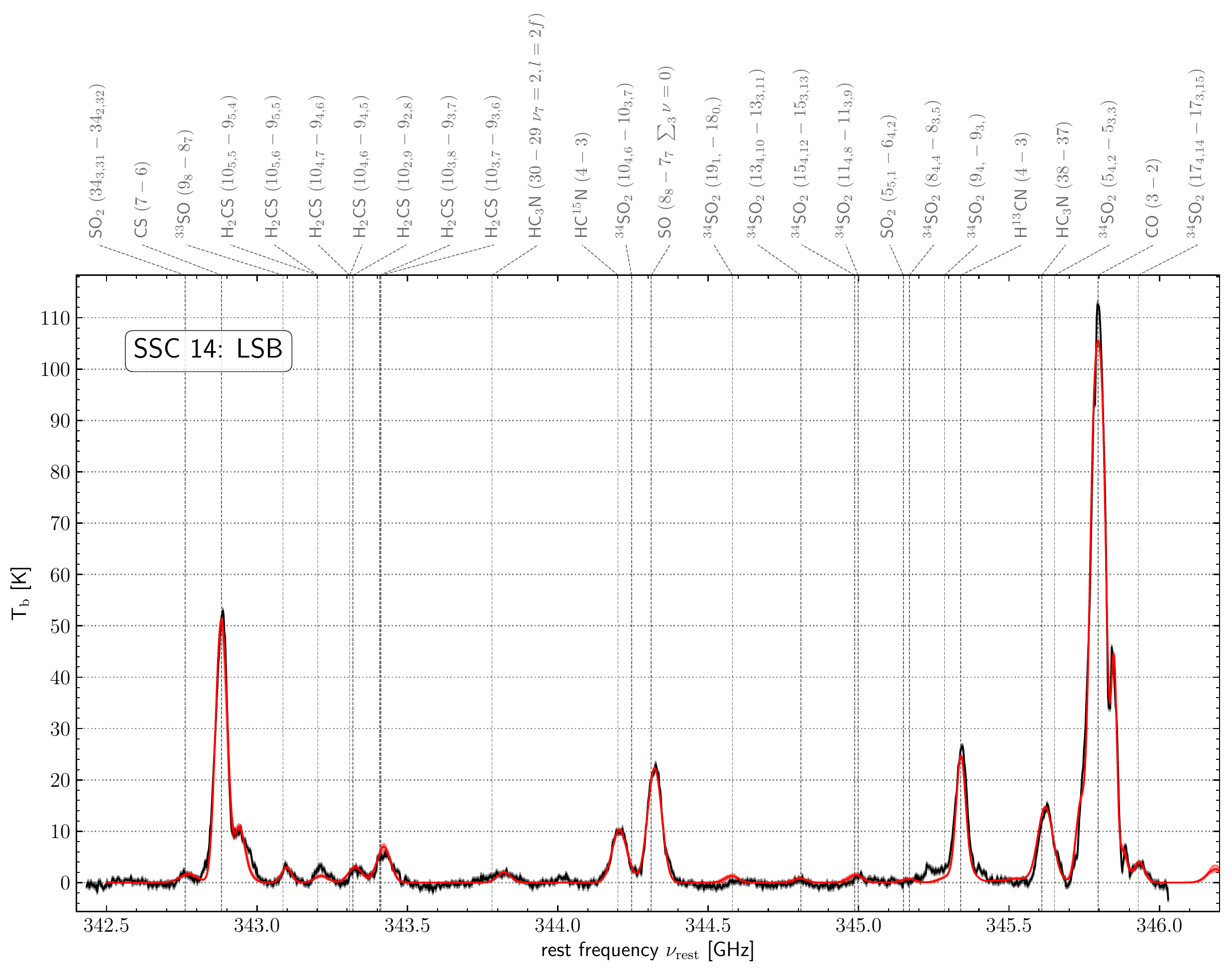}
\figsetplot{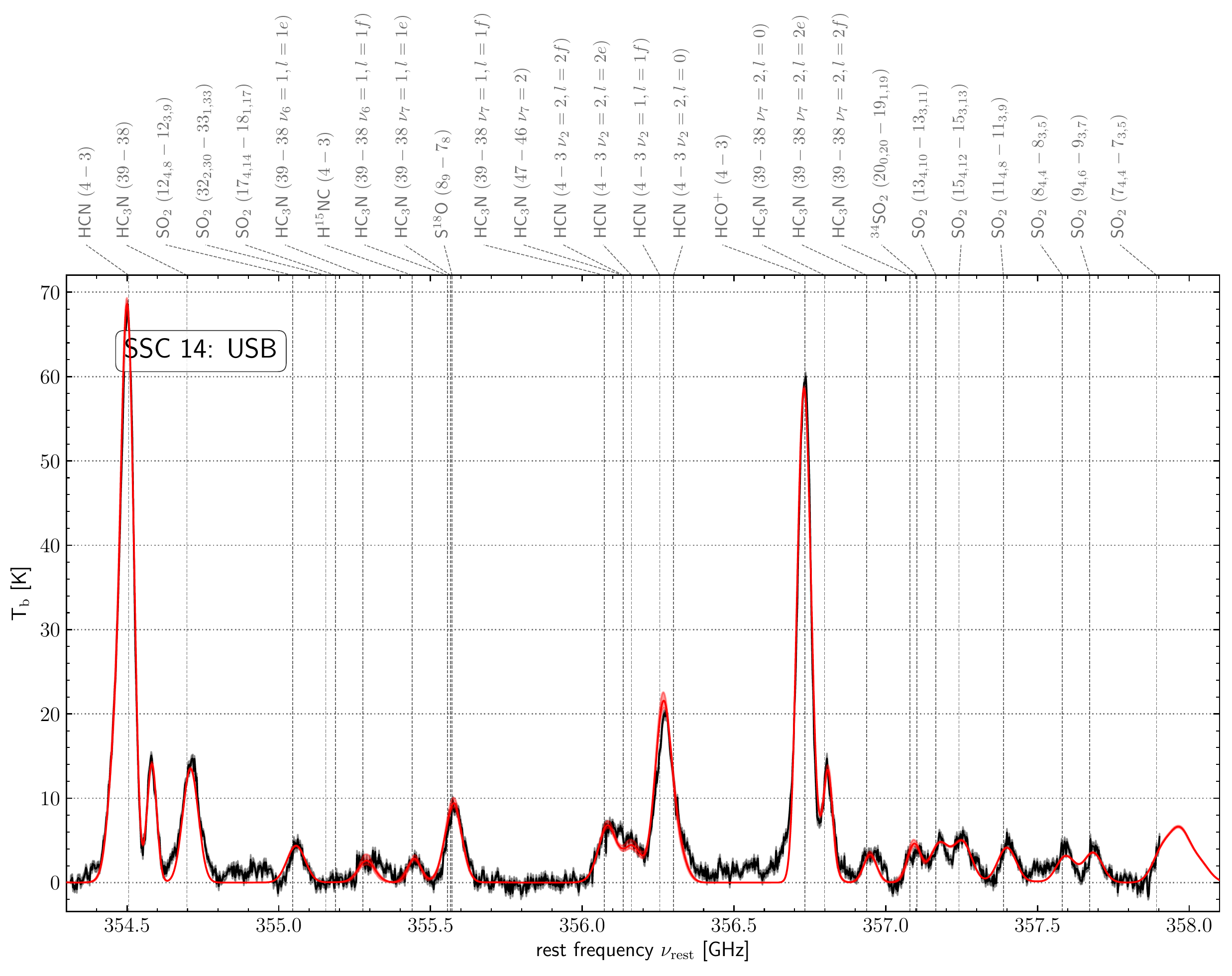}
\figsetgrpnote{Spectra of the LSB (\emph{top}) and USB (\emph{bottom}) in SSC~14. The observed spectrum (black) sits on top of a grey band indicating the $16^\mathrm{th}$ to $84^\mathrm{th}$ percentiles of the noise added for error estimation in the fit (cf. Section~\ref{section: error estimation})). The red lines represent the median fit obtained by \xclass using the fitting procedure described in Section~\ref{section: spectral fitting procedure}. The Monte Carlo-estimated errors are shown as a red band ($16^\mathrm{th}$ to $84^\mathrm{th}$ percentiles) that is hardly visible due to its small size relative to the strong spectral lines.}
\figsetgrpend
    
\figsetend

% sample figure from the figure set
\begin{figure*}
    \centering
    \includegraphics[width=0.72\textwidth]{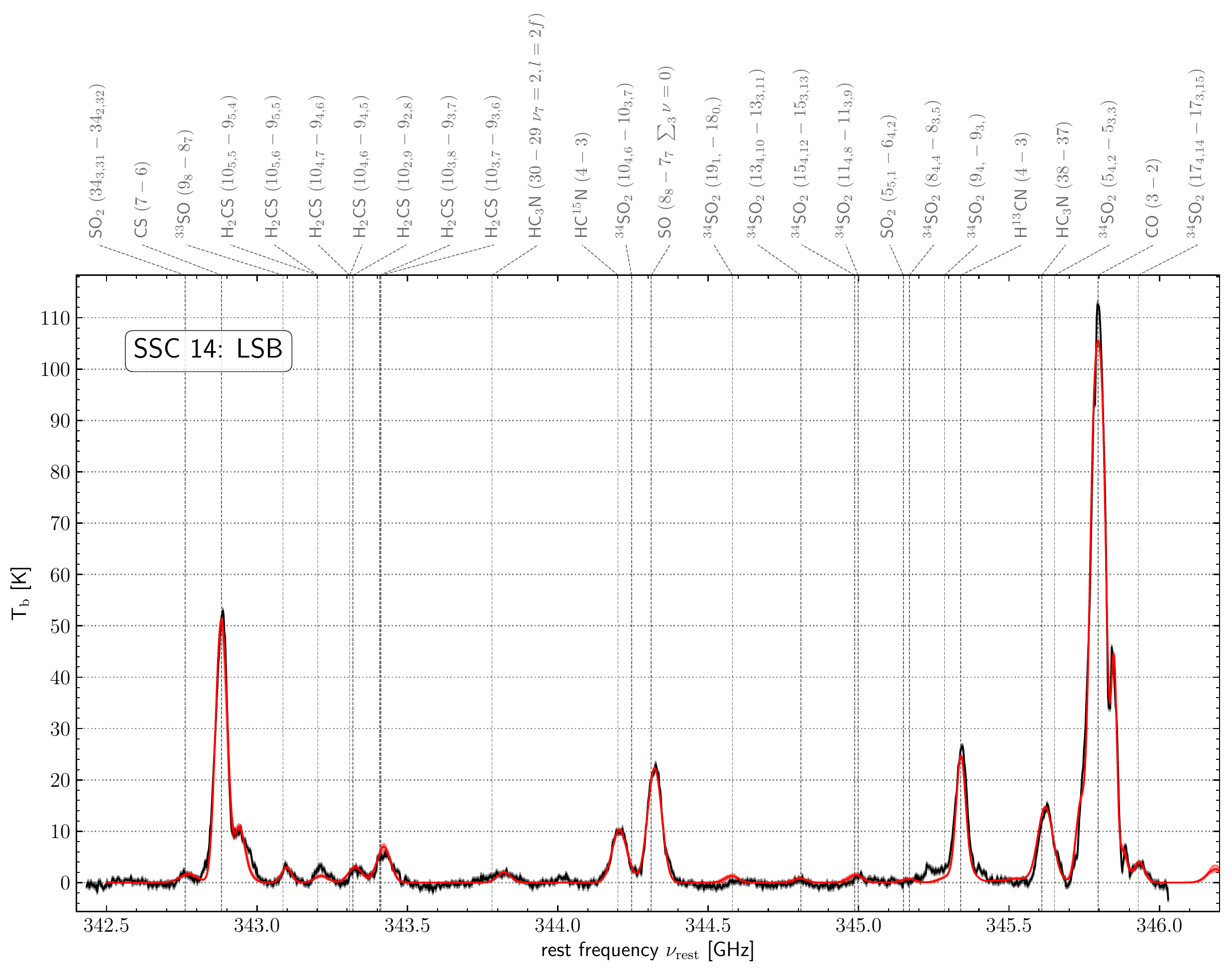}
    \includegraphics[width=0.72\textwidth]{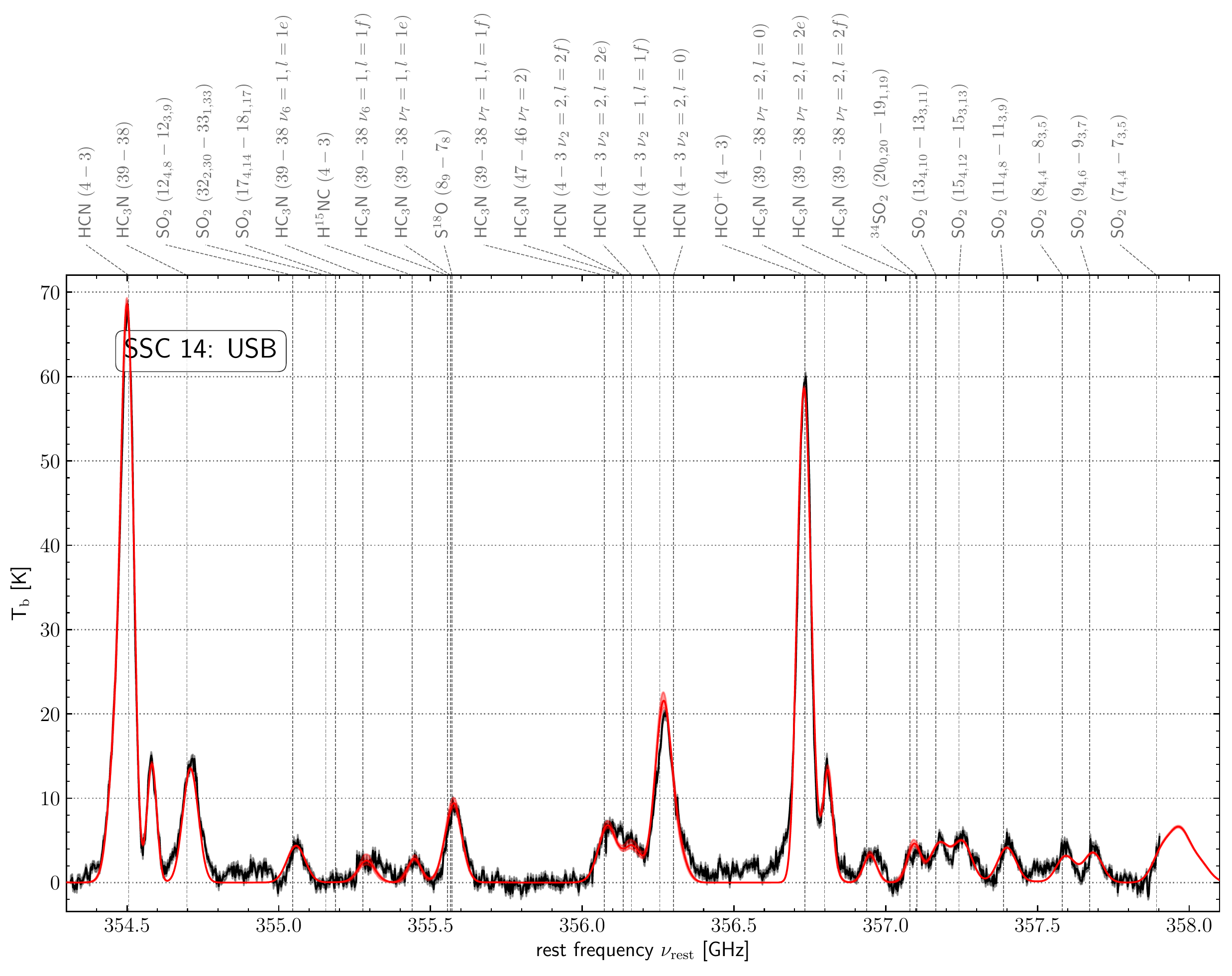}
    \caption{Spectra of SSC~14 in the LSB (\emph{top}) and USB (\emph{bottom}). This figure and Figure Set 2 in the online journal show a zoom into the bottom two panels of Figure~\ref{figure: fitted bands}. The observed spectrum (black) sits on top of a grey band indicating the $16^\mathrm{th}$ to $84^\mathrm{th}$ percentiles of the noise added for error estimation in the fit (cf. Section~\ref{section: error estimation})). The red lines represent the median fit obtained by \xclass using the fitting procedure described in Section~\ref{section: spectral fitting procedure}. The Monte Carlo-estimated errors are shown as a red band ($16^\mathrm{th}$ to $84^\mathrm{th}$ percentiles) that is hardly visible due to its small size relative to the strong spectral lines. 
    The complete figure set (Figure Set 2) of 14 images is available in the online journal. All spectra are available in electronic form as the Data behind the Figure.}
    \label{figure: sample spectrum}
\end{figure*}

%%%%%%%%%%%%%%%%%%%%%%%%%%%%%%%%%%%%%%%%%%%%%%%%%%%%%%%%%%%%%%%%%%%%%%%%%%%%%%%%%%%%%%%%%%%%%%%%%%%%

\section{Spectral Line Fitting with \xclass}\label{section: spectral line fitting}

As lines of the same or different species are blended in the spectra, an integration over a line over a fixed velocity range will overestimate the actual intensity of the line. We therefore optimize model spectra to the observed spectra using the eXtended CASA Line Analysis Software Suite \citep[\xclass\footnote{\url{https://xclass.astro.uni-koeln.de/Home}},][]{2018ascl.soft10016M}. This method requires some physical assumptions about the emission but yields a physical interpretation of the spectra, for instance column density or excitation temperature. It further allows us to de-blend the spectra and unambiguously derive observational properties such as integrated intensity.
Table~\ref{table: intensities} lists the 55 lines that we fit for.

\subsection{Treatment of multiple spectral components}
The \co, \hcn, \hco and \cs lines show multiple peaks of varying strength in the spectra whereas all other lines do not show noticeable deviations from Gaussian shapes within noise limits. From the line profiles alone it is impossible to decide if a double peak seen in a line is the result of multiple emission components or absorption. For consistency, we assume that the spectra are composed of emission components only and try to model them with as few components as necessary. In \co, five or six components are required while in \hcn, \hco and \cs up to three components suffice to achieve a good fit. Upon closer inspection of the fit results below, it seems likely that the appearance of multiple components in a few cases is at least partially caused by absorption.
In the following, ``main component'' denotes the velocity component closest to the source's systemic velocity (as defined by the brightest \cs peak and consistent with the single component species, cf. section~\ref{section: spectra}) which is usually also the brightest component.

\subsection{Spectral fitting procedure}
\label{section: spectral fitting procedure}

The observed spectral lines originate from (rotationally or vibrationally) excited molecular gas with excitation temperature $\mathrm{T}_\mathrm{ex}$ and column density N. The dependence of observed line amplitude, width and shape on $\mathrm{T}_\mathrm{ex}$ and N is complex and requires solving the radiative transfer. This task is simplified substantially by software tools and the availability of cataloged line properties. In this work, we use \xclass for this task. \xclass models input data by solving the radiative transfer equation for an isothermal object in one dimension and optimizes the model using the \textsc{MAGIX} optimizer to fit the data.
The molecular parameters required for this task are obtained from the CDMS catalog.

The complexity of fitting up to 55 lines with $n \times 55$ parameters is challenging and cannot reliably be done in one go. We instead employ a three step method to obtain the best possible results: (1) an unconstrained, independent \xclass fit of each species individually for each source, followed by (2) a joint fit of all detected species to the whole data range for each source, and (3) refinement fits for selected species while keeping all other species fixed.

In the first run, we fit the species completely independently or include in the fit as few other lines as possible. This allows us to explore a large parameter range with a simple fit. 
A joint fit a of whole band is much more complicated due to the many, often blended, lines in each band and only succeeds if the parameter range is limited sufficiently by the first run. We restrict the fit parameters in the second run to the $16^\mathrm{th} - 84^\mathrm{th}$ percentile range of the respective parameter of the first run.
In the third run, we fit for the excitation temperature of the temperature sensitive species while keeping all other species fixed. 
This approach limits the degrees of freedom in each run to a manageable level. 

The parameters we fit for are column density, linewidth and line centroid. We fix the source size to unity which means the source completely covers the beam, a reasonable estimate for slightly unresolved SSCs \Leroy{p}. The species in our sample are typically detected with only a single line, so we have to assume an excitation temperature or infer it from the data. The latter cannot be done for all SSCs (see discussion in section \ref{section: ISM temperature}), so we fix $\mathrm{T_{rot}}=130$\,K for the purely rotationally excited species and $\mathrm{T_{vib}}=300$\,K for the ro-vibrational transitions. The former corresponds to the warm ISM component found by \citet{2013ApJ...779...33M} and \citet{Gorski:2017es} on lower spatial resolution. Using this value, our analysis is consistent with \Leroy{t}, who used the same value. As it turns out (Section~\ref{section: ISM temperature}), the mean SO$_2$ rotational temperature of the SSCs where this measurement is possible is 127\,K, hence $\mathrm{T_{rot}}=130$\,K is a good assumption. 
The assumption of $\mathrm{T_{vib}}=300$\,K is an educated guess and supposedly on the lower side for an SSC environment \citep[compared to less extreme Galactic hot cores as in][]{1983ApJ...274..184G,1999A&A...341..882W}.

In the third run, we fit SO$_2$ and H$_2$CS for the most unstable parameter, the excitation temperature $\mathrm{T_{rot}}$, in the range $25-1000$\,K. 
The detected SO$_2$ lines span a range in $\mathrm{E_{lower}}$ of $\sim40$\,K to $>300$\,K, 
the detected H$_2$CS lines cover $\mathrm{E_{lower}} = 50-770$\,K.
Both species should therefore allow for an estimate of their excitation temperatures.
Other species than these two do not cover enough range in the energy states or detected lines for successful fits with free temperature parameters.

For our XCLASS fits, we find that an under-/overestimation in the fixed excitation temperature ($\mathrm{T_{rot}}$, $\mathrm{T_{vib}}$) causes over-/underestimation in column density by similar factors. Fitted column densities must thus be understood with a systematic error of the same order than the assumed excitation temperature (see discussion in Section~\ref{section: ISM temperature}). 
For column density ratios of two species, this systematic effect affects both species similarly if the excitation temperature is equally under-/overestimated. 
Two species with similar excitation temperature are therefore similarly over-/underestimated by the fixed $\mathrm{T_{ex}}$ and the column density ratio is only weakly affected by the choice of $\mathrm{T_{ex}}$.
On the same note, relative comparisons between sources are meaningful if $\mathrm{T_{ex}}$ is similar between them.

\subsection{Error estimation}
\label{section: error estimation}

We estimate fit errors with a Monte Carlo scheme with 100 variations of the spectra to fit. For each variation, we add random draws of Gaussian noise to the observed spectra and thus sample the fit robustness against noise in the spectra which is the dominant source of error.
Errors are reported as the differences between median and $16^\mathrm{th}$ ($84^\mathrm{th}$) percentiles\footnote{For Gaussian distributions, this range corresponds to $\pm1\sigma$.} for lower (upper) error.

Details regarding the handling of blended lines, parameters and parameter ranges (especially a discussion of the choice of $\mathrm{T_{ex}}$ and its implications), fit algorithms and the Monte Carlo error estimation are described in Appendix~\ref{appendix: xclass}.

\subsection{Derivation of integrated intensity}\label{section: integrated intensity}
\xclass does not offer the possibility to also report integrated intensity since it is built on physical instead of observational quantities. We therefore use the final \xclass models to de-blend the spectra and derive integrated intensities for lines within the observed frequency range of the fitted species. This is done through Gaussian fits to the de-blended spectra. As such the integrated intensities inherit uncertainties due to blended lines and the Monte Carlo error estimation from the \xclass fits.

\subsection{Fitted spectra}\label{section: fitted spectra}

The observed spectra with the joint fit (third run) for all 14 SSCs in both sidebands are shown in Figure~\ref{figure: fitted bands}. Figure~\ref{figure: sample spectrum} provides an enlarged view of the spectra for SSC~14. 
The corresponding enlarged spectra for SSCs~1-13 can be found in Figure Set 2 in the online journal.
In general, the model spectra fit the observed data well with only small deviations. Spectral ranges with many blended lines are less well fit than isolated bright lines but the fit reliably disentangles the relative contribution of the species. 
The fit can deviate from the data slightly because \xclass fits all lines of a species simultaneously and respects the relations between lines. A particular line may thus be under-/overestimated if another line of the same species is found at lower/higher relative intensity. 
Some spectra, especially of the bright sources like SSC~14, show more spectral features than fitted, which will be studied in future work.
The fits prove to provide robust parameter estimates as indicated by the Monte Carlo error estimation. Noise fluctuation and repetition of the fit only marginably affect the results. Hence, the dominant source of error is the flux calibration and continuum subtraction (only relevant in the USB for SSCs~2, 3, 4, 8, 13, 14). Note that
the LTE approximation used by \xclass is also a severe source of uncertainty if the conditions in the SSCs are not close to LTE conditions.

%%%%%%%%%%%%%%%%%%%%%%%%%%%%%%%%%%%%%%%%%%%%%%%%%%%%%%%%%%%%%%%%%%%%%%%%%%%%%%%%%%%%%%%%%%%%%%%%%%%%
% intensity table
%%%%%%%%%%%%%%%%%%%%%%%%%%%%%%%%%%%%%%%%%%%%%%%%%%%%%%%%%%%%%%%%%%%%%%%%%%%%%%%%%%%%%%%%%%%%%%%%%%%%

\begin{figure}
\begin{rotatetable*}
% \floattable
\begin{deluxetable*}{lclCRRRRRRRRRRRRRR}
    \tabletypesize{\footnotesize}
    \tablecaption{Sample of the fitted integrated intensities in K\,\kms. The full table is available in the machine-readable format including error intervals in the online journal. The shown sample of the strongest spectral lines provides guidance regarding its form and content.
    \label{table: intensities}}
    \tablehead{\colhead{molecule} & \multicolumn{2}{c}{transition} & \colhead{comp.$^\dagger$} & \multicolumn{14}{c}{SSC no.}\\ \cline{2-3} \cline{5-18}
        & \colhead{rotational} & \colhead{vibrational} & & \colhead{1}&\colhead{2}&\colhead{3}&\colhead{4}&\colhead{5}&\colhead{6}&\colhead{7}&\colhead{8}&\colhead{9}&\colhead{10}&\colhead{11}&\colhead{12}&\colhead{13}&\colhead{14}}
    \startdata
CO             & 3--2                     & $\nu=0$         & 0 &  810 &  684 &  167 &  744 & 1607 &  199 &  149 & 1361 &  188 & 3004 & 1451 & 4132 &  161 &  127 \\
CO             & 3--2                     & $\nu=0$         & 1 & 1741 & 2328 & 1692 &  788 & 1014 & 3509 &  249 &  394 & 2919 &  418 & 2978 & 2962 & 3552 &  197 \\
CO             & 3--2                     & $\nu=0$         & 2 & 1737 & 1698 &  568 & 2654 & 2048 & 1862 &  324 &  702 & 2308 & 1462 & 1148 & 1749 & 1114 &  806 \\
CO             & 3--2                     & $\nu=0$         & 3 &  288 &  245 & 2187 &  724 & 1452 &  302 & 2678 &  496 &  212 & 2103 &  959 &  575 &  904 & 5486 \\
CO             & 3--2                     & $\nu=0$         & 4 &  551 &  367 &  544 &  501 & 1030 &  491 & 3327 & 3578 & 1096 &  747 & 1148 &  762 &  406 &  517 \\
CO             & 3--2                     & $\nu=0$         & 5 &  ... &  ... &  ... &  ... &  ... &  ... &  572 &  232 & 1282 &  ... & 2593 &  ... &  ... &  ... \\
HCO$^+$        & 4--3                     & $\nu=0$         & 0 &  366 &  736 &  630 &  321 & 2460 &  638 &  527 &  812 &  327 &  884 &  463 &  616 &  604 &  246 \\
HCO$^+$        & 4--3                     & $\nu=0$         & 1 &  278 &  212 &  333 &  352 &  ... &  499 &  157 &  647 &  840 &  295 &  600 &  391 &  153 & 2591 \\
HCO$^+$        & 4--3                     & $\nu=0$         & 2 &  ... &  ... &  ... &   43 &  ... &  ... &  209 &  280 &  208 &  ... &  715 &  317 &  104 &  ... \\
HCN            & 4--3                     & $\nu=0$         & 0 &  354 &  727 &  547 &  234 & 2230 & 1021 &  263 &  798 &  406 &  736 &  556 &  678 &  267 &  477 \\
HCN            & 4--3                     & $\nu=0$         & 1 &  201 &  127 &  286 &  324 &  ... &  ... &  209 &  569 &  537 &  290 &  581 &  404 &  398 & 2399 \\
HCN            & 4--3                     & $\nu=0$         & 2 &  ... &  ... &  ... &  ... &  ... &  ... &  178 &  289 &  137 &  ... &  527 &  280 &  124 &  ... \\
HCN            & 4--3                     & $\nu_2=2$, l=2f & 0 &    6 &   25 &   12 &   28 &   28 &  ... &  ... &    9 &  ... &  ... &  ... &  ... &   41 &  141 \\
HCN            & 4--3                     & $\nu_2=2$, l=2e & 0 &    6 &   25 &   12 &   28 &   28 &  ... &  ... &    9 &  ... &  ... &  ... &  ... &   41 &  136 \\
HCN            & 4--3                     & $\nu_2=2$, l=0  & 0 &    9 &   36 &   17 &   40 &   39 &  ... &  ... &   13 &  ... &  ... &  ... &  ... &   59 &  197 \\
HCN            & 4--3                     & $\nu_2=1$, l=1f & 0 &   55 &  127 &   93 &   79 &  283 &  ... &  ... &  212 &  ... &  ... &   34 &  ... &  288 & 1179 \\
H$^{13}$CN     & 4--3                     & $\nu=0$         & 0 &   32 &   53 &   41 &   13 &  370 &   34 &  ... &  177 &   51 &   11 &   92 &   69 &  124 &  697 \\
HC$^{15}$N     & 4--3                     & $\nu=0$         & 0 &  ... &   61 &   52 &   22 &  167 &  ... &  ... &   76 &  ... &  ... &  ... &  ... &  130 &  534 \\
H$^{15}$NC     & 4--3                     & $\nu=0$         & 0 &   17 &   39 &   31 &  ... &   71 &  ... &  ... &   27 &  ... &  ... &  ... &  ... &   37 &  135 \\
CS             & 7--6                     & $\nu=0$         & 0 &   97 &  103 &   61 &  209 & 1452 &  118 &  126 &  654 &  337 &  238 &  498 &  551 &  282 & 2116 \\
CS             & 7--6                     & $\nu=0$         & 1 &  172 &  268 &  132 &  ... &  ... &  ... &   40 &  ... &  ... &   79 &  ... &  ... &  238 &  285 \\
CS             & 7--6                     & $\nu=0$         & 2 &  ... &  ... &  151 &  ... &  ... &  ... &  ... &  ... &  ... &  ... &  ... &  ... &   54 &  ... \\
    \enddata
    \vspace{1.5\baselineskip}
    \tablenotetext{\dagger}{Components do not necessarily correspond to each other. For instance, the undetected marks (...) for component 5 of \co in SSC~1 merely indicate that five components fit the spectrum sufficiently well whereas in SSC~7 six components are required.}
    \tablecomments{The typical intensity error is 13.0\% and consists primarily of uncertainty due to line crowding. Additionally, the systematic flux uncertainty applies with $\lesssim 5\%$ for these observations according to the ALMA specifications.
    }
\end{deluxetable*}
\end{rotatetable*}
\end{figure}

%%%%%%%%%%%%%%%%%%%%%%%%%%%%%%%%%%%%%%%%%%%%%%%%%%%%%%%%%%%%%%%%%%%%%%%%%%%%%%%%%%%%%%%%%%%%%%%%%%%%
% fit table
%%%%%%%%%%%%%%%%%%%%%%%%%%%%%%%%%%%%%%%%%%%%%%%%%%%%%%%%%%%%%%%%%%%%%%%%%%%%%%%%%%%%%%%%%%%%%%%%%%%%

\floattable
\begin{deluxetable*}{rlccCRRRRR}
    \tablecaption{Sample of the parameters fitted with XCLASS. The full table is available in the machine-readable format including error intervals in the online journal. The shown sample for SSC~1 provides guidance regarding its form and content.
    \label{table: column densities}}
    \tablehead{\colhead{SSC} & \colhead{molecule} & \colhead{vibration} & \colhead{component} & \colhead{$\log \mathrm{N}$} & \colhead{$\sigma$} & \colhead{$\mu$} & \colhead{T$_\mathrm{kin}$} & \colhead{$\tau_\mathrm{int}$} \\
                & & & & [\mathrm{cm}^{-2}] & [\mathrm{km}\,\mathrm{s}^{-1}] & [\mathrm{km}\,\mathrm{s}^{-1}] & [\mathrm{K}] & \\
                & & & (1) & (2) & (3) & (4) & (5) & (6)}
    \startdata
1 & CO           & $\nu=0$      & 0 & 17.85 & 40.0 &  -44.6 & 130 &  3.32 \\
1 & CO           & $\nu=0$      & 1 & 18.23 & 33.3 &  -22.1 & 130 &  7.89 \\
1 & CO           & $\nu=0$      & 2 & 18.24 & 29.0 &   20.3 & 130 &  8.08 \\
1 & CO           & $\nu=0$      & 3 & 17.39 & 30.0 &   68.7 & 130 &  1.15 \\
1 & CO           & $\nu=0$      & 4 & 17.68 & 39.6 &  106.2 & 130 &  2.22 \\
1 & HCO$^+$      & $\nu=0$      & 0 & 14.40 & 39.4 &  -34.3 & 130 &  1.50 \\
1 & HCO$^+$      & $\nu=0$      & 1 & 14.28 & 22.7 &   24.2 & 130 &  1.15 \\
1 & HCN          & $\nu=0$      & 0 & 14.62 & 37.4 &  -37.1 & 130 &  1.44 \\
1 & HCN          & $\nu=0$      & 1 & 14.38 & 20.5 &   24.6 & 130 &  0.82 \\
1 & HCN          & $\nu_2=2$    & 0 & 16.29 & 31.9 &   -5.3 & 300 &  0.04 \\
1 & HCN          & $\nu_2=1$    & 0 & 15.64 & 30.5 &    7.5 & 300 &  0.18 \\
1 & H$^{13}$CN   & $\nu=0$      & 0 & 13.59 & 20.3 &   -3.9 & 130 &  0.13 \\
1 & H$^{15}$NC   & $\nu=0$      & 0 & 13.37 & 18.1 &   10.0 & 130 &  0.07 \\
1 & CS           & $\nu=0$      & 0 & 14.53 & 33.9 &  -21.7 & 130 &  0.38 \\
1 & CS           & $\nu=0$      & 1 & 14.78 & 32.7 &    3.3 & 130 &  0.67 \\
1 & HC$_3$N      & $\nu=0$      & 0 & 14.66 & 50.0 &   -0.3 & 130 &  0.43 \\
1 & HC$_3$N      & $\nu_6=1$    & 0 & 14.95 & 26.5 &   -0.7 & 300 &  0.05 \\
1 & SO           & $\nu=0$      & 0 & 15.04 & 21.6 &    8.8 & 130 &  0.51 \\
1 & S$^{18}$O    & $\nu=0$      & 0 & 14.85 & 48.6 &    9.9 & 130 &  0.14 \\
1 & SO$_2$       & $\nu=0$      & 0 & 15.52 & 35.1 &    1.9 &  84 &  3.53 \\
    \enddata
    \tablecomments{(1) The numbers assigned to the fitted components do not necessarily correspond to each other.\\
    (2) Molecular column density.\\
    (3) Velocity dispersion.\\
    (4) Line centroid position with respect to the systemic velocity of each SSC (cf. \ref{section: spectra}).\\
    (5) Kinetic temperature is fitted for SO$_2$ and H$_2$CS only and fixed to 130\,K and 300\,K for rotational and ro-vibrational species, respectively.\\
    (6) Total line opacity, i.e. integrated over the line.\\
    The typical column density error is 14.3\% and consists primarily of uncertainty due to line crowding. Additionally, the systematic flux uncertainty applies with $\lesssim 5\%$ for these observations according to the ALMA specifications.
    }
\end{deluxetable*}

%%%%%%%%%%%%%%%%%%%%%%%%%%%%%%%%%%%%%%%%%%%%%%%%%%%%%%%%%%%%%%%%%%%%%%%%%%%%%%%%%%%%%%%%%%%%%%%%%%%%
\section{Discussion}\label{section: discussion}
%%%%%%%%%%%%%%%%%%%%%%%%%%%%%%%%%%%%%%%%%%%%%%%%%%%%%%%%%%%%%%%%%%%%%%%%%%%%%%%%%%%%%%%%%%%%%%%%%%%%

The ALMA dataset allows us to study the ISM properties in the (proto-)SSCs in the core of a starburst. In the following, we focus on line ratios and selected properties for these objects.

\subsection{Chemical composition of the SSCs}

Table~\ref{table: intensities} lists the detected lines with their integrated intensity for all 14 SSCs examined in this study.
SSC~14 is the most chemically rich proto-cluster in NGC~253 where we detect 19 species with 55 spectral lines. In SSC~2, 3 and 13 all species but $^{34}$SO$_2$ are detected. Only the bright species CO, HCN, HCO$^+$ and CS are detected in SSC~7 which is the poorest chemical repertoire detected in our sample.
The vast differences in detection rate is affected by brightness: SSCs 5 and 14 are significantly brighter in HCN, HCO$^+$ and CS than the other SSCs, which are of comparable brightness in these lines. There is also an effect of intrinsic chemical richness and excitation environment as e.g. SSCs $8-12$ are located within $\sim20$\,pc (projected) of each other and are comparable in total brightness in CS, HCN and HCO$^+$ but differ significantly in the amount of detected species and spectral lines.
SSC~14 is still the chemically richest SSC when considering its brightness. Even at 20\% of its actual HCN or HCO$^+$ brightness (comparable to the HCN and HCO$^+$ total brightness of SSCs~$1-4$, $6-7$, $9-13$), all species but $^{34}$SO and the faint HC$_3$N$^*$ would still be detected. The detected chemical richness of an SSC does not correlate with its location.

The 14 SSCs we study have been covered by \citet{2015ApJ...801...63M} in ALMA band~3 ($86-115$\,GHz) observations at $\sim 50$\,pc resolution and overlap with the star-forming clumps identified by \citet{2017ApJ...849...81A} in $\sim 7$\,pc ALMA band~7 ($340.2-343.4$\,GHz, $350.6-355.7$\,GHz, $362.2-365.2$\,GHz) observations. They also correspond to the ``super hot cores'' identified by \citet{2020MNRAS.491.4573R} through their analysis of the HC$_3$N vibrationally excited emission, which we discuss in Section~\ref{section: HC3N}. Our results of the chemical richness of the SSCs are qualitatively consistent with these studies. A detailed comparison with \citet{2015ApJ...801...63M} proves difficult because their resolution of $\sim 50$\,pc is not enough to resolve the SSCs and thus blends sources with varying chemical richness (e.g. their position~6 corresponds to SSCs~8-13). \citet{2017ApJ...849...81A} identify their clump~1 (corresponding to our SSC~14) as the most chemically rich clump, clumps~3 and 5 (SSCs~10+12 and 9) as the poorest and clumps~2, 4, 6, 7, and 8 (SSCs~13, 8, 5, 4, 2+3) as intermediate types. Our detection rates match that general picture, however, we place the SSCs corresponding to a few of the intermediate richness clumps towards higher or lower richness. Hence, high spatial resolution is required to separate out sources of strongly varying chemical richness.

The detection fraction of species does not significantly correlate with the (integrated) intensity of one of the bright lines CO, HCN, HCO$^+$ and CS. Peak intensities as well as integrated intensities of the strongest component in CO are comparable within a factor of two among the SSCs. In CS, HCN and HCO$^+$, peak intensity of the main component and also integrated intensity are correlated across lines. This might be due to CO tracing different gas than CS, HCN and HCO$^+$ as expected from their different critical densities and tracer properties or most likely CO is be optically thicker and saturated for all SSCs.

Especially in CO, but also in CS, HCN and HCO$^+$, multiple components of similar peak intensity are present (SSCs~1, 2, 3, 5, 10, 11, 12) while single components are detected in other species. This is already seen in the lower resolution study by \citet{2017ApJ...849...81A} in their clumps 3 (SSCs 2 and 3) and 8 (SSCs 10 and 12). In SSCs~1, 2, 3, 10 and potentially SSC~4, the center of the dip between double peaks in CO, HCN and HCO$^+$ coincides closely ($<5$\,\kms) with the peak position of other detected lines. 
Potential explanations are temperature gradients in the SSCs, self-absorption with in the gas or absorption against a background source.
Similar absorption features are often seen in HCN and HCO$^+$ in Galactic molecular clumps and sometimes associated to inflow and outflow motions \citep[e.g.][]{2016A&A...585A.149W}. Very similar line profiles are seen in the Galactic Center proto-SSC Sgr~B2 and attributed to self-absorption \citep{2017ApJ...835...76M}.
In SSC~1 and 10, the potential absorption is strong and approaches zero in HCN and HCO$^+$. We note, however, that none of these features drops below zero in our continuum-subtracted spectra, which would be a clear indication of absorption into the continuum.
These spectral features could also be interpreted structurally such that in CO, HCN and HCO$^+$, we see layers of surrounding gas at higher/lower velocity whereas in the other molecules we see the central region of the cloud. Absorption is more likely, though, especially when considering the lower critical densities and expected higher opacities of CO, HCN and HCO$^+$ relative to the other species which make CO, HCN and HCO$^+$ more prone to be affected by absorption. It is still noteworthy that only SSCs~1, 2, 3, (4) and 10 show noticeable dip (absorption) features although other sources are of similar or even higher brightness.
If these spectral features are indeed caused by absorption, the reported intensities and column densities are underestimated. Derived ratios must then be considered limits (lower/upper depending on the ratio) in SSCs~1, 2, 3, and 10.

%%%%%%%%%%%%%%%%%%%%%%%%%%%%%%%%%%%%%%%%%%%%%%%%%%%%%%%%%%%%%%%%%%%%%%%%%%%%%%%%%%%%%%%%%%%%%%%%%%%%

\floattable
\begin{deluxetable*}{r|cccc|cccc}
    \tablewidth{\linewidth}
    \tablecaption{Dense gas fractions specified by line intensity ratios and column density ratios.\label{table: ratios dense gas}}
    \tablehead{\colhead{SSC} & \multicolumn{4}{c}{line intensity ratio R$_\mathrm{I}$} & \multicolumn{4}{c}{column density ratio R$_\mathrm{N}$}\\
        \colhead{} & \colhead{CO/HCN} & \colhead{CO/CS} & \colhead{HCN/HCO$^+$} & \colhead{CS/HCN} & \colhead{CO/HCN} & \colhead{CO/CS} & \colhead{HCN/HCO$^+$} & \colhead{CS/HCN}}
    \startdata
1 & $ 4.9^{+ 0.2}_{- 0.6}$ & $10.1^{+ 9.3}_{- 2.2}$ & $ 1.0^{+ 0.1}_{- 0.0}$ & $ 0.5^{+ 0.1}_{- 0.2}$ & $ 4100^{+  190}_{-  500}$ & $ 2900^{+ 2700}_{-  650}$ & $  1.7^{+  0.1}_{-  0.0}$ & $  1.4^{+  0.4}_{-  0.7}$\\
2 & $ 3.3^{+ 0.8}_{- 0.9}$ & $ 9.0^{+ 9.1}_{- 3.9}$ & $ 1.0^{+ 0.0}_{- 0.0}$ & $ 0.4^{+ 0.2}_{- 0.2}$ & $ 2700^{+  650}_{-  710}$ & $ 2600^{+ 2300}_{- 1200}$ & $  1.7^{+  0.1}_{-  0.1}$ & $  1.1^{+  0.7}_{-  0.5}$\\
3 & $ 4.0^{+ 0.1}_{- 0.1}$ & $14.5^{+ 1.5}_{- 1.4}$ & $ 0.9^{+ 0.1}_{- 0.0}$ & $ 0.3^{+ 0.0}_{- 0.0}$ & $ 3300^{+  140}_{-   72}$ & $ 4100^{+  430}_{-  380}$ & $  1.5^{+  0.1}_{-  0.0}$ & $  0.8^{+  0.1}_{-  0.1}$\\
4 & $ 8.0^{+ 1.4}_{- 1.8}$ & $12.5^{+ 0.8}_{- 2.5}$ & $ 0.9^{+ 0.2}_{- 0.3}$ & $ 0.6^{+ 0.1}_{- 0.1}$ & $ 7700^{+ 1300}_{- 2000}$ & $ 4200^{+  270}_{-  520}$ & $  1.6^{+  0.4}_{-  0.4}$ & $  1.9^{+  0.3}_{-  0.3}$\\
5 & $ 0.9^{+ 0.0}_{- 0.0}$ & $ 1.4^{+ 0.0}_{- 0.0}$ & $ 0.9^{+ 0.0}_{- 0.0}$ & $ 0.7^{+ 0.0}_{- 0.0}$ & $  760^{+    9}_{-   17}$ & $  400^{+    4}_{-    7}$ & $  1.5^{+  0.0}_{-  0.0}$ & $  1.9^{+  0.0}_{-  0.0}$\\
6 & $ 3.4^{+ 0.2}_{- 0.3}$ & $29.1^{+ 3.2}_{- 2.4}$ & $ 1.6^{+ 0.3}_{- 0.2}$ & $ 0.1^{+ 0.0}_{- 0.0}$ & $ 2900^{+  150}_{-  260}$ & $ 8700^{+ 1000}_{-  780}$ & $  2.8^{+  0.6}_{-  0.3}$ & $  0.3^{+  0.0}_{-  0.0}$\\
7 & $12.7^{+ 2.5}_{- 2.8}$ & $26.1^{+ 4.7}_{- 2.1}$ & $ 0.5^{+ 0.3}_{- 0.1}$ & $ 0.5^{+ 0.1}_{- 0.2}$ & $11000^{+ 2200}_{- 2300}$ & $ 7400^{+ 1200}_{-  630}$ & $  0.8^{+  0.5}_{-  0.2}$ & $  1.4^{+  0.3}_{-  0.5}$\\
8 & $ 4.5^{+ 7.6}_{- 1.2}$ & $ 5.5^{+ 0.1}_{- 0.1}$ & $ 0.9^{+ 0.6}_{- 0.5}$ & $ 0.8^{+ 1.4}_{- 0.2}$ & $ 4500^{+ 7500}_{- 1300}$ & $ 1900^{+   41}_{-   42}$ & $  1.6^{+  1.1}_{-  0.8}$ & $  2.4^{+  3.7}_{-  0.7}$\\
9 & $ 5.5^{+ 4.4}_{- 1.7}$ & $ 8.5^{+ 0.9}_{- 1.5}$ & $ 0.6^{+ 0.4}_{- 0.2}$ & $ 0.6^{+ 0.7}_{- 0.2}$ & $ 5000^{+ 4300}_{- 1600}$ & $ 2700^{+  320}_{-  460}$ & $  1.1^{+  0.7}_{-  0.4}$ & $  1.8^{+  2.1}_{-  0.5}$\\
10 & $ 4.0^{+ 0.6}_{- 1.6}$ & $11.9^{+ 9.2}_{- 4.6}$ & $ 0.8^{+ 0.0}_{- 0.0}$ & $ 0.3^{+ 0.1}_{- 0.2}$ & $ 3700^{+  360}_{- 1700}$ & $ 3600^{+ 2600}_{- 1600}$ & $  1.4^{+  0.0}_{-  0.0}$ & $  1.0^{+  0.2}_{-  0.5}$\\
11 & $ 5.0^{+ 4.9}_{- 1.8}$ & $ 5.9^{+ 1.9}_{- 1.1}$ & $ 0.8^{+ 0.5}_{- 0.4}$ & $ 0.8^{+ 0.7}_{- 0.3}$ & $ 4600^{+ 4700}_{- 1800}$ & $ 1800^{+  620}_{-  440}$ & $  1.4^{+  0.8}_{-  0.6}$ & $  2.5^{+  2.1}_{-  0.9}$\\
12 & $ 6.3^{+ 1.0}_{- 2.3}$ & $ 7.8^{+ 1.2}_{- 2.9}$ & $ 1.1^{+ 0.1}_{- 0.1}$ & $ 0.8^{+ 0.0}_{- 0.0}$ & $ 5700^{+  990}_{- 2400}$ & $ 2400^{+  430}_{- 1000}$ & $  1.9^{+  0.1}_{-  0.1}$ & $  2.4^{+  0.1}_{-  0.1}$\\
13 & $ 9.3^{+ 1.4}_{- 5.1}$ & $11.4^{+ 6.6}_{- 5.4}$ & $ 0.7^{+ 0.4}_{- 0.2}$ & $ 0.7^{+ 0.3}_{- 0.3}$ & $ 8400^{+ 1200}_{- 3100}$ & $ 3500^{+ 2000}_{- 1800}$ & $  1.2^{+  0.7}_{-  0.3}$ & $  2.0^{+  0.9}_{-  0.8}$\\
14 & $ 2.3^{+ 0.0}_{- 0.1}$ & $ 2.6^{+ 0.2}_{- 0.1}$ & $ 0.9^{+ 0.0}_{- 0.0}$ & $ 0.9^{+ 0.0}_{- 0.1}$ & $ 2400^{+   48}_{-   74}$ & $  960^{+   68}_{-   30}$ & $  1.6^{+  0.0}_{-  0.0}$ & $  2.5^{+  0.1}_{-  0.2}$\\
    \enddata
    \tablecomments{The ratios above are derived from the following rotational transitions: \co, \hcn, \hco and \cs. Errors of 0.0 are due to rounding. The line intensity ratios R$_\mathrm{I}$ are derived from Gaussian fitting as explained in Section~\ref{section: integrated intensity} while the column density ratios R$_\mathrm{N}$ result directly from XCLASS (see Section~\ref{section: spectral fitting procedure} for details).}
\end{deluxetable*}

\floattable
\begin{deluxetable*}{r|ccccccc}
    \tablewidth{\linewidth}
    \tablecaption{Line intensity ratios of selected species.\label{table: ratios other}}
    \tablehead{\colhead{SSC} & \colhead{HC$^{15}$N/H$^{15}$NC} & \colhead{HCN/H$^{13}$CN} & \colhead{HCN/HC$^{15}$N} & \colhead{SO/S$^{18}$O} & \colhead{HCN/HC$_3$N} & \colhead{SO/SO$_2$} & \colhead{CS/SO$_2$}}
    \startdata
 1 &                    ... & $ 6.4^{+ 1.5}_{- 1.0}$ &                    ... & $ 1.7^{+ 0.8}_{- 0.3}$ & $ 3.5^{+ 0.8}_{- 0.7}$ & $ 1.7^{+ 0.3}_{- 0.2}$ & $ 5.0^{+ 1.6}_{- 2.6}$\\
 2 & $ 1.5^{+ 0.5}_{- 0.3}$ & $13.5^{+ 0.9}_{- 1.6}$ & $11.8^{+ 1.3}_{- 1.9}$ & $14.2^{+ 1.5}_{- 7.9}$ & $ 5.5^{+ 1.1}_{- 0.5}$ & $ 2.3^{+ 0.1}_{- 0.1}$ & $ 3.9^{+ 2.3}_{- 1.9}$\\
 3 & $ 1.6^{+ 0.5}_{- 0.5}$ & $ 6.8^{+ 0.4}_{- 0.3}$ & $ 5.5^{+ 1.6}_{- 0.8}$ & $1218^{+  78}_{-  99}$ & $ 4.0^{+ 0.5}_{- 0.3}$ & $ 1.4^{+ 0.1}_{- 0.1}$ & $ 1.6^{+ 0.4}_{- 0.3}$\\
 4 &                    ... & $17.0^{+ 4.5}_{- 7.0}$ & $10.3^{+ 3.6}_{- 3.9}$ & $ 8.5^{+ 4.7}_{- 3.2}$ & $ 2.9^{+ 0.8}_{- 1.3}$ & $ 1.5^{+ 0.1}_{- 0.2}$ & $ 3.5^{+ 0.3}_{- 0.3}$\\
 5 & $ 2.3^{+ 0.1}_{- 0.2}$ & $ 6.0^{+ 0.2}_{- 0.2}$ & $13.4^{+ 0.7}_{- 0.9}$ & $67.2^{+ 2.4}_{-17.6}$ & $ 9.9^{+ 0.5}_{- 0.4}$ & $ 4.4^{+ 0.2}_{- 0.2}$ & $14.6^{+ 0.5}_{- 0.4}$\\
 6 &                    ... & $30.5^{+ 6.1}_{- 3.7}$ &                    ... &                    ... &                    ... &                    ... &                    ...\\
 7 &                    ... &                    ... &                    ... &                    ... &                    ... &                    ... &                    ...\\
 8 & $ 3.0^{+ 0.8}_{- 0.6}$ & $ 3.2^{+ 2.1}_{- 2.2}$ & $ 6.9^{+ 5.0}_{- 4.6}$ & $15.8^{+ 6.9}_{- 5.3}$ & $ 3.9^{+ 3.2}_{- 2.4}$ & $ 4.9^{+ 0.3}_{- 0.4}$ & $11.1^{+ 0.5}_{- 0.7}$\\
 9 &                    ... & $ 7.7^{+ 3.8}_{- 3.2}$ &                    ... &                    ... &                    ... &                    ... &                    ...\\
10 &                    ... & $66.0^{+ 2.3}_{- 9.1}$ &                    ... &                    ... & $13.3^{+ 3.9}_{- 2.7}$ & $ 3.0^{+ 0.6}_{- 0.5}$ & $ 8.6^{+ 2.6}_{- 3.7}$\\
11 &                    ... & $ 6.2^{+ 4.1}_{- 3.1}$ &                    ... &                    ... & $ 5.9^{+ 3.4}_{- 2.6}$ & $ 6.3^{+ 1.6}_{- 1.4}$ & $26.9^{+ 6.6}_{- 5.5}$\\
12 &                    ... & $ 9.7^{+ 1.5}_{- 1.3}$ &                    ... &                    ... &                    ... & $ 1.5^{+ 0.5}_{- 0.3}$ & $15.1^{+ 1.3}_{- 1.8}$\\
13 & $ 3.4^{+ 1.3}_{- 1.6}$ & $ 3.2^{+ 0.5}_{- 0.5}$ & $ 3.0^{+ 0.6}_{- 0.6}$ & $ 499^{+ 128}_{-  57}$ & $ 2.1^{+ 0.3}_{- 0.3}$ & $ 2.6^{+ 0.1}_{- 0.1}$ & $ 2.6^{+ 0.8}_{- 0.7}$\\
14 & $ 4.0^{+ 0.5}_{- 0.5}$ & $ 3.5^{+ 0.1}_{- 0.1}$ & $ 4.5^{+ 0.2}_{- 0.2}$ & $94.1^{+ 108}_{-42.5}$ & $ 3.0^{+ 0.1}_{- 0.1}$ & $ 4.3^{+ 0.3}_{- 0.1}$ & $ 8.1^{+ 0.6}_{- 0.6}$\\
    \enddata
    \tablecomments{The ratios above are derived from the following rotational transitions: \hcn, \mbox{H$^{13}$CN(4--3)}, HC$^{15}$N(4--3), H$^{15}$NC(4--3), HC$_{3}$N(38--37), \cs, SO($8_8$--$7_7$), S$^{18}$O ($8_9$--$7_8$) and SO$_2$($11_{4,8}$--$11_{3,9}$).}
\end{deluxetable*}

%%%%%%%%%%%%%%%%%%%%%%%%%%%%%%%%%%%%%%%%%%%%%%%%%%%%%%%%%%%%%%%%%%%%%%%%%%%%%%%%%%%%%%%%%%%%%%%%%%%%

\subsection{Dense gas}\label{section: dense gas}

HCN, HCO$^+$ and other molecules are often referred to as dense gas tracers, but it remains unclear what ``dense gas'' is and how well it is traced by these species.
In the extra-galactic community, HCN and HCO$^+$ transitions are typically considered to trace densities similar to their critical densities, and thus they are used to label molecular gas at $n > 10^5$\,\pcm3 as ``dense''. Historically, these species used to be among the few molecular species detectable in nearby galaxies due to their brightness.
Newer studies in the Milky Way, however, show that HCN emission also arises from less dense regions down a few 100\,\pcm3 and therefore well below the critical density \citep[e.g.][]{2015PASP..127..299S,2017A&A...605L...5K,2017A&A...599A..98P}.
In this section, we refer to HCN, HCO$^+$, HNC and CS as \emph{dense gas tracers} in the sense of tracers of \emph{molecular gas denser than that traced by CO}.

Table~\ref{table: ratios dense gas} lists the line ratios of said dense gas tracers with each other and CO. The involved species are detected with multiple components, some of which deviate from the systemic velocity of the SSC inferred from less complex lines as discussed above. For table~\ref{table: ratios dense gas}, we select the spectral component closest to said systemic velocity to focus on the SSCs instead of foreground (or background) emission.
We do not show CO/HCO$^+$ because it is similar to CO/HCN as shown by the HCN/HCO$^+$ ratio.

\subsubsection{Ratios with CO}

Keeping in mind aforementioned caveats about dense gas tracers, the ratios of HCN, HCO$^+$ and CS with CO translate to a proxy of the dense gas fraction \citep[e.g.,][]{2018ApJ...858...90G}. HCN most likely arises from gas with $n>10^5$\,\pcm3, but can also be excited below these densities by infrared pumping or electron collisions \citep[e.g.][]{1995ApJ...438..695J,1997ApJ...484..656P,2007ApJ...666..156K}. \cs has a critical density of $\sim 3 \times 10^7$\,\pcm3 and emits most effectively at $>60$\,K thus tracing warm dense gas.

Both the CO/HCN and CO/CS ratios are lower than what has been found globally in e.g. LIRGs \citep[CO(1--0)/HCN(1--0) = 12.5-100, CO(1--0)/CS(3--2) = 9-100]{Baan:2008hx} or a wide sample of galaxies \citep{2004ApJS..152...63G}. \citet{2004ApJS..152...63G} list CO(1--0)/HCN(1--0) = 17 on galaxy scales in NGC~253. \citet{Sakamoto:2011et} found CO(3--2)/HCN(4--3) $=12-17$ in molecular complexes at $\sim 30$\,pc resolution.
At similar resolution of 32\,pc, \citet{2015ApJ...801...63M} report CO(1--0)/HCN(1--0) of $\sim 6.0$ to $\sim 10$ across 6 locations in the central starburst disk.

Since we observe the higher $J$-transitions CO(3--2) and HCN(4--3) instead of the (1--0) transistions in the literature, the line ratios are not directly comparable until excitation has been considered. 
For the typical column densities, line widths and temperatures observed in our data, a simple estimation with the \textsc{radex} \citep{2007A&A...468..627V} radiative transfer code\footnote{Run using the online version at \url{http://var.sron.nl/radex/radex.php}.} suggests CO(3--2)/HCN(4--3) to be about twice the CO(1--0)/HCN(1--0) line ratio. 
Instead, we observe $\sim 1.5-2$ times lower CO(3--2)/HCN(4--3) ratios compared to the \citet{2015ApJ...801...63M} CO(1--0)/HCN(1--0) and even lower relative to other literature values. 
The difference between our line ratios and the literature therefore cannot be explained by the higher excitation state of the lines observed here. Instead, the difference must be caused by selection and/or density effects.
This may be due to the fact that we select on regions of actively star-forming dense gas while the resolution of previous studies yield GMC scale averages.

The very low CO/HCN and CO/CS ratios in SSC~5 are noteworthy. The peak intensity differs by a factor of $\sim 1.5$ but the linewidth of the main component appears much more narrow in CO, so both end up at similar integrated intensity. In this source, CO shows a wide distribution of peaks, some of which might be caused by absorption, whereas CS, HCN and HCO$^+$ have only one peak very close to Gaussian in shape. The other CO components in SSC~5 do not contain relevant amounts of dense gas, apparently.

Similar issues must be kept in mind for the other SSCs. We try to fit the component matching the other species most closely in velocity but this assumption may not be exactly valid because of the unknown influence of absorption. Due to these effects, we caution against overinterpretating the line ratios involving CO and to a lesser degree also CS, HCN and HCO$^+$.
Depending on the unknown strength and frequency of absorption, sources may be affected to a varying degree.

\subsubsection[HCN/HCO+]{HCN/HCO$^+$}\label{section: HCN/HCO+}

The HCN/HCO$^+$ line ratio is proposed to differentiate starbursts from AGN on an empirical basis \citep{2001ASPC..249..672K,Krips:2008dg}. The theoretical background of this tracer property is not yet clear but probably related to high temperatures and/or mechanical heating \citep{2013PASJ...65..100I,2012A&A...537A..44A}. Seyferts usually show HCN/HCO$^+ >1$ while starbursts often have ratios $\lesssim 1$ due to HCO$^+$ enhancement by PDRs \citep{2001ASPC..249..672K,Krips:2008dg,2009AJ....137.3581I,2013PASJ...65..100I,Izumi:2016js,2015A&A...573A.116M,2015ApJ...801...63M,2015ApJ...814...39P,2016ApJ...825...44I}.
Exceptions to the usually good diagnostic power of HCN/HCO$^+$ are known in the literature \citep[e.g.][]{2015ApJ...814...39P}.

The HCN/HCO$^+$ line ratios in all 14 SSCs are close to unity and thus within the transition regime between AGN and starburst. SSC~6 is the only source with a ratio significantly greater than unity whereas all other sources have ratios below or consistent with unity. Any hypothetical contribution of an highly obscured AGN should show in the SSCs near the stellar kinematic center \citep[$\alpha,\delta = 00^h 47^m 33.134^s, -25^\circ 17^\prime 19.68\arcsec$;][]{MullerSanchez:2010dr} and the continuum source TH2 \citep[$\alpha,\delta = 00^h 47^m 31.2^s, -25^\circ 17^\prime 17\arcsec$;][]{1997ApJ...488..621U} proposed to be an AGN, both of which are close to SSCs~8-12. Signs of AGN activity have not been detected in NGC~253 and SSC~6 is located $\sim 50$\,pc (projected) from the kinematic center, so this cannot explain the slightly enhanced ratio. NGC~253 or at least the SSCs within the starburst are hence an exception to the aforementioned tracer properties of the HCN/HCO$^+$ ratio. 
If the HCN/HCO$^+$ intensity ratio reflects the abundance ratio, HCN/HCO$^+ \sim 1$ would indicate that HCO$^+$ is not (yet) enhanced as some studies suggest for a starburst environment \citep[e.g.][]{Krips:2008dg}. Potentially, the SSCs are still too young to have enhanced the HCO$^+$ abundance in PDRs through feedback. Following this line of argument, the SSCs must be either too young to have developed PDRs or the PDR is still embedded in the cluster and does not affect the observable gas at the outer shells of the cloud (cf. opacity discussion below and in section~\ref{section: optical depth}).

Line ratios of order unity are also expected if both lines are subject to significant opacity and thermalised. As discussed in section~\ref{section: optical depth}, opacities in HCN are high at $\tau \gtrsim 1$ and similar values are expected for HCO$^+$ in the same source. The measured line ratios are therefore most likely set by opacity effects rather than abundance variations.

On larger scales, HCN/HCO$^+ \sim 1$ was found before. \citet{2015ApJ...801...63M} report HCN/HCO$^+ \sim 1.1$ for ten regions across the central molecular zone, and \citet{2007ApJ...666..156K} and \citet{JimenezDonaire:2017eg} find HCN/HCO$^+ = 1.1 \pm 0.2$ ($1.1 \pm 0.1$, respectively) over the central region of NGC~253. 
The combined HCN (HCO$^+$) intensity of the SSCs amounts to only 0.15\% (0.13\%) of the total HCN (HCO$^+$) reported in \citet{2007ApJ...666..156K} in the 4--3 state. This means the galactic ratios are not dominated or influenced to a relevant degree by emission associated with the SSCs.
Therefore, the SSCs do not deviate significantly from their large scale surrounding in terms of their HCN/HCO+ line ratio.

\subsubsection{CS/HCN}\label{section: CS/HCN}

The CS/HCN ratio was proposed as a potential starburst/AGN tracer by tracing PDR/XDR environments where enhanced CS/HCN corresponds to XDR conditions \citep{2007A&A...461..793M,2013PASJ...65..100I,Izumi:2016js}. \citet{2005ApJ...618..259M} suggest that CS is a PDR tracer because it better correlated with other known molecular PDR tracers (C$_2$H).

In NGC~253, the CS/HCN ratio scatters significantly across SSCs and does not correlate with the other dense gas tracer ratios in Table~\ref{table: ratios dense gas}. It seems to correlate with the chemical richness, expressed as the number of detected species, in an SSC. 
There are no significant correlations between CS/HCN and SSC properties listed by \Leroy{t} (virial mass, gas mass, stellar mass, surface and volume density, freefall time, pressure).
According to the mentioned PDR/XDR models by \citet{2007A&A...461..793M,2013PASJ...65..100I,Izumi:2016js}, the obtained CS/HCN ratios are consistent with PDR chemistry under the reasonable assumption that the density of the emitting gas is within factors of a few of $10^5$\,\pcm3. For XDR chemistry, CS/HCN$>1$ would be expected.

\subsubsection{HCN/HNC}\label{section: HCN/HNC}

HCN/HNC $<1$ is usually a sign of cold (10-20\,K) dark clouds but can also occur through mid-IR pumping and XDRs in (U)LIRGs, starbursts or active galaxies \citep[e.g.][]{1992A&A...256..595S,2007A&A...464..193A,Baan:2008hx}. Some (U)LIRGS show bright HNC where the HNC might even be brighter than HCN \citep{1995A&A...295..571H,2002A&A...381..783A,2015A&A...579A.101A}. Such low HCN/HNC ratios are possible through chemical reactions at moderate temperatures and densities, infrared pumping but also high optical depth in HCN \citep{2002A&A...381..783A}. \citet{2014ApJ...787...74G} argue that HNC and HCN have a great potential as a temperature tracer because the formation and destruction paths are temperature dependent with increasing HCN/HNC for increasing temperature. This line ratio can also be a proxy for the evolutionary stage on a starburst where HCN/HNC $\gg 1$ indicate shock-dominated regions common in early starbursts \citep{2015A&A...579A.101A}. \citet{Loenen:2008fb} show a PDR/XDR model and show that the HCN/HNC line ratio can differentiate between PDR and XDR. In the case of PDRs, this model can further discern between density regimes based on the HCN/HCO$^+$ ratio.
There seems to be no final consensus about the tracer properties of the HCN/HNC ratio in the literature. It seems likely, however, that this ratio is related to PDR/XDR chemistry.

Table~\ref{table: ratios other} lists the line ratios of HCN and its isomer HNC. This ratio is measured in the $^{15}$N species and is thus less affected by optical depth effects (cf. Section~\ref{section: optical depth}). The line ratios in SSC~2 and 3 are almost consistent with unity while the SSCs 5, 8, 13 and 14 are significantly higher at $\sim 2-4$. In the literature, H$^{13}$CN/HN$^{13}$C$ \sim 2.5$ is reported for ten $\sim2$\arcsec regions across the center of NGC~253 \citep{2015ApJ...801...63M} which should be comparable to first order since both $^{13}$C and $^{15}$N are significantly less abundant than the main isotopologues. SSCs 5, 8, 13 and 14 are thus similar in HC$^{15}$N/H$^{15}$NC to the large scale ratio while SSC~2 and 3 deviate towards lower ratios.

The generally low ratios in SSC~2 and 3, and to a lesser degree also in SSC~5, 8, 13 and 14 might be explained by radiative pumping. 
This is further substantiated by the detection of vibrationally excited HCN in these sources, which are excited via infrared pumping (cf. Section~\ref{section: vibrational excitation}).
In the model of \citet{2014ApJ...787...74G}, the observed HCN/HNC ratios imply temperatures $<35$\,K which is unlikely in a starbursting environment (also see section~\ref{section: ISM temperature} on temperature measurements)\footnote{If the reaction barrier of 200\,K is used. Their alternative model with 1200\,K reaction barrier does not show a temperature dependence of the HCN/HNC ratio.}. 
In the \citet{Loenen:2008fb} model, the measured HCN/HNC implies that none of the SSCs contains an XDR but all are characterized by PDR chemistry at $n \sim 10^5$\,\pcm3 as we discuss in Section~\ref{section: energy source}.

\subsubsection{SSC energy source}\label{section: energy source}

\begin{figure*}
    \centering
    \includegraphics[width=0.58\linewidth]{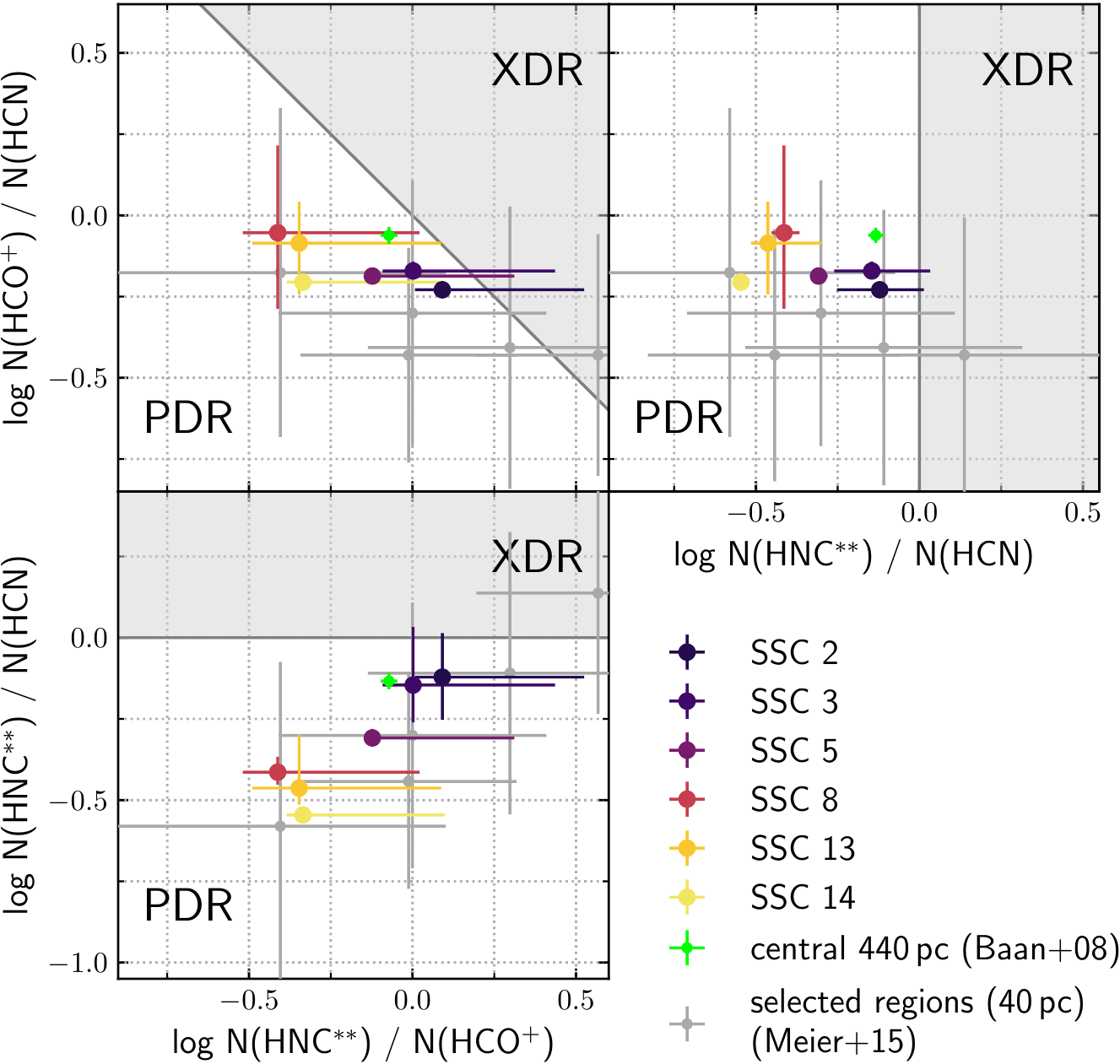}
    \caption{PDR--XDR chart according to \citet{Loenen:2008fb} and \citet{Baan:2008hx} for ratios of HCN, HCO$^+$ and HNC column densities. In our observations, we do not observe H$^{14}$NC but H$^{15}$NC, so we estimate H$^{14}$NC using the isotope ratio $^{14}$N/$^{15}$N derived from H$^{14}$NC and H$^{15}$NC of each SSC. In the plot, this is marked by $^{**}$ in the label. The correction is possible for only 6 out of 14 SSCs due to the detection rate of H$^{15}$NC. The \citet{Baan:2008hx} estimate for the central region (26\arcsec, 440\,pc) of NGC~253 extends over all SSCs and surrounding gas. The 40\,pc (2.4\arcsec) regions of \citet{2015ApJ...801...63M} focus on selected positions within NGC~253 and also cover the SSCs. However, they are derived from the respective (1--0) transitions.
    }
    \label{figure: XDR PDR column densities}
\end{figure*}

\citet{Loenen:2008fb} and \citet{Baan:2008hx} proposed diagnostic diagrams to infer the excitation environment in the nuclear regions of galaxies. They list different regimes for line ratios and column density ratios of HCN, HCO$^+$, HNC and CS related to PDR/XDR chemistry. These are based on PDR/XDR models by \citet{2005A&A...436..397M} and \citet{2006ApJ...650L.103M,2007A&A...461..793M} with the assumption of steady-state chemistry. 

We use these diagnostics to infer the probable excitation mechanisms in the SSCs. Figure~\ref{figure: XDR PDR column densities} shows the relevant column density ratios of HCN, HCO$^+$ and HNC. H$^{14}$NC (4--3) at 362.63\,GHz falls outside our observed band but the we can correct the detected H$^{15}$NC using the $^{14}$N/$^{15}$N isotope ratio obtained from HCN and HC$^{15}$N.
The SSCs where all necessary lines are detected (SSCs~2, 3, 5, 8, 13, 14) all fall in the PDR regime of the model.
SSCs~2 and 3 are barely compatible with XDR conditions within $1\sigma$ errors. 
For PDRs, the HCO$^+$/HCN ratio can allow to differentiate density regimes \citep{Loenen:2008fb} where the measured $log \left( \mathrm{HCO}^+/\mathrm{HCN} \right)\sim 0$ indicates volume density $n \sim 10^5$\,\pcm3. \citet{Baan:2008hx} further utilize CS/HCN as a column density tracer with CS/HCN$\geq1$ at $N_H \gtrsim 10^{22}$\,\pcm2 and CS/HCN$\leq1$ for lower column densities. In the SSCs~2, 3, 5, 8, 13, 14 that we can relate to PDRs, the line ratio CS/HCN$<1$ in all cases implying high column densities $>10^{22}$\,\pcm2 as is expected for this high-mass, dense environment. 
The H$_2$ column density derived from the CO intensity in Table~\ref{table: intensities} \citep[using $X_\mathrm{CO} = 0.5 \times 10^{20}$\,\pcm2\,(K\,\kms)$^{-1}$;][]{2001ApJS..135..183P,2015ApJ...801...25L} is indeed between $\sim 5.1 \times 10^{22}$\,\pcm2 (SSC~5) and $\sim 2.7 \times 10^{23}$\,\pcm2 (SSC~14).

Figure~\ref{figure: XDR PDR column densities} also shows the ratios obtained by \citet{Baan:2008hx} for the (3--2) transitions of the involved species. Their results from single dish observations cover the central 26\arcsec ($\sim 440$\,pc) and thus average all SSCs and further nuclear molecular gas. 
Reassuringly, their position for NGC~253 in the diagnostic diagrams is close to the average of our detected SSCs. The $\lesssim 0.2$\,dex mismatch is not significant but, if real, most likely caused by the nuclear gas outside the SSCs.
Our measurements are furthermore consistent with \citet{2015ApJ...801...63M} who report the (1--0) transitions of HCN, HNC, HCO$^+$ for 10 selected regions of 2.4\arcsec ($\sim40$\,pc). Since they, too, do not observe HNC directly, we infer the HNC column density and line intensity from HN$^{13}$C and the $^{12/13}$C ratio obtained from H$^{12}$CN and H$^{13}$CN. Note that \citet{2015ApJ...801...63M} estimate their column densities assuming optically thin LTE conditions whereas we consider optical depth.

As a consistency check, we also construct Figure~\ref{figure: XDR PDR column densities} with observational (instead of modelled) data using the line intensity ratios and find the the same qualitative result of chemistry consistent with a PDR environment. Details are discussed in Appendix~\ref{appendix: energy source intensity}.

Finding PDR chemistry in the SSCs on parsec scales is consistent with the large scale study by \citet{2009ApJ...706.1323M} who confirm a photo-dominated chemistry for their detected typical PDR tracers the central $400-500$\,pc of NGC~253.

Obviously, the two PDR/XDR models discussed in Section~\ref{section: CS/HCN} and here cannot simultaneously apply as they attribute different tracer properties to CS/HCN. However, it is reassuring that both models consistently favor PDR conditions in all SSCs.
For embedded (proto-)SSCs, PDR-dominated chemistry is expected although large amounts of young O/B-stars could potentially emit enough short-wavelength radiation to create a mild XDR.
As mentioned before, SSCs~$8-12$ are located close ($<0.5$\arcsec, $<8.5$\,pc) in projection to the stellar kinematic center \citep[$\alpha, \delta = 00^h47^m13.179^s, -25^\circ17^\prime17.13\arcsec$][]{MullerSanchez:2010dr} and the continuum source TH2 \citep[$\alpha,\delta = 00^h 47^m 31.2^s, -25^\circ 17^\prime 17\arcsec$;][]{1997ApJ...488..621U} proposed to be an AGN. At this location, they could be influenced by a potential AGN that may exhibit XDR properties. NGC~253 hosts a SMBH of $\sim 8 \times 10^6$\,\Msun \citep{2006ApJ...644..914R,2014ApJ...789..124D} but shows no clear sign of even a low luminosity AGN \citep{MullerSanchez:2010dr,Gunthardt:2015ba}. XDR chemistry in nearby SSCs could provide hints towards a low luminosity, obscured AGN. As discussed above, both models at hand, however, exclude this possibility.

Finally, it must be noted that our finding that all SSCs (where possible) are consistent with PDR chemistry suggests that X-rays are not the main source of energy input. It does not imply the inverse, that UV heating is the dominant energy input. Mechanical heating from e.g. shocks associated with accretion or outflows can dominate the energy budget while UV radiation only contributes.
In fact, \citet{2006ApJS..164..450M} detect typical PDR tracers in the central 200\,pc of NGC~253 but conclude that the heating is dominated by large-scale low velocity shocks as they show by comparing NGC~253 to other starbursts and the Galactic center.
This tension was solved by \citet[see their figure~10]{2015ApJ...801...63M} who could spatially separate energy sources in early ALMA observations at 3\arcsec ($\sim 50$\,pc) resolution. They draw a picture of an inner nuclear disk (where the SSCs are located) dominated by PDRs but with a contribution of wide-spread shocks. The inner nuclear disk is surrounded by an outer nuclear disk in which shocks dominate similar to bar shock regions in other galaxies. 
This picture is consistent with the new data presented here.

%%%%%%%%%%%%%%%%%%%%%%%%%%%%%%%%%%%%%%%%%%%%%%%%%%%%%%%%%%%%%%%%%%%%%%%%%%%%%%%%%%%%%%%%%%%%%%%%%%%%

\subsection{Isotopic ratios and optical depth}\label{section: optical depth}

Table~\ref{table: ratios other} lists line ratios of the HCN isotopologues H$^{13}$CN and HC$^{15}$N with HCN. As the isotopologues are less abundant and therefore less bright, so less optically thick, such a line ratio can act as a proxy for the optical depth.
Although \xclass fits for opacity, we here follow the classic approach and estimate it from line ratios. We find that the opacities fitted by \xclass are often highly uncertain and strongly influenced by e.g. noise spikes or (partially) blended lines. Therefore, classical line ratios provide a more robust estimate of opacities.

We detect HCN/H$^{13}$CN with a large spread of $\sim$ one order of magnitude in the range $3.2-66$.
In HCN/HC$^{15}$N the ratios scatter less between $3.0-11.8$. 
Both line ratios show a similar pattern with low ratios in SSCs~13, 14, intermediate ratios in SSCs~3 and high ratios in SSC~2, 4. In SSCs~5,8, the HCN/H$^{13}$CN ratios are significantly lower than HCN/HC$^{15}$N considering the relative range of ratios observed.

Our HCN/H$^{13}$CN ratios scatter around the respective large scale ratios in the center of NGC~253 with a bias towards lower ratios. \citet{JimenezDonaire:2017eg} found HCN/H$^{13}$CN = $17 \pm 1$ and \citet{2015ApJ...801...63M} obtained HCN/H$^{13}$CN = $10-15$. Both studies cover the entire center of NGC~253 at 170\,pc (single aperture) and 35\,pc (probing 10 regions) resolution, respectively, whereas we focus on the SSCs at parsec scales.

In the simplest case where HCN is optically thick, H$^{13}$CN optically thin and the gas is perfectly mixed at a constant density, the optical depth
\begin{equation}
    \tau^{^{13}C} = T_b^{^{13}\mathrm{C}}/T_b^{^{12}\mathrm{C}}
    \label{equation: tau thin}
\end{equation}
depends on the observed intensity ratio and 
\begin{equation}
    \tau^{^{12}\mathrm{C}} = \tau^{^{13}\mathrm{C}} \times [^{12}\mathrm{C}]/[^{13}\mathrm{C}]
    \label{equation: tau thick}
\end{equation}
follow from the abundance ratio \citep[e.g.,][]{JimenezDonaire:2017eg}. Further assumptions are equal beam filling factors and a common excitation temperature for both isotopologues. The same arguments apply to HC$^{15}$N and S$^{18}$O. This simple model of optically thick main species and optically thin isotopologues is often assumed in nearby galaxies. Given the dense and compact SSCs, it provides an estimate of optical depth but may not be exactly valid.

For the abundance ratios $^{12/13}$C = [$^{12}$C]/[$^{13}$C], $^{14/15}$N = [$^{14}$N]/[$^{15}$N] and $^{16/18}$O = [$^{16}$O]/[$^{18}$O] we turn to the literature to break the degeneracy of optical depth and abundance ratio.
We adopt $^{12/13}$C = $40\pm20$ \citep[among nearby galaxies including NGC~253,][]{2014A&A...565A...3H,2010A&A...522A..62M,2019A&A...624A.125M,2019arXiv190606638T},
$^{14/15}$N = $200\pm1$\,dex \citep[and references therein, see table~1 for an overview]{2019MNRAS.486.4805V} and
$^{16/18}$O = $130\pm40$ \citep[NGC~253;][]{2019A&A...624A.125M}.

With the arguments above we obtain $\tau_\mathrm{H^{13}CN} = 0.05 - 0.3$ and $\tau_\mathrm{HCN} = 2.0 - 12$.
SSCs 1, 3, 5, 8, 9, 11, 13 and 14 would have quite high optical depths in the $^{12}$C ($\tau_\mathrm{H^{12}CN}>5$) as well as the $^{13}$C lines ($\tau_\mathrm{H^{13}CN} \gtrsim 0.15$) while in the other SSCs $\tau_\mathrm{H^{12}CN} \sim 1-5$ are moderately thick. SSC~10 with $\tau_\mathrm{H^{12}CN} \sim 0.6$ is marginally optically thick.
For the $^{14/15}$N ratio, $\tau_\mathrm{HC^{15}N} = 0.08 - 0.3$ or $\tau_\mathrm{HC^{14}N} = 15-66$. 
In $^{16}$O and $^{18}$O derived from SO, we obtain $\tau_\mathrm{S^{18}O} < 0.1$ and $\tau_\mathrm{S^{16}O} \lesssim 15$.
A very low ratio in SSC~1 and corresponding high opacity ($\tau_\mathrm{S^{18}O} = 0.6$, $\tau_\mathrm{S^{16}O} = 78$) likely originates from confusion between S$^{18}$O and HC$_3$N lines.

In order to match the two estimates for $\tau_{\mathrm{HCN}}$, $^{12/13}$C on the high side and $^{14/15}$N on the low side of extra-galactic measurements would be required.
\citet{JimenezDonaire:2017eg} estimate the optical depth at $\tau_\mathrm{H^{12}CN} = 2.5$ and $\tau_\mathrm{H^{13}CN} = 0.07$ averaged over the whole center of NGC~253. The HCN opacities of SSCs~2, 4, 6 and 12 do not stand out above this large scale environment while the other SSCs show higher optical depths in HCN. As indicated above, this trend in optical depth correlates with cluster age (cf. Section~\ref{section: other lines}).

Given the in parts high optical depths it must be noted that different species do not have to originate from the same location in the SSC cloud. If the gas density is not constant, emission from the more optically thick line originates from outer layers of the cloud. Chemical variations may thus be reflected in one line but not the other in all line ratios of this study. For instance, chemical enrichment by stellar winds of young stars within the forming cluster may affect only the inner regions of the SSC parent cloud.

%%%%%%%%%%%%%%%%%%%%%%%%%%%%%%%%%%%%%%%%%%%%%%%%%%%%%%%%%%%%%%%%%%%%%%%%%%%%%%%%%%%%%%%%%%%%%%%%%%%%

\subsection{Other lines}\label{section: other lines}

We detect a multitude of molecular species and spectral lines in these deep observations. In the following section, we focus on a few of them.

%%%%%%%%%%%%%%%%%%%%%%%%%%%%%%%%%%%%%%%%%%%%%%%%%%%%%%%%%%%%%%%%%%%%%%%%%%%%%%%%%%%%%%%%%%%%%%%%%%%%

\subsubsection[HC3N]{HC$_3$N}\label{section: HC3N}

HC$_3$N is a tracer for warm and dense gas \citep[e.g.][]{2018ApJS..236...40T}. It has multiple bending modes ($\nu_5$, $\nu_6$, $\nu_7$) at IR frequencies which makes it sensitive to a strong IR field and allows HC$_3$N to act as a hot dust tracer when IR cannot be observed directly due to extinction \citep{2020MNRAS.491.4573R}. UV radiation and CRs can easily destroy the molecule, so it traces shielded IR irradiated dense gas \citep[e.g.][]{2010A&A...515A..71C}. HC$_3$N abundances are enhanced in hot environments due to evaporation from dust grain mantles. Galactic detections are thus typically in hot cores \citep[e.g. Sgr~B2][]{2000A&A...361.1058D} but it was also detected in other galaxies, such as NGC~253, NGC~4418 or IC~342 \citep[e.g.][]{2011A&A...528A..30C,2011A&A...525A..89A,2005ApJ...618..259M,2007A&A...475..479A,2011AJ....142...32M}. Most often the 10--9 transition at $\sim 90$\,GHz is detected \citep[e.g.][]{2011A&A...528A..30C} but detections up to $J_{upper} >30$ are given in the literature as well as detection of the vibrational states \citep[$\nu_6$, $\nu_7$; e.g.][]{2010A&A...515A..71C,2011A&A...527A..36M,2011A&A...528A..30C}.

HC$_3$N lines are often the brightest, or at least among the brightest, lines after CO, HCN, HCO$^+$ and CS in our spectral window. This directly implies a high IR radiation field, because due to their high critical densities ($>10^8$\,\pcm3), HC$_3$N transitions cannot be purely collisionally excited; they also need to be pumped in the IR. Furthermore, we detect the (38--37) and (39--38) lines, so the temperatures must be high in order to excite these transitions (cf. Section~\ref{section: ISM temperature}). 
The ubiquitous, highly excited HC$_3$N and the observation that many SSCs contain PDRs (Section~\ref{section: dense gas}) pose constraints on the PDR or the dense gas geometry.
High UV fluxes in a PDR can quickly dissociate HC$_3$N. Therefore, the UV field illuminating the PDRs must be either weak, or the HC$_3$N is well shielded from the photodissociating radiation, or alternatively the HC$_3$N emission is spatially separate from the PDRs.
As discussed in Section~\ref{section: energy source}, the PDR vs. XDR discrimination only indicates the relative, not the absolute strength and does not cover mechanical heating as an energy source. In this context, the observation of bright HC$_3$N implies a weak PDR and dominant mechanical heating likely by gas accretion onto the forming SSCs and proto-stellar outflows.
On the other hand, if the HC$_3$N and the PDR gas traced by the HCN/HCO$^+$/HNC emission are spatially separated, the PDR can either be driven from the inside by the SSC in the center of the surrounding molecular cloud or from the outside by neighboring sources in an outer shell of the cloud. The first case implies a radially stratified model of onion-like layers with HCN/HCO$^+$/HNC in the inner irradiated region and an outer layer of HC$_3$N. In the second case, HC$_3$N is shielded inside the cloud from external radiation but HCN/HCO$^+$/HNC can be well mixed with HC$_3$N.
Without additional tracers of mechanical heating and cloud structure we currently cannot distinguish these possibilities.
At much lower resolution and for the central $400-500$\,pc, \citet{2009ApJ...706.1323M} show that the PDR tracers originate in the outer layers of UV-illuminated clouds, similar to the aforementioned onion-layer structure. Our study, however, focuses on particular sources at 200 times higher resolution which may not share this large scale structure.

The ratio of HCN over HC$_3$N might be interpreted as a ``super dense gas'' fraction or very high density to high density gas ratio. Such a ratio is, as all line ratios are, only meaningful if HCN and HC$_3$N originate from the same region. Under that assumption, HCN/HC$_3$N implies the highest fraction of very dense gas can be found in SSC~13 whereas SSC~10 would contain little very high density gas (c.f., Table \ref{table: ratios other}).
Increasing HCN/HC$_3$N ratios should occur when a molecular cloud gets disrupted by feedback as HC$_3$N is dissociated before HCN, and even earlier the favourable conditions for IR pumping are shut down as the self-shielding of the cloud diminishes. Towards the end of a molecular cloud lifetime increasing HCN/HC$_3$N should thus be an age tracer, and the evolutionary sequence would be $13,4,14,1,8,3,2$ from young to old. \citet{2020MNRAS.491.4573R} recently dated\footnote{Their method can only date SSCs with a significant fraction of proto-stellar contribution and thus age dating is possible only until the cluster reaches the zero age main sequence.} the super hot cores within the SSCs by the fraction of proto-stellar (inferred from the IR radiation field pumping the HC$_3$N vibrational transitions) to stellar luminosity (inferred from the ionizing luminosity), which results in an almost inverted sequence ($2,3,13,1,8,14,4$). This suggests that the HCN/HC$_3$N ratio is not a reliable age tracer, at least for the young, deeply embedded SSCs in this study. Potential causes of the mismatch are structural effects, i.e. HCN and HC$_3$N not originating from the same region, or all the SSCs are too young to cause detectable effects on the HCN/HC$_3$N ratio. The latter is supported by the fact that the oldest SSCs (zero age main sequence SSCs $6,7,9,10,11,12$, according to \citealt{2020MNRAS.491.4573R}) are consistently found at higher HCN/HC$_3$N than the younger proto-SSCs ($1,2,3,4,8,13,14$) by factors of $\sim1-4$.

HCN/HC$_3$N ratios of $3-4$ are commonly reported in nearby galaxies on scales of hundreds of pc for low-J and high-J HC$_3$N lines \citep[NGC~4418, IC~342;][]{2005ApJ...618..259M,2007A&A...475..479A,2010ApJ...725L.228S,2011AJ....142...32M}. Apart from SSC~13, the other SSCs show considerably larger ratios than this nearby galaxy average.
In the center of NGC~253, HCN/HC$_3$N$\sim10$ has been measured by \citet{Lindberg:2011kk} over 25\arcsec (425\,pc). Hence, HC$_3$N/HCN in the SSCs does deviate in both directions by factors of a few from the large scale average.

The detection of vibrationally excited HC$_3$N is among the first extragalactic of such detections \citep{2010A&A...515A..71C,2011A&A...527A..36M,2011A&A...528A..30C}. It is especially noteworthy that we detect high-J vibrationally excited lines which requires high temperatures and a strong IR field. With these observations, we can confirm that HC$_3$N is associated with dense starforming gas and a starbursting environment.

%%%%%%%%%%%%%%%%%%%%%%%%%%%%%%%%%%%%%%%%%%%%%%%%%%%%%%%%%%%%%%%%%%%%%%%%%%%%%%%%%%%%%%%%%%%%%%%%%%%%

\subsubsection{Sulfur chemistry}\label{section: sulfur chemistry}

Chemical studies suggest that the fractional abundance of CS is sensitive to both the abundances of sulfur and oxygen \citep{1982ApJS...48..321G}. SO$_2$ is related to turbulent gas near stellar activity \citep{Minh:2016dx}. In undisturbed gas, sulfur is thought to be depleted onto ice grain mantles and sulfur-bearing species may act as chemical clocks in the evolution of SF. SO and SO$_2$ form from grain-evaporated H$_2$S and abundances increase with time until at later times most of the sulfur is captured in CS, H$_2$CS and OCS \citep{1998A&A...338..713H}. The SO/SO$_2$ ratio may act as a crude clock with lower ratios towards later times \citep{1997ApJ...481..396C}. SO$_2$ is also used as a tracer for low velocity outflows in stellar cores \citep{Wright:1996hi,Liu:2012ex}.

SO/SO$_2$ and CS/SO$_2$ line ratios are given in Table~\ref{table: ratios other} for the ground state of SO, \cs and SO$_2$ $11_{4,8}-11_{3,9}$.
The SO/SO$_2$ ratios varies by a factor of $\sim5$ across SSCs and $\sim 7$ in CS/SO$_2$. Both ratios follow the same trend with low ratios in SSCs~1, 3 and 4, and highest ratios in SSCs~5, 8 and 11. If SO/SO$_2$ relates to age, the former are older while the latter are younger. This sequence does not correlate with the HC$_3$N age dating and \citet{2020MNRAS.491.4573R}.

\citet{2005ApJ...620..210M} studied the sulfur chemistry in NGC~253 at $200-500$\,pc resolution which provides an average value for all SSCs and surrounding gas. They find fractional abundances CS/SO$_2$ = 5 and SO$_2$/SO = 1 which is considerably lower than our average line ratios over the detected SSCs of CS/SO$_2$ = 9.2 and SO$_2$/SO = 3.1. It must be noted that \citet{2005ApJ...620..210M} observed less excited gas with transitions up to CS(5--4), SO($4_3-3_2$) and SO$_2$($8_{2,6}-8_{1,7}$) whereas we cover transitions above those and up to CS(7--6), SO($8_8-7_7$) and SO$_2$($17_{4,14}-17_{3,15}$). Therefore, excitation effects may shift the line ratios. Given the uncertainties, the enhanced ratios in the SSCs could be explained by a factor of two depletion of SO$_2$ or, more likely, factors of $2-3$ enhancement of SO and CS relative to the large scale average from \citet{2005ApJ...620..210M}.

%%%%%%%%%%%%%%%%%%%%%%%%%%%%%%%%%%%%%%%%%%%%%%%%%%%%%%%%%%%%%%%%%%%%%%%%%%%%%%%%%%%%%%%%%%%%%%%%%%%%

\subsubsection{Vibrationally excited species}\label{section: vibrational excitation}

The vibrationally excited lines of HCN and HC$_3$N have high critical densities (e.g. HCN $\nu_2=1, l=1f$ $n_\mathrm{crit}>10^{10}$\,\pcm3) which makes purely collisional excitation unlikely. Instead radiative excitation, particularly in the IR, is required (e.g. at 14\,\mum for HCN $\nu_2=1, l=1f$, \citealt{1986ApJ...300L..19Z}). These vibrational lines have been frequently observed towards nuclei of nearby (U)LIRGs \citep{2010ApJ...725L.228S,2013ApJ...764...42S,2015A&A...584A..42A,2016ApJ...825...44I,2017ApJ...849...29I,2018A&A...609A..75F} but also in Galactic hot cores \citep{2011A&A...529A..76R,2011A&A...527A..68R,2011A&A...536A..33R,2015A&A...584A..42A}. In buried galactic nuclei powered by AGNs or starbursts, the dust emission can become optically thick and very effectively pump vibrationally excited HCN and HC$_3$N lines similar to a greenhouse \citep[e.g.][]{2019ApJ...882..153G}. However, no AGN, not even a low luminosity AGN, has been found in NGC~253 yet \citep{MullerSanchez:2010dr,Gunthardt:2015ba}.

\floattable
\begin{deluxetable*}{r|cccc}
    \tablecaption{Integrated intensity ro-vibrational over rotational excitation fraction for selected HCN and HC$_3$N lines. The ro-vibrational line used for the respective ratio is given in the second row.\label{table: ratios vibrational excitation}}
    \tablehead{\colhead{SSC} & \colhead{HCN/HCN*} & \colhead{HCN/HCN*} & \colhead{HC$_3$N/HC$_3$N*} & \colhead{HC$_3$N/HC$_3$N*}\\
    \colhead{} & \colhead{$\nu_2=1$, $l=1f$} & \colhead{$\nu_2=2$, $l=2f$} & \colhead{$\nu_7=1$, $l=1f$} & \colhead{$\nu_7=2$, $l=2f$}}
    \startdata
1 & $3.7^{+0.5}_{-0.3}$ & $30.1^{+15.4}_{-9.9}$ & ... & ...\\
2 & $5.7^{+0.3}_{-0.4}$ & $28.6^{+3.9}_{-3.4}$ & $1.6^{+0.2}_{-0.3}$ & $2.1^{+0.3}_{-0.4}$\\
3 & $3.1^{+0.3}_{-0.2}$ & $24.2^{+8.6}_{-4.8}$ & $1.1^{+0.2}_{-0.1}$ & $3.1^{+0.9}_{-0.6}$\\
4 & $2.9^{+0.7}_{-1.2}$ & $8.4^{+2.4}_{-3.3}$ & $2.9^{+0.8}_{-0.5}$ & $3.1^{+0.8}_{-0.5}$\\
5 & $7.9^{+0.1}_{-0.2}$ & $81.4^{+13.6}_{-10.5}$ & $2.2^{+0.1}_{-0.2}$ & $5.1^{+1.0}_{-0.7}$\\
6 & ... & ... & ... & ...\\
7 & ... & ... & ... & ...\\
8 & $2.7^{+1.8}_{-1.8}$ & $58.7^{+50.6}_{-38.4}$ & $2.2^{+1.0}_{-0.3}$ & ...\\
9 & ... & ... & ... & ...\\
10 & ... & ... & ... & ...\\
11 & $16.2^{+9.3}_{-7.5}$ & ... & $6.1^{+1.4}_{-0.7}$ & ...\\
12 & ... & ... & ... & ...\\
13 & $1.4^{+0.2}_{-0.2}$ & $9.7^{+1.6}_{-1.5}$ & $1.4^{+0.1}_{-0.1}$ & $2.2^{+0.2}_{-0.2}$\\
14 & $2.0^{+0.1}_{-0.0}$ & $17.0^{+2.4}_{-2.9}$ & $1.9^{+0.1}_{-0.1}$ & $4.8^{+0.2}_{-0.2}$\\
    \enddata
\end{deluxetable*}

For HCN and HC$_3$N, we detect vibrationally excited lines in 6 out of 14 SSCs for HCN$^*$ ($\nu_2=1$, $\nu_2=2$) and 8/14 ($\nu_6=1$), 8/14 ($\nu_7=1$), 6/14 ($\nu_7=2$) in HC$_3$N$^*$ (cf. Table~\ref{table: intensities}), so we can estimate the fraction of ro-vibrational vs. purely rotational excitation.
The ro-vibrational lines are very crowded, however, and blend into compact blocks of emission in some SSCs. All vibrational bending modes ($f$ and $e$) for each $\nu$ level are detected with one line only which makes it difficult to disentangle the spectra. The theoretical relationship between vibrational modes within the same ro-vibrational species as implemented in \xclass constrains the relative line strengths but systematic effects might remain.
Table~\ref{table: ratios vibrational excitation} lists the rotational over ro-vibrational line ratios of the $f$ bending modes.
In \hcn $\nu_2=1$, $l=1f$ we find line ratios of HCN/HCN$^*$ typically in the range $\sim 1.5-8$ where detected. This corresponds to vibrationally excited rotational state fractions of $\sim 10-40$\%. The higher value of $\sim 16$ in SSC~11 indicates close to negligible relative importance of vibrational excitation of only 5\%. The higher \hcn $\nu_2=2$, $l=1f$ state is negligible relative to HCN and \hcn $\nu_2=1$, $l=1f$ according to the high intensity ratios of $>8$.

For HC$_3$N vibrational excitation is a more important mechanism relative to purely rotational excitation: where detected, the line ratios between purely rotational and vibrationally excited variant are low at typically $\sim 1 - 3$ corresponding to fractional contributions of the vibrationally excited lines of $25\%-50\%$.
These high fractions of emission in vibrationally excited lines hints at strong IR (around $\sim 14$\,\mum) fields in the SSCs. Such an IR environment can be expected for forming SSCs as they are still deeply embedded \Leroy{p} and undetected in optical and near-IR \citep{2017ApJ...835..265W}. Emission by the forming stars is trapped inside the SSC by high opacity without leaking outside yet. \Leroy{t} argue for high optical depths in the IR due to their detection of 350\,GHz emission at $\tau \sim 0.1$ which implies $\tau = 5-10$ at the peak of the dust SED ($20-30$\,\mum). Such opacities are more than enough to create the aforementioned conditions for 14\,\mum trapping and IR pumping.

\citet{2020MNRAS.491.4573R} recently investigated the super hot cores in the SSCs using HC$_3$N. Their spectral window partially overlaps with our setup but we detect further vibrationally excited HC$_3$N lines as listed in Table~\ref{table: intensities}.
From our detections, we can confirm their observation that SSCs~6 and 7 lack HC$_3$N even in the ground state and SSCs~9,10 and 12 do not show vibrationally excited HC$_3$N. However, we detect two HC$_3$N$^*$ species in SSC~11. Qualitatively, our HC$_3$N detection rates confirm the conclusion of \citet{2020MNRAS.491.4573R} that SSCs~6, 7, 9, 10, 11 and 12 are older than SSCs~1, 2, 3, 4, 5, 8, 13 and 14.

%%%%%%%%%%%%%%%%%%%%%%%%%%%%%%%%%%%%%%%%%%%%%%%%%%%%%%%%%%%%%%%%%%%%%%%%%%%%%%%%%%%%%%%%%%%%%%%%%%%%

\subsection{ISM temperature}\label{section: ISM temperature}

\floattable
\begin{deluxetable}{rcccc}
    \tablecaption{Excitation temperatures obtained by \xclass fitting.\label{table: temperatures}}
    \tablehead{\colhead{SSC} & \colhead{H$_2$CS} & \colhead{SO$_2$}}
    \startdata
 1 &                 ... &  84$^{+ 25}_{- 14}$\\
 2 & 103$^{+ 60}_{- 23}$ & 114$^{+ 11}_{- 14}$\\
 3 &              $>242$ &  91$^{+ 13}_{-  7}$\\
 4 &              $>160$ & 191$^{+ 35}_{- 32}$\\
 5 &              $>237$ & 134$^{+ 20}_{- 26}$\\
 6 &                 ... &                 ...\\
 7 &                 ... &                 ...\\
 8 & 248$^{+166}_{- 75}$ & 129$^{+ 11}_{- 23}$\\
 9 &                 ... &                 ...\\
10 &                 ... &  47$^{+ 72}_{- 11}$\\
11 &                 ... &                 ...\\
12 &                 ... &                 ...\\
13 &              $>226$ & 122$^{+ 25}_{- 21}$\\
14 & 141$^{+ 15}_{- 16}$ & 228$^{+ 15}_{- 20}$\\
    \enddata
\end{deluxetable}

\begin{figure*}
    \centering
    \includegraphics[width=0.65\linewidth]{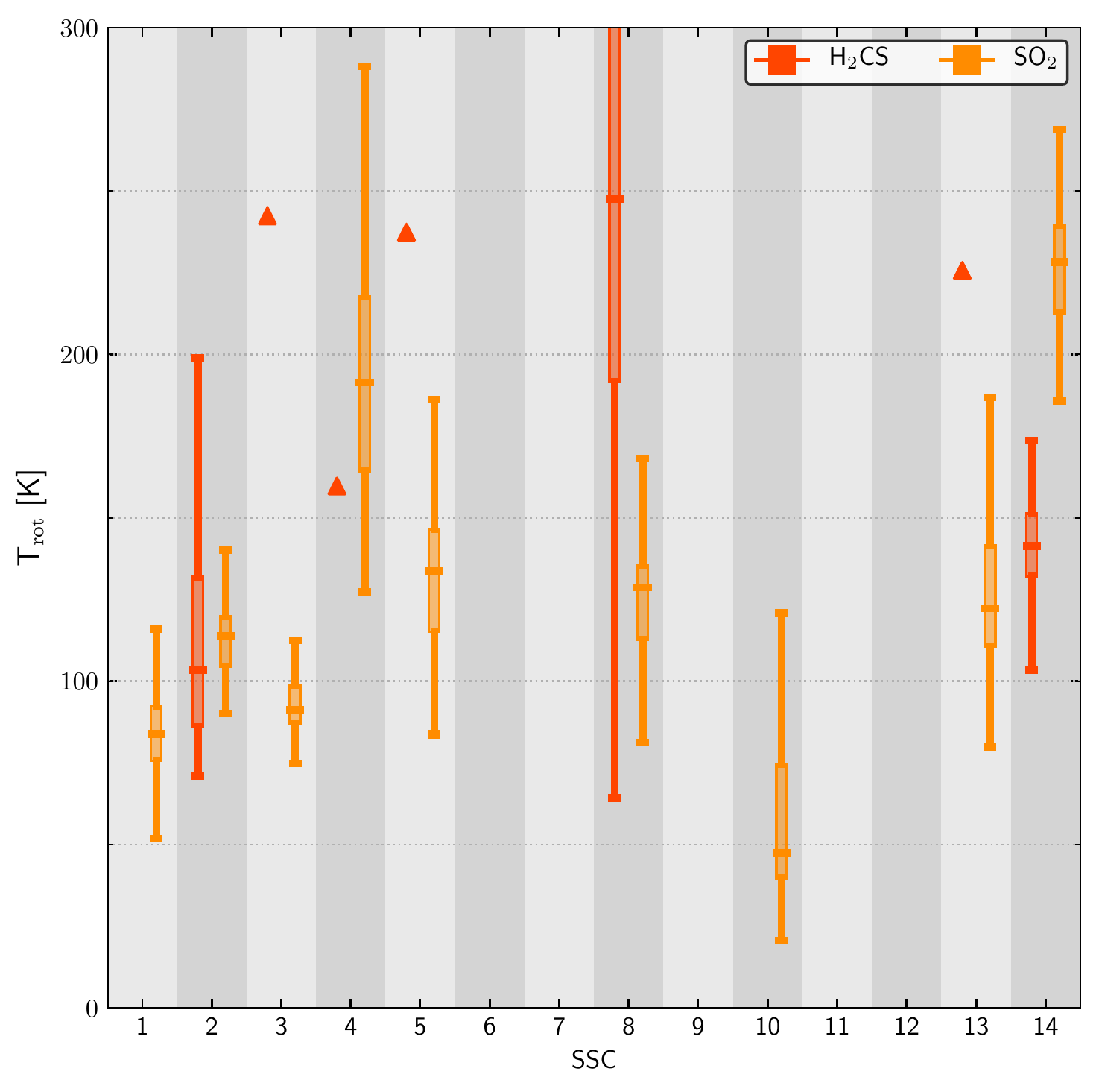}
    \caption{Comparison of the rotational temperatures derived from SO$_2$ and H$_2$CS. Boxplots show the distribution of the 100 fit iterations: a vertical line extends over the full range of values, the second and third quartile are represented by a box with the median as a horizontal line. Triangles represent lower limits (16$^{th}$ percentile) in the cases where the temperature is weakly constrained.}
    \label{figure: temperatures}
\end{figure*}

The rotational temperatures derived by \xclass for H$_2$CS and SO$_2$ are given in Table~\ref{table: temperatures} and also plotted in Figure~\ref{figure: temperatures} for comparison.

SO$_2$ is detected in only 11 out of 14 SSCs but confines the temperature to typically $\pm 20$\,K ($16^\mathrm{th}$ to $84^\mathrm{th}$ percentile) if detected.
In SSCs~11 and 12, the SO$_2$ lines are detected but do not allow for a robust temperature estimation as the SNR is too low.
The detected H$_2$CS lines only allow for a robust temperature estimation in SSCs~2,8,14 while in SSCs~3, 4, 5 and 13, $\mathrm{T_{rot}}$ it is weakly constrained and we report lower limits of the obtained XCLASS results.

The rotational temperatures inferred from SO$_2$ and H$_2$CS (where possible as discussed above) do only partially agree (SSCs~2, 4, 8). In SSCs~3, 5 and 13, the lower limits on H$_2$CS temperature are substantially higher than the inferred SO$_2$ temperature. In SSC~14, $\mathrm{T_{rot}}$ is $\sim 85$\,K higher in SO$_2$ than in H$_2$CS. It must be stressed here that further radiative transfer modelling is required to translate $\mathrm{T_{rot}}$ to kinetic gas temperature $\mathrm{T_{kin}}$ and allow for fair comparison across species. Rotational temperatures provide a lower limit to the kinetic gas temperature. 
For a physical interpretation, it needs to be considered that SO$_2$ is fitted with optical depths of 3 to $>5$ in all SSCs. All reported temperatures thus correspond to outer gas layers of an SSC rather than its core temperature which might be different due to internal heating (e.g. feedback).

The mean rotational temperature over the SSCs with good estimates is 127\,K in SO$_2$. The common excitation temperature of 130\,K assumed for all other rotational species is therefore a good common estimate.
The molecular gas in SSC~14 is most likely significantly hotter than in the other SSCs. This is in line with the rich chemistry (high species detection rate) and in most regards more extreme nature of this source. Similarly, SSC~10, the coolest SSC in the sample is among the faintest SSCs in most lines and has low line detection rates. SSC~4 shows surprisingly high temperatures given that it is average in all other quantities in this study and \Leroy{t}. The other SSCs are similar in temperature at $\sim 90$\,K (SSC~1, 3) or $\sim 115-130$\,K (SSC~2, 5, 8, 13).

In the Central Molecular Zone (CMZ) of the Milky Way, an environment similar to the center of NGC~253, \citet{2018ApJS..236...40T} found a strong positive correlation of HC$_3$N (10--9) with temperature. If this correlation extends to high-J lines of HC$_3$N, it would imply a high temperature in SSC~14, somewhat elevated temperatures in SSC~5 and 13, but lower temperatures in SSCs 1, 2, 3, 4, 8, 9, 10, 11 and 12. For SSC~6 and 7, neither of the two HC$_3$N lines are detected.
Such an HC$_3$N -- temperature correlation is not present in our data. SSC~14 is indeed hot and bright in HC$_3$N, but there are no elevated temperatures in SSC~5 and 13. The high temperature in SSC~4 is matched by HC$_3$N line brightness and column density marginally different from SSC~2 or 3 with much lower temperatures.
Hence, a potential HC$_3$N -- temperature correlation does not extend to the high-J lines ($\mathrm{J_u}=38$ with $\mathrm{T_{ex}}=307$\,K and $\mathrm{J_u}=39$ with $\mathrm{T_{ex}}=323$\,K) that we observed here.

Our results of an average $\mathrm{T_{rot}}\sim 130$\,K is consistent with \citet{2020MNRAS.491.4573R}, the only other temperature measurement on cluster scales, who find HC$_3$N rotational temperatures of $107\pm22$\,K to $125\pm45$\,K and dust temperatures of $200-375$\,K. Considering that $\mathrm{T_{rot}}$ is a lower limit to $\mathrm{T_{kin}}$, the molecular gas is at similar temperatures to the dust and probably thermally coupled.

%%%%%%%%%%%%%%%%%%%%%%%%%%%%%%%%%%%%%%%%%%%%%%%%%%%%%%%%%%%%%%%%%%%%%%%%%%%%%%%%%%%%%%%%%%%%%%%%%%%%

\section{Summary}\label{section: summary}

We present high-resolution ALMA observations of the SSCs in the starbursting center of NGC~253. Our spectral setup in band~7 covers a wealth of molecular species and pushes at resolving the compact (proto-) super star clusters with 2.5\,pc spatial resolution. In spectra focused on the SSCs, we detect up to 55 lines of 19 species. Modelling the spectra with \xclass allows us to independently study observation-based line ratios and modelled physical quantities.

The SSCs differ significantly in chemical complexity between 4 and 19 detected species. In CO, HCN, HCO$^+$ and CS, we detect multiple components and potential signs of self-absorption in HCN and HCO$^+$ in four SSCs. The other species are associated in velocity with the same CO/HCN/HCO$^+$/CS component or the absorption component of present.

The line ratios CO/HCN, CO/HCO$^+$ of $\sim 1-10$ are low implying high dense gas fractions. HCN/HCO$^+$ is consistent with unity in all but one clusters and most likely caused by high opacity. CS/HCN scatters significantly across SSCs and does not correlate with other properties aside from the number of detected species. Its tracer properties remain unclear.

All SSCs favor PDR chemistry over XDR chemistry as indicated by combinations of the line ratios of HCN, HCO$^+$, HNC and CS in comparison to models \citep{Loenen:2008fb,Baan:2008hx}. According to these models, our data favors densities of $\sim 10^5$\,\pcm3. The SSCs close to the central SMBH at projected distances $<8.5$\,pc are inconsistent with XDR chemistry induced by a potential low luminosity AGN. NGC~253's putative AGN continues to be elusive.

Opacities derived from HCN and HC$^{13}$N fall in the high optical depth regime with $\tau \gtrsim 1$ to $\tau >10 $ in HCN and up to $\tau = 0.3$ in H$^{13}$CN and HC$^{15}$N.

We detect bright HC$_3$N in highly excited states in many SSCs which implies high IR radiation fields and gas temperatures. This is at odds with finding PDR chemistry as the UV flux in PDRs can dissociate HC$_3$N. Potential solutions to this discrepancy are that mechanical heating dominates the energy input over a weak UV field or detected HC$_3$N and HCN/HCO$^+$/HNC/CS emission originates from different locations in the cloud governed by opacity.

Vibrationally excited lines of HCN and HC$_3$N are frequently detected (in 6-8 of 14 SSCs) in our observations. The fraction of vibrational excitation of a rotational state can be significant in some SSCs at order $10-30$\%. The excitation of these lines is likely caused by strong IR radiation fields that are trapped by a greenhouse effect due to high continuum opacities.

The gas in the SSCs is hot as indicated by SO$_2$ rotational temperatures of $\sim 130$\,K on average.

The presented observations demonstrate the power of ALMA to zoom into some of the most actively starforming regions in the local universe.

%%%%%%%%%%%%%%%%%%%%%%%%%%%%%%%%%%%%%%%%%%%%%%%%%%%%%%%%%%%%%%%%%%%%%%%%%%%%%%%%%%%%%%%%%%%%%%%%%%%%

\software{CASA \citep{McMullin:2007tj}, astropy \citep{Collaboration:2013cd,Collaboration:2018ji}, \xclass \citep{2018ascl.soft10016M}}

\acknowledgements
The authors would like to thank the anonymous referee for a detailed report that helped to significantly improve this paper.

The work of AKL is partially supported by NASA ADAP grants NNX16AF48G and NNX17AF39G and National Science Foundation under grants No.~1615105, 1615109, and 1653300.
The work of EACM is supported by the National Science Foundation under grant No. AST-1813765.

This paper makes use of the following ALMA data: ADS/JAO.ALMA \#2015.1.00274.S. ALMA is a partnership of ESO (representing its member states), NSF (USA) and NINS (Japan), together with NRC (Canada), NSC and ASIAA (Taiwan), and KASI (Republic of Korea), in cooperation with the Republic of Chile. The Joint ALMA Observatory is operated by ESO, AUI/NRAO and NAOJ.
The National Radio Astronomy Observatory is a facility of the National Science Foundation operated under cooperative agreement by Associated Universities, Inc.

%%%%%%%%%%%%%%%%%%%%%%%%%%%%%%%%%%%%%%%%%%%%%%%%%%%%%%%%%%%%%%%%%%%%%%%%%%%%%%%%%%%%%%%%%%%%%%%%%%%%
\clearpage
\appendix
%%%%%%%%%%%%%%%%%%%%%%%%%%%%%%%%%%%%%%%%%%%%%%%%%%%%%%%%%%%%%%%%%%%%%%%%%%%%%%%%%%%%%%%%%%%%%%%%%%%%

\section{Details of \xclass fitting}\label{appendix: xclass}

\subsection{Handling of blended lines in the first fit run}
Joint fitting of multiple species with insufficient constraints increases the number of degrees of freedom to a point where the fitter does not converge reliably anymore. 
We therefore fit the species listed in Table~\ref{table: intensities} independently where possible or include potential blended lines in the fit if necessary. 
For CO, CS, HCN, HCO+, H2CS, H$^{13}$CN, HC$^{15}$N, H$^{15}$NC, SO, $^{33}$SO, SO$_2$ and HCN ($\nu_2=1$) completely independent fitting is possible with appropriately selected fit range. The other species HCN ($\nu_2=2$'), H$C_3$N ($\nu=0$'),
H$C_3$N ($\nu_6=1$'), H$C_3$N ($\nu_6=2$'), H$C_3$N ($\nu_7=1$'), H$C_3$N ($\nu_7=2$'), $^{34}$SO ($\nu=0$'), S$^{18}$O ($\nu=0$') and $^{34}$SO2 ($\nu=0$') must be fitted jointly with lines of other species.

\subsection{\xclass fit parameters}
\xclass models the spectra based on the molecular parameters of the species to be fitted and can directly solve for physical quantities such as rotational (vibrational) temperature and column density. Further fit parameters are linewidth and centroid of the line. As we work with single pixel spectra, we leave the additional \emph{source size parameter} fixed at unity. This assumes the source to completely fill the beam ($0.13\arcsec \times 0.17\arcsec$, $\sim 2.5$\,pc) as is indicated by the SSC sizes of $\sim 1.5-4$\,pc obtained by \Leroy{t}.

\xclass allows to fit for \emph{excitation temperature} even when only one line of a species is detected due to the effect on the line shape (e.g. flattening due to opacity). 
However, with a single transition the temperature cannot be well constrained and the results scatter wildly. The fitted excitation temperature in such a case strongly depends on the line shape that is easily influenced by random noise fluctuations.
For the species with only a single line detected, we therefore need to fix the temperature. In the case of multiple detected lines (SO$_2$, H$_2$CS), the temperature also scatters considerably and sometimes even provides unphysical results (e.g. higher than the molecular binding energy) when the fit fails to converge successfully. Hence, we fix the rotational temperature $\mathrm{T_{rot}} = 130$\,K which is the temperature of the warm ISM component found by \citet{2013ApJ...779...33M} and \citet{Gorski:2017es} at lower spatial resolution. Assuming this temperature keeps our analysis consistent with \Leroy{t} who also assumed 130\,K. The observed CO peak brightness temperature $\mathrm{T_b} = 60-130$\,K may act as a proxy for $\mathrm{T_{rot}}$ under certain assumptions (optically thick emission, beam filling factor unity). Since we measure $\mathrm{T_{rot}} \sim 130$\,K (Section~\ref{section: ISM temperature}), one of these assumptions is not met. The excitation temperature $\mathrm{T_{vib}}$ of vibrationally excited species is certainly higher but difficult to estimate. Line ratios of vibrational states with differing $\mathrm{E_{upper}}$ (or $\mathrm{E_{lower}}$) could place limits on $\mathrm{T_{vib}}$ but in many SSCs no vibrationally excited species are detected. We therefore use a common fixed excitation temperature of $\mathrm{T_{vib}} = 300$\,K. This value is supposedly on the lower side of the actual excitation temperatures and thus causes the column densities of the vibrationally excited states to be on the higher side. The observed emission intensity is influenced by temperature and column density because higher excitation and more emitting molecules provide stronger line emission. In SO$_2$, the most reliable temperature tracer in our sample, changes in temperature and column density are inversely correlated at ratios of $0.8-1.0$ in the SSCs with successful temperature estimation (cf. section~\ref{section: ISM temperature}). This means any under-/overestimation of the fixed temperatures by a factor $x$ directly translate to an $x$ times under-/overestimation in column density. A factor of 2 variation in the chosen excitation temperature ($65\,\mathrm{K} < \mathrm{T_{rot}} < 260\,\mathrm{K}$ and $150\,\mathrm{K} < \mathrm{T_{vib}} < 600\,\mathrm{K}$) is well plausible in the SSCs. Hence, the derived column densities should be understood with a systematic error of a factor of two.

We apply loose limits on column density ($10^{12} - 10^{25}$\,\pcm2), linewidth ($5-80$\,\kms) and line centroid ($-10 - 10$\,\kms relative to the first manual estimate in Section~\ref{section: spectra}). For the second run these are limited to the $16^\mathrm{th} - 84^\mathrm{th}$ percentile ranges of the first run.

\subsection{Fit algorithm}
\xclass offers a choice of algorithms that can be daisychained to allow for faster and more robust exploration of the parameter space depending on the dataset. For this dataset, the fitting generally works well and is robust against repetition of the fit and variations of the initial guesses. For some species in a few SSCs, it is necessary to adjust initial guesses or boundaries of the fit parameters to allow the algorithm to find a solution.
We use a combination of two algorithms in sequence to assure the solver finds the global minimum of the fit and then converges to this minimum. We achieve this by a combination of 50 iterations of the ``Genetic'' algorithm followed by 50 iterations of the ``Levenberg-Marquardt'' algorithm. For a detailed description of the fit algorithms, we refer to \citep{2018ascl.soft10016M}.

\subsection{Error estimation}
This \xclass fitting procedure reliably finds the best fit but does not estimate errors of the fit parameters. We therefore bootstrap the errors using a Monte-Carlo scheme: We draw 100 versions of Gaussian noise and add it to the observed spectra which are then fitted as described above. The added Gaussian noise is set up with standard deviation 0.46\,K, the measured RMS noise in the data. This scheme tests the robustness of the fit to noise fluctuations in the data and thus the statistical error of the fit. Systematic errors such as the flux uncertainty of $\lesssim5$\% for ALMA observations (ALMA Technical Handbook) apply additionally. Of the 100 fit variations plus a fit to the unaltered spectra, we report the median and $16^\mathrm{th}$ to $84^\mathrm{th}$ percentiles range\footnote{The range $16^\mathrm{th}$ to $84^\mathrm{th}$ percentiles corresponds to $-1\sigma$ to $+1\sigma$ for Gaussian distributions.} as best estimate and respective error margin for each parameter.

%%%%%%%%%%%%%%%%%%%%%%%%%%%%%%%%%%%%%%%%%%%%%%%%%%%%%%%%%%%%%%%%%%%%%%%%%%%%%%%%%%%%%%%%%%%%%%%%%%%%

\section{SSC energy source using line intensity ratios}\label{appendix: energy source intensity}

\begin{figure*}
    \centering
    \includegraphics[width=0.58\linewidth]{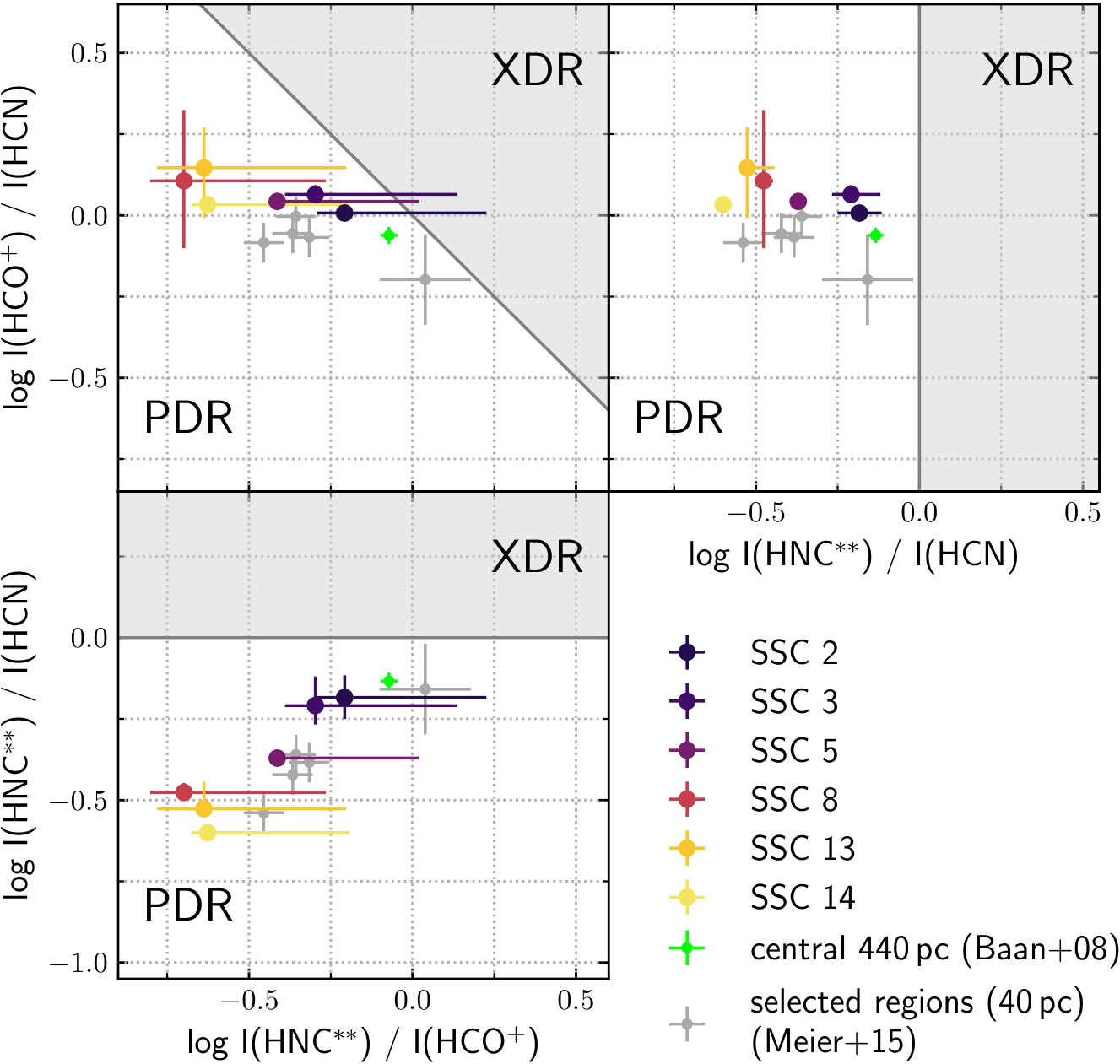}
    \caption{PDR--XDR chart according to \citet{Loenen:2008fb} and \citet{Baan:2008hx} for ratios of HCN, HCO$^+$ and HNC intensities. In our observations, we do not observe H$^{14}$NC but H$^{15}$NC which we estimate using the isotope ratio $^{14}$N/$^{15}$N from H$^{14}$NC and H$^{15}$NC. In the plots, this is marked by $^{**}$ in the labels. The correction is possible for only 6 out of 14 SSCs due to the detection rate of H$^{15}$NC. The \citet{Baan:2008hx} estimate for the central region of NGC~253 extends over all SSCs and surrounding gas.
    }
    \label{figure: XDR PDR intensity}
\end{figure*}

As discussed in Section~\ref{section: energy source}, the model by \citet{Loenen:2008fb} and \citet{Baan:2008hx} provides a tool to estimate the excitation environment in the SSCs which can then be interpreted for the potential energy sources.
Since we derive column densities with XCLASS, we can directly use physical quantities for this analysis instead of observational quantities. As a test, however, we also construct the ratio diagrams for line ratios in Figure~\ref{figure: XDR PDR intensity}. Note that in this case the correction of the observed H$^{15}$NC to H$^{14}$NC is only an approximation because we do not consider optical depth but only the observed line intensity ratios.
Nonetheless, we arrive at the same conclusion that the chemistry in the SSCs is powered by PDRs rather than XDRs. The exact placement in the ratio--ratio planes is slightly different, though, with deviations of $\sim 0.1-0.2$\,dex. SSCs~2 and 3 are not compatible with XDR conditions which was barely the case for the column density ratios (Figure~\ref{figure: XDR PDR column densities}).

The density estimation deviates between estimation from line ratios and column density ratios. The CS/HCN line ratios are consistently $<1$ suggesting $N_H < 10^{22}$\,\pcm2. This is in tension with the fact that the SSCs are undetected at IR wavelength and thus must be hidden behind large columns of gas and dust.
The assumption made by \citet{Baan:2008hx} that integrated line ratios have the same diagnostic value as column density ratios is thus not valid, at least in the case of NGC~253's SSCs.

As in Section~\ref{section: energy source}, we overplot the measurement from \citet[central 440\,pc]{Baan:2008hx} and the 10 selected regions from \citet[][40\,pc]{2015ApJ...801...63M}. As opposed to the column density ratios (Figure~\ref{figure: XDR PDR column densities}, the large scale measurement of \citet{Baan:2008hx} deviates from the SSCs towards higher HNC/HCO$^+$ and slightly lower HCO$^+$/HCN but is still close to the ratios in SSCs~2 and 3. Arguably, this is caused by the large amount of molecular gas outside the SSCs that we focus on.
The regions by \citep{2015ApJ...801...63M} are consistent with our measurements in HNC/HCN vs. HNC/HCO$^+$ but are offset towards lower HCO$^+$/HCN by $\sim 0.1$ as seen in the upper two panels. This is most likely a systematic effects in the measured HCN or HCO$^+$ intensities.

%%%%%%%%%%%%%%%%%%%%%%%%%%%%%%%%%%%%%%%%%%%%%%%%%%%%%%%%%%%%%%%%%%%%%%%%%%%%%%%%%%%%%%%%%%%%%%%%%%%%

\clearpage
\bibliographystyle{aasjournal}
\bibliography{bibliography.bib}

%%%%%%%%%%%%%%%%%%%%%%%%%%%%%%%%%%%%%%%%%%%%%%%%%%%%%%%%%%%%%%%%%%%%%%%%%%%%%%%%%%%%%%%%%%%%%%%%%%%%

\end{document}